\documentstyle[osa]{revtex}

\begin{document}
\title{Dirac-Schr\"odinger equation for quark-antiquark bound states and derivation
of its interaction kernel}
\author{Jun-Chen Su$^{1,2}$}
\maketitle
\address{1. Department of Physics, Harbin Institute of Technology, Harbin\\
150006,\\
People's Republic of China\\
2.Center for Theoretical Physics, Physics College, Jilin University,\\
Changchun 130023, People's Republic of China}

\begin{abstract}
The four-dimensional Dirac-Schr\"odinger equation satisfied by
quark-antiquark bound states is derived from Quantum Chromodynamics.
Different from the Bethe-Salpeter equation, the equation derived is a kind
of first-order differential equations of Schr\"odinger-type in the position
space. Especially, the interaction kernel in the equation is given by two
different closed expressions. One expression which contains only a few types
of Green's functions is derived with the aid of the equations of motion
satisfied by some kinds of Green's functions. Another expression which is
represented in terms of the quark, antiquark and gluon propagators and some
kinds of proper vertices is derived by means of the technique of irreducible
decomposition of Green's functions. The kernel derived not only can easily
be calculated by the perturbation method, but also provides a suitable basis
for nonperturbative investigations. Furthermore, it is shown that the
four-dimensinal Dirac-Schr\"odinger equation and its kernel can directly be
reduced to rigorous three-dimensional forms in the equal-time Lorentz frame
and the Dirac-Schr\"odinger equation can be reduced to an equivalent
Pauli-Schr\"odinger equation which is represented in the Pauli spinor space.
To show the applicability of the closed expressions derived and to
demonstrate the equivalence between the two different expressions of the
kernel, the t-channel and s-channel one gluon exchange kernels are chosen as
an example to show how they are derived from the closed expressions. In
addition, the connection of the Dirac-Schr\"odinger equation with the
Bethe-Salpeter equation is discussed.
\end{abstract}

\section{Introduction}

It is the common recognition that the Bethe-Salpeter (B-S) equation which
was proposed early in Refs.[1, 2] is a rigorous formalism for relativistic
bound states. The prominent features of the equation are: (1) The equation
is derived from the quantum field theory and hence set up on the firm
dynamical basis; (2) The interaction kernel in the equation contains all the
interactions taking place in the bound states and therefore the equation
provides a possibility of exactly solving the problem of relativistic bound
states; (3) The equation is elegantly formulated in a manifestly
Lorentz-covariant form in the Minkowski space which allows us to discuss the
equation in any coordinate frame. However, there are tremendous difficulties
in practical applications of the equation, particularly, for solving the
nuclear force in the nuclear physics and the quark confinement in hadron
physics. One of the difficulties arises from the fact that the kernel in the
equation was not given a closed form in the past. The kernel usually is
defined as a sum of B-S (two-particle) irreducible Feynman diagrams each of
which can only be individually determined by a perturbative calculation.
This definition is, certainly, not suitable to investigate the subjects such
as the nuclear force and the quark confinement which must necessarily be
solved by a nonperturbative method. This is why, as said in Ref. [3],'' The
Bethe-Salpeter equation has not led to a real breakthrough in our
understanding of the quark-quark force''. Opposite to the conventional
concept as commented in Ref. [4] that ''The kernel $K$ can not be given in
closed form expression'', we have derived a closed expression of the B-S
kernel for quark-antiquark bound states in a recent publication [5]. The
expression derived contains only a few types of Green's functions which not
only are easily calculated by the perturbation method, but also suitable to
be investigated by a certain nonperturbation approach. Another difficulty of
solving the B-S equation was attributed to the four-dimensional nature of
the equation because the relative time (or the relative energy) would lead
to unphysical solutions. Therefore, many efforts were made in the past to
recast the four-dimensional equation in three-dimensional ones in either
approximate manners or exact versions such as the instantaneous
approximation [6], the quasipotential approach [7-12] and the equal-time
formalism [13-16].

As one knows, the four-dimensionally covariant B-S equation for a
two-fermion system is ordinarily formulated in a second-order differential
equation with respect to the space-time variables in the position space.
This kind of equation has been shown to have unphysical solutions with the
negative norm. It was pointed out in Ref. [17] that'' The appearance of the
negative-norm B-S amplitude is a quite common phenomenon in the B-S
equation''. A similar phenomenon was encountered in the Klein-Gordon (K-G)
equation [18,19] 
\begin{equation}
(\Box _x+m^2)\psi (x)=0  \eqnum{1.1}
\end{equation}
which was originally viewed as the wave equation satisfied by the single
free fermion states. It is well-known that the K-G equation, as a
second-order differential equation, has a solution with negative
probability. This is because the wave function of the equation is determined
not only by its initial value $\psi (0)$, but also by the initial value of
the time-differential$\frac{\partial \psi }{\partial t}\mid _{t=0}$ which
would possibly cause the solution to have negative probability [18, 19].
Nevertheless, the Dirac equation [18, 19] 
\begin{equation}
(i{\bf \partial }_x-m)\psi (x)=0  \eqnum{1.2}
\end{equation}
where ${\bf \partial }_x{\bf =}\gamma ^\mu \partial _\mu ^x$ has not the
negative norm solution because it is a first-order differential equation. As
widely recognized, the Dirac equation gives a correct description of the
single free fermion states.

Analogous to the case of single fermion, the relativistic states for a
two-fermion system may also be formulated by a set of first-order
differential equations just as the Hamilton equation in Mechanics and the
Maxwell equation in Electrodynamics which are equivalent to the second-order
differential equations, i.e. the Lagrange equation and the D'Alembert
equation respectively. Motivated by this idea, it was proposed in the
literature [20-28] that the quark-antiquark bound system may be described by
two coupled Dirac equations which are constructed in accord with the Dirac's
Hamiltonian constraint formalism [29] such that [23-25] 
\begin{equation}
\lbrack i{\bf \partial }_{x_1}-m_1-V_1(x_1,x_2)]\psi (x_1,x_2)=0  \eqnum{1.3}
\end{equation}
\begin{equation}
(i{\bf \partial }_{x_2}-m_2-V_2(x_1,x_2)]\psi (x_1,x_2)=0  \eqnum{1.4}
\end{equation}
where $\psi (x_1,x_2)$ denotes the two-fermion wave function, $m_1$ and $m_2$
are the masses of quark and antiquark, $V_1$ and $V_2$ stand for the
effective potentials which are determined by the requirement of satisfying
the Lorentz-invariance, the charge conjugation symmetry and a certain
constraint (or say, compatibility) conditions. With a constraint imposed on
the relative time, the above equations will be reduced to a
three-dimensional eigenvalue equation.

As emphasized in the previous literature [23-25], Eqs. (1.3) and (1.4) are
built up within the framework of relativistic quantum mechanics and the
interaction potentials are given in a phenomenological way although they are
inspired by the quantum field theory and, as demonstrated in Ref. [23], are
linked with the corresponding B-S equation. Obviously, in order to
understand the Dirac-type equations for the two-fermion system more
precisely, it is necessary to give such equations an extensive investigation
and an exact formulation from the viewpoint of quantum field theory. This
just is the purpose of this paper. In this paper, we limit ourself to
discuss the quark and antiquark ($q\overline{q})$ bound states. The results
certainly suit to other two-fermion bound systems. First we derive two
first-order differential equations for the quark-antiquark bound states from
Quantum Chromodynamics (QCD) which describe the evolution of the bound state
with the total (center of mass) time and the relative time respectively.
These equations will be called Dirac-Schr\"odinger (D-S) equation because
the Dirac equation is, in essence, the Schr\"odinger equation in the
relativistic case which is identified with itself as the uniquely correct
equation of describing the evolution of a quantum state with time in the
quantum theory. Next, we concentrate our main attention on the interaction
kernel appearing in the D-S equation. We are devoted to deriving a closed
and explicit expression of the interaction kernel. The kernel will be
derived by two different methods: one is to utilize equations of motion
satisfied by the $q\overline{q}$ four-point Green's function and some other
four-point Green's functions in which the gluon field is involved; another
is to employ the technique of irreducible decomposition of the Green's
functions involved in the D-S equation. The first method is similar to that
proposed previously in Ref. [14]. The kernel derived by this method has a
compact expression which contains only a few types of Green's functions. The
kernel derived by the second method is expressed in terms of the quark,
antiquark and gluon propagators and some kinds of three, four and five-line
proper vertices and therefore exhibits a more specific structure of the
kernel. Especially, the kernel derived can not only be easily calculated by
the perturbation method, but also provides a suitable basis for
nonperturbative investigations. The D-S equation and its interaction kernel
mentioned above are Lorentz-covariant. We will show how this equation and
its kernel are reduced to the exact three-dimensional forms given previously
in Ref. [15] in the equal-time Lorentz frame. It is well-known that the D-S
equation is represented in the Dirac spinor space. This equation actually is
a coupled set of sixteen scalar equations. In practical applications,
sometimes it is more convenient to reduce the D-S equation to the Pauli
spinor space following the procedure proposed in Ref. [30]. By this
procedure, we will obtain an equivalent Pauli-Schr\"odinger (P-S) equation
represented in the Pauli spinor space from the D-S equation. In the P-S
equation, the interaction Hamiltonian is explicitly given in a series
expression which has an one-to-one correspondence with the perturbative
expansion of the S-matrix. To illustrate the applicability of the kernels
derived and the equivalence between the aforementioned two differerent
expressions of the kernel, we will show how the one-gluon exchange kernels
in the D-S equation and the corresponding interaction Hamiltonian in the P-S
equation can be derived from the closed expressions. Finally, we will
discuss the relation between the D-S equation and the corresponding B-S
equation.

The remainder of this paper is arranged as follows. In section 2, we will
first derive two Dirac-type equations satisfied by the $q\overline{q}$
four-point Green's function. From these equations, the D-S equations obeyed
by the B-S amplitudes will be derived by making use of the Lehmann
representation of the Green's function [31]. Then, we will show how the
four-dimensional D-S equation is reduced to the three-dimensional one. In
section 3, the first explicit expression of the interaction kernel in the
D-S equation will be derived by virtue of the equations of motion satisfied
by the Green's functions involved in the Dirac-like equations. And, it will
be shown how the closed expression of the exact three-dimensional kernel can
be written out from the four-dimensional one. In section 4, the second
expression of the interaction kernel will be derived by means of the
technique of irreducible decomposition of the Green's functions. In section
5, the D-S equation will be reduced to the corresponding P-S equation.
Section 6 will be used to give a brief derivation and description of the
one-gluon exchange kernels. The last section serve to discuss the relation
between the D-S equation and the corresponding B-S equation and to make some
remarks. In Appendix A, we will describe the derivation of the equations of
motion satisfied by the Green's functions which are necessary to be used in
the derivation of the D-S equation and its interaction kernel. In Appendix
B, the irreducible decomposition of the relevant Green's functions will be
performed for the purpose of deriving the second expression of the kernel.

\section{Derivation of the Dirac-Schr\"odinger equation}

The Dirac-Schr\"odinger (D-S) equation satisfied by the $q\overline{q}$
bound states may be derived from the corresponding equation for the $q%
\overline{q}$ four-point Green's function which is defined in the Heisenberg
picture as follows [32] 
\begin{equation}
{\cal G}(x_{1,}x_2;y_1,y_2)_{\alpha \beta \rho \sigma }=\langle 0^{+}\left|
T\{N[{\bf \psi }_\alpha (x_1){\bf \psi }_\beta ^c(x_2)]N[\overline{{\bf \psi 
}}_\rho (y_1)\overline{{\bf \psi }}_\sigma ^c(y_2)]\}\right| 0^{-}\rangle 
\eqnum{2.1}
\end{equation}
where ${\bf \psi }(x)$ and ${\bf \psi }^c(x)$ are the quark and antiquark
field operators respectively, $\overline{{\bf \psi }}(x)$ and $\overline{%
\text{ }{\bf \psi }}^c(x)$ are their corresponding Dirac conjugates [18] 
\begin{equation}
{\bf \psi }^c(x)=C\overline{{\bf \psi }}^T(x),\overline{{\bf \psi }}^c(x)=-%
{\bf \psi }^T(x)C^{-1}  \eqnum{2.2}
\end{equation}
here $C=i\gamma ^2\gamma ^0$ is the charge conjugation operator, $\mid
0^{\pm }\rangle $ denote the physical vacuum states, $T$ symbolizes the
time-ordering product and $N$ designates the normal product which is defined
by 
\begin{equation}
N[{\bf \psi }_\alpha (x_1){\bf \psi }_\beta ^c(x_2)]=T[{\bf \psi }_\alpha
(x_1){\bf \psi }_\beta ^c(x_2)]-\langle 0^{+}\left| T[{\bf \psi }_\alpha
(x_1){\bf \psi }_\beta ^c(x_2)]\right| 0^{-}\rangle  \eqnum{2.3}
\end{equation}
It is emphasized here that the above normal product can only be viewed as a
definition in the Heisenberg picture. With the definition shown in Eq.
(2.3), the Green's function in Eq. (2.1) may be represented as 
\begin{equation}
{\cal G}(x_{1,}x_2;y_1,y_2)_{\alpha \beta \rho \sigma
}=G(x_{1,}x_2;y_1,y_2)_{\alpha \beta \rho \sigma }+S_F^{*}(x_1-x_2)_{\alpha
\beta }\overline{S}_F^{*}(y_1-y_2)_{\rho \sigma }  \eqnum{2.4}
\end{equation}
where 
\begin{equation}
G(x_{1,}x_2;y_1,y_2)_{\alpha \beta \rho \sigma }=\langle 0^{+}\left| T\{{\bf %
\psi }_\alpha (x_1){\bf \psi }_\beta ^c(x_2)\overline{{\bf \psi }}_\rho (y_1)%
\overline{{\bf \psi }}_\sigma ^c(y_2)\}\right| 0^{-}\rangle  \eqnum{2.5}
\end{equation}
is the ordinary $q\overline{q}$ four-point Green's function [18], 
\begin{equation}
\begin{tabular}{l}
$S_F^{*}(x_1-x_2)_{\alpha \beta }=\frac 1i\langle 0^{+}\left| T\{{\bf \psi }%
_\alpha (x_1){\bf \psi }_\beta ^c(x_2)\}\right| 0^{-}\rangle $ \\ 
$=S_F(x_1-x_2)_{\alpha \gamma }(C^{-1})_{\gamma \beta
}=S_F^c(x_2-x_1)_{\beta \lambda }C_{\lambda \alpha }$%
\end{tabular}
\eqnum{2.6}
\end{equation}
and 
\begin{equation}
\begin{tabular}{l}
$\overline{S}_F^{*}(y_1-y_2)_{\rho \sigma }=\frac 1i\langle 0^{+}\left| T\{%
\overline{{\bf \psi }}_\rho (y_1)\overline{{\bf \psi }}_\sigma
^c(y_2)\}\right| 0^{-}\rangle $ \\ 
$=C_{\sigma \tau }S_F(y_2-y_1)_{\tau \rho }=(C^{-1})_{\rho \delta
}S_F^c(y_1-y_2)_{\delta \sigma }$%
\end{tabular}
\eqnum{2.7}
\end{equation}
in which 
\begin{equation}
S_F(x_1-x_2)_{\alpha \gamma }=\langle 0^{+}\left| T\{{\bf \psi }_\alpha (x_1)%
\overline{{\bf \psi }}_\gamma (x_2)\}\right| 0^{-}\rangle  \eqnum{2.8}
\end{equation}
and 
\begin{equation}
S_F^c(y_1-y_2)_{\delta \sigma }=\langle 0^{+}\left| T\{{\bf \psi }_\delta
^c(y_1)\overline{{\bf \psi }}_\sigma ^c(y_2)\}\right| 0^{-}\rangle 
\eqnum{2.9}
\end{equation}
are the ordinary quark and antiquark propagators respectively [18]. It is
clear that the propagators defined in Eqs. (2.6) and (2.7) are nonzero only
for the quark and the antiquark which are of the same flavor. For the quark
and antiquark of different flavors, the Green's function defined in Eq.
(2.1) is reduced to the ordinary form shown in Eq. (2.5) since the second
term on the right hand side (RHS) of Eq. (2.4) vanishes. In the case of the
quark and antiquark of the same flavor, the normal product in Eq. (2.1)
plays a role of excluding the contraction between the quark field and the
antiquark one from the Green's function. Physically, this avoids the $q%
\overline{q}$ annihilation to break stability of a bound state. It would be
pointed out that use of $\psi ^c(x)$ other than $\overline{\psi }(x)$ to
represent the antiquark field in this paper has an advantage that the
antiquark field would behave as a quark one in the D-S equation so that the
quark-antiquark equation formally is the same as the corresponding two-quark
equation in the case that the quark and antiquark have different flavors.

The equations of motion which describe the variation of the $q\overline{q}$
four-point Green's function $G(x_{1,}x_2;y_1,y_2)$ with the coordinates $x_1$
and $x_2$ may easily be derived from the QCD generating functional as
described in Appendix A. The results are 
\begin{equation}
\begin{tabular}{l}
$(i{\bf \partial }_{x_1}-m_1)_{\alpha \gamma }G(x_{1,}x_2;y_1,y_2)_{\gamma
\beta \rho \sigma }=\delta _{\alpha \rho }\delta
^4(x_1-y_1)S_F^c(x_2-y_2)_{\beta \sigma }$ \\ 
$+C_{\alpha \beta }\delta ^4(x_1-x_2)\overline{S}_F^{*}(y_1-y_2)_{\rho
\sigma }-(\Gamma ^{a\mu })_{\alpha \gamma }G_\mu ^a(x_1\mid
x_1,x_2;y_1,y_2)_{\gamma \beta \rho \sigma }$%
\end{tabular}
\eqnum{2.10}
\end{equation}
\begin{equation}
\begin{tabular}{l}
$(i{\bf \partial }_{x_2}-m_2)_{\beta \lambda }G(x_{1,}x_2;y_1,y_2)_{\alpha
\lambda \rho \sigma }=\delta _{\beta \sigma }\delta
^4(x_2-y_2)S_F(x_1-y_1)_{\alpha \rho }$ \\ 
$+C_{\alpha \beta }\delta ^4(x_1-x_2)\overline{S}_F^{*}(y_1-y_2)_{\rho
\sigma }-(\overline{\Gamma }^{b\nu })_{\beta \lambda }G_\nu ^b(x_2\mid
x_1,x_2;y_1,y_2)_{\alpha \lambda \rho \sigma }$%
\end{tabular}
\eqnum{2.11}
\end{equation}
in which 
\begin{equation}
(\Gamma ^{a\mu })_{\alpha \gamma }=g(\gamma ^\mu T^a)_{\alpha \gamma },\text{
}(\overline{\Gamma }^{b\nu })_{\beta \lambda }=g(\gamma ^\nu \overline{T}%
^b)_{\beta \lambda }  \eqnum{2.12}
\end{equation}
where $g$ is the coupling constant, $T^a=\frac{\lambda ^a}2$ and $\overline{T%
}^a=-\lambda ^{a*}/2$ are the quark and antiquark color matrices
respectively,

\begin{equation}
\begin{tabular}{l}
$G_\mu ^a(x_i\mid x_1,x_2;y_1,y_2)_{\alpha \beta \rho \sigma }$ \\ 
$=\left\langle 0^{+}\left| T\{{\bf A}_\mu ^a(x_i){\bf \psi }_\alpha (x_1)%
{\bf \psi }_\beta ^c(x_2)\overline{{\bf \psi }}_\rho (y_1)\overline{{\bf %
\psi }}_\sigma ^c(y_2)\}\right| 0^{-}\right\rangle $%
\end{tabular}
\eqnum{2.13}
\end{equation}
with $i=1,2$ are the new four-point Green's function including a gluon field
in it and the propagators were defined in Eqs. (2.6)-(2.9). It would be
noted here that the terms related to $\overline{S}_F^{*}(y_1-y_2)$ in Eqs.
(2.10) and (2.11) are absent when the quark and the antiquark have different
flavors. The equations of motion satisfied by the Green's function defined
in Eq. (2.1) may be found by substituting the relation in Eq. (2.4) into
Eqs. (2.10) and (2.11) and by making use of the following equations as
mentioned in Appendix A 
\begin{equation}
\begin{array}{c}
\lbrack (i{\bf \partial }_{x_1}-m_1)_{\alpha \gamma
}S_F^{*}(x_1-x_2)_{\gamma \beta }=-C_{\alpha \beta }\delta
^4(x_1-x_2)-(\Gamma ^{a\mu })_{\alpha \gamma }\Lambda _\mu ^{a*}(x_1\mid
x_1,x_2)_{\gamma \beta } \\ 
\lbrack (i{\bf \partial }_{x_2}-m_2)_{\beta \lambda
}S_F^{*}(x_1-x_2)_{\alpha \lambda }=-C_{\alpha \beta }\delta ^4(x_1-x_2)-(%
\overline{\Gamma }^{b\nu })_{\beta \lambda }\Lambda _\nu ^{b*}(x_2\mid
x_1,x_2)_{\alpha \lambda }
\end{array}
\eqnum{2.14}
\end{equation}
where 
\begin{equation}
\Lambda _\mu ^{a*}(x_i\mid x_1,x_2)_{\alpha \beta }=\frac 1i\langle
0^{+}\left| T\{{\bf A}_\mu ^a(x_i){\bf \psi }_\alpha (x_1){\bf \psi }_\beta
^c(x_2)\}\right| 0^{-}\rangle  \eqnum{2.15}
\end{equation}
with $i=1,2$ are a kind of quark-antiquark-gluon Green's function. The
results are 
\begin{equation}
\begin{tabular}{l}
$(i{\bf \partial }_{x_1}-m_1)_{\alpha \gamma }{\cal G}_{\gamma \beta \rho
\sigma }(x_{1,}x_2;y_1,y_2)=\delta _{\alpha \rho }\delta
^4(x_1-y_1)S_F^c(x_2-y_2)_{\beta \sigma }$ \\ 
$-(\Gamma ^{a\mu })_{\alpha \gamma }{\cal G}_\mu ^a(x_1\mid
x_1,x_2;y_1,y_2)_{\gamma \beta \rho \sigma }$%
\end{tabular}
\eqnum{2.16}
\end{equation}
and

\begin{equation}
\begin{tabular}{l}
$(i{\bf \partial }_{x_2}-m_2)_{\beta \lambda }{\cal G}_{\alpha \lambda \rho
\sigma }(x_{1,}x_2;y_1,y_2)=\delta _{\beta \sigma }\delta
^4(x_2-y_2)S_F(x_1-y_1)_{\alpha \rho }$ \\ 
$-(\overline{\Gamma }^{b\nu })_{\beta \lambda }{\cal G}_\nu ^b(x_2\mid
x_1,x_2;y_1,y_2)_{\alpha \lambda \rho \sigma }$%
\end{tabular}
\eqnum{2.17}
\end{equation}
where 
\begin{equation}
\begin{tabular}{l}
${\cal G}_\mu ^a(x_i\mid x_1,x_2;y_1,y_2)_{\alpha \beta \rho \sigma
}=\langle 0^{+}\left| T[\{N[{\bf A}_\mu ^a(x_i){\bf \psi }_\alpha (x_1){\bf %
\psi }_\beta ^c(x_2)]N[\overline{{\bf \psi }}_\rho (y_1)\overline{{\bf \psi }%
}_\sigma ^c(y_2)]\}\right| 0^{-}\rangle $ \\ 
$=G_\mu ^a(x_i\mid x_1,x_2;y_1,y_2)_{\alpha \beta \rho \sigma }+\Lambda _\mu
^{a*}(x_i\mid x_1,x_2)_{\alpha \beta }\overline{S}_F^{*}(y_1-y_2)_{\rho
\sigma }$%
\end{tabular}
\eqnum{2.18}
\end{equation}
here $i=1,2$. In the above, the normal products defined in the same way as
that in Eq. (2.3). In comparison of Eqs. (2.16) and (2.17) with Eqs. (2.10)
and (2.11), we see, the terms related to $\overline{S}_F^{*}(y_1-y_2)$ in
Eqs. (2.10) and (2.11) disappear in Eqs. (2.16) and (2.17). Therefore, the
equations (2.16) and (2.17) formally are the same as given in the case that
the quark and the antiquark are of different flavors.

Multiplying Eqs. (2.16) and (2.17) with the matrices $\gamma _1^0$ and $%
\gamma _2^0$ respectively, we have 
\begin{equation}
\begin{tabular}{l}
$\lbrack i\frac \partial {\partial t_1}-h^{(1)}(\overrightarrow{x_1}%
)]_{\alpha \gamma }{\cal G}_{\gamma \beta \rho \sigma }(x_1,x_2;y_1,y_2)$ \\ 
$=(\gamma _1^0)_{\alpha \rho }\delta ^4(x_1-y_1)S_F^c(x_2-y_2)_{\beta \sigma
}+{\cal G}_{\alpha \beta \rho \sigma }^{(1)}(x_1,x_2;y_1,y_2)$%
\end{tabular}
\eqnum{2.19}
\end{equation}
and 
\begin{equation}
\begin{tabular}{l}
$\lbrack i\frac \partial {\partial t_2}-h^{(2)}(\overrightarrow{x_2}%
)]_{\beta \lambda }{\cal G}_{\alpha \lambda \rho \sigma }(x_1,x_2;y_1,y_2)$
\\ 
$=(\gamma _2^0)_{\beta \sigma }\delta ^4(x_2-y_2)S_F(x_1-y_1)_{\alpha \rho }+%
{\cal G}_{\alpha \beta \rho \sigma }^{(2)}(x_1,x_2;y_1,y_2)$%
\end{tabular}
\eqnum{2.20}
\end{equation}
where 
\begin{equation}
h^{(i)}(\overrightarrow{x_i})=-i\overrightarrow{\alpha _i}\cdot \nabla _{%
\overrightarrow{x_i}}+m_i\gamma _i^0  \eqnum{2.21}
\end{equation}
is the i-th free fermion Hamiltonian, 
\begin{equation}
\begin{array}{c}
{\cal G}_{\alpha \beta \rho \sigma }^{(1)}(x_1\mid x_1,x_2;y_1,y_2)=(\Omega
_1^{a\mu })_{\alpha \gamma }{\cal G}_\mu ^a(x_1\mid x_1,x_2;y_1,y_2)_{\gamma
\beta \rho \sigma } \\ 
{\cal G}_{\alpha \beta \rho \sigma }^{(2)}(x_2\mid x_1,x_2;y_1,y_2)=(\Omega
_2^{b\nu })_{\beta \lambda }{\cal G}_\nu ^b(x_2\mid x_1,x_2;y_1,y_2)_{\alpha
\lambda \rho \sigma }
\end{array}
\eqnum{2.22}
\end{equation}
here 
\begin{equation}
\Omega _1^{a\mu }=-g\gamma _1^0\gamma _1^\mu T_1^a,\text{ }\Omega _2^{b\nu
}=-g\gamma _2^0\gamma _2^\nu \overline{T}_2^b  \eqnum{2.23}
\end{equation}

As will be proved in section 4, the Green's functions ${\cal G}_\mu
^a(x_i\mid x_1,x_2;y_1,y_2)$ are B-S reducible, therefore, we can write 
\begin{equation}
{\cal G}^{(i)}(x_i\mid x_1,x_2;y_1,y_2)=\int
d^4z_1d^4z_2K^{(i)}(x_1,x_2;z_1,z_2){\cal G}(z_1,z_2;y_1,y_2)  \eqnum{2.24}
\end{equation}
where $K^{(i)}(x_1,x_2;z_1,z_2)$ $(i=1,2)$ are just the interaction kernels.
With the expression given in the above, Eqs. (2.19) and (2.20) can be
represented as 
\begin{equation}
\begin{tabular}{l}
$\lbrack i\frac \partial {\partial t_1}-h^{(1)}(\overrightarrow{x_1}%
)]_{\alpha \gamma }{\cal G}_{\gamma \beta \rho \sigma
}(x_1,x_2;y_1,y_2)=(\gamma _1^0)_{\alpha \rho }\delta
^4(x_1-y_1)S_F^c(x_2-y_2)_{\beta \sigma }$ \\ 
$+\int d^4z_1d^4z_2K^{(1)}(x_1,x_2;z_1,z_2)_{\alpha \beta \lambda \tau }%
{\cal G}_{\lambda \tau \rho \sigma }(z_1,z_2;y_1,y_2)$%
\end{tabular}
\eqnum{2.25}
\end{equation}
and 
\begin{equation}
\begin{tabular}{l}
$\lbrack i\frac \partial {\partial t_2}-h^{(2)}(\overrightarrow{x_2}%
)]_{\beta \lambda }{\cal G}_{\alpha \lambda \rho \sigma
}(x_1,x_2;y_1,y_2)=(\gamma _2^0)_{\beta \sigma }\delta
^4(x_2-y_2)S_F(x_1-y_1)_{\alpha \rho }$ \\ 
$+\int d^4z_1d^4z_2K^{(2)}(x_1,x_2;z_1,z_2)_{\alpha \beta \lambda \tau }%
{\cal G}_{\lambda \tau \rho \sigma }(z_1,z_2;y_1,y_2)$%
\end{tabular}
\eqnum{2.26}
\end{equation}

From the above two equations, we may obtain two equivalent equations: one
describes the evolution of the Green's function ${\cal G}(z_1,z_2;y_1,y_2)$
with the center of mass time; another describes the evolution of the Green's
function with the relative time. The first equation is given by summing up
the two equations in Eqs. (2.25) and (2.26). Introducing the cluster
coordinates 
\begin{equation}
\begin{tabular}{l}
$X=\eta _1x_1+\eta _2x_2,$ $x=x_1-x_2;$ \\ 
$Y=\eta _1y_1+\eta _2y_2,$ $y=y_1-y_2,$ \\ 
$\eta _i=\frac{m_i}{m_1+m_2},i=1,2$%
\end{tabular}
\eqnum{2.27}
\end{equation}
the equation may be represented in the matrix notation as 
\begin{equation}
\begin{tabular}{l}
$\lbrack i\frac \partial {\partial t}-H_0(\overrightarrow{X},\overrightarrow{%
x})]{\cal G}(X-Y,x,y)$ \\ 
$=S(X-Y,x,y)+\int d^4Zd^4zK(X-Z,x,z){\cal G}(Z-Y,z,y)$%
\end{tabular}
\eqnum{2.28}
\end{equation}
where 
\begin{equation}
S(X-Y,x,y)=\delta ^4(x_1-y_1)\gamma _1^0S_F^c(x_2-y_2)+\delta
^4(x_2-y_2)\gamma _2^0S_F(x_1-y_1)  \eqnum{2.29}
\end{equation}
\begin{equation}
H_0(\overrightarrow{X},\overrightarrow{x})=h^{(1)}(\overrightarrow{x}%
_1)+h^{(2)}(\overrightarrow{x}_2)  \eqnum{2.30}
\end{equation}
is the total free Hamiltonian, 
\begin{equation}
K(X-Z,x,z)=\sum_{i=1}^2K^{(i)}(X-Z,x,z)  \eqnum{2.31}
\end{equation}
is the total interaction kernel and $t=X_0$ is the center of mass time. In
the above, the translation-invariance of the Green's function and the
interaction kernel has been considered.

The second equation mentioned above is given by subtracting the equation in
Eq. (2.26) with weight $\eta _1$ from the equation in Eq. (2.25) with weight 
$\eta _2$%
\begin{equation}
\begin{tabular}{l}
$\lbrack i\frac \partial {\partial \tau }-\overline{H}_0(\overrightarrow{X},%
\overrightarrow{x})]{\cal G}(X-Y,x,y)$ \\ 
$=\overline{S}(X-Y,x,y)+\int d^4Zd^4z\overline{K}(X-Z,x,z){\cal G}(Z-Y,z,y)$%
\end{tabular}
\eqnum{2.32}
\end{equation}
where 
\begin{equation}
\overline{S}(X-Y,x,y)=\eta _2\delta ^4(x_1-y_1)\gamma
_1^0S_F^c(x_2-y_2)-\eta _1\delta ^4(x_2-y_2)\gamma _2^0S_F(x_1-y_1) 
\eqnum{2.33}
\end{equation}
\begin{equation}
\overline{H}_0(\overrightarrow{X},\overrightarrow{x})=\eta _2h^{(1)}(%
\overrightarrow{x}_1)-\eta _1h^{(2)}(\overrightarrow{x}_2)  \eqnum{2.34}
\end{equation}
which is the relative Hamiltonian, 
\begin{equation}
\overline{K}(X-Z,x,z)=\eta _2K^{(1)}(X-Z,x,z)-\eta _1K^{(2)}(X-Z,x,z) 
\eqnum{2.35}
\end{equation}
which is the relative kernel and $\tau =x_0$ is the relative time.

By virtue of the well-known Lehmann representation of the Green's function $%
{\cal G}(z_1,z_2;y_1,y_2)$ [31], one may derived the equations satisfied by
the B-S amplitude from the above equations. The Lehmann representation as
shown below can easily be written out by the procedure of inserting the
complete set of the $q\overline{q}$ bound states into the Green's function $%
{\cal G}(z_1,z_2;y_1,y_2)$ denoted in Eq. (2.1), then considering the
translation-invariance property of the Green's function and finally
employing the integral representation of the step function, 
\begin{equation}
\begin{tabular}{l}
${\cal G}(X-Y,x,y)=\sum\limits_n\frac i{(2\pi )^4}\int d^4Q_ne^{-iQ_n(X-Y)}$
\\ 
$\times \frac 1{2\omega _n}\{\frac{\chi _{Q_n}(x)\overline{\chi }_{Q_n}(y)}{%
Q_n^0-\omega _n+i\epsilon }-\frac{\chi _{-Q_n}^{+}(y)\overline{\chi }%
_{-Q_n}^{+}(x)}{Q_n^0+\omega _n-i\epsilon }\}$%
\end{tabular}
\eqnum{2.36}
\end{equation}
where 
\begin{equation}
\begin{array}{c}
\chi _{Q_n}(X,x)=e^{-iQ_nX}\chi _{Q_n}(x)=\langle 0^{+}\left| N\{{\bf \psi }%
(x_1){\bf \psi }^c(x_2)\}\right| n\rangle \\ 
\overline{\chi }_{Q_n}(Y,y)=e^{iQ_nY}\overline{\chi }_{Q_n}(y)=\langle
n\left| N\{\overline{{\bf \psi }}(y_1)\overline{{\bf \psi }}^c(y_2)\}\right|
0^{-}\rangle
\end{array}
\eqnum{2.37}
\end{equation}
are the B-S amplitudes describing the bound state and $\omega _n$ is the
energy of the state $\mid n\rangle $. Upon substituting Eq. (2.36) into Eqs.
(2.28) and (2.32), then taking the limit: $\lim_{Q_n^0\rightarrow \omega
_n}(Q_n^0-\omega _n)$ and finally performing the integration: $\int
d^4Ye^{-iPY}$, one can find that 
\begin{equation}
\lbrack i\frac \partial {\partial t}-H_0(\overrightarrow{X},\overrightarrow{x%
})]\chi _{P\varsigma }(X,x)=\int d^4Yd^4yK(X-Y,x,y)\chi _{P\varsigma }(Y,y) 
\eqnum{2.38}
\end{equation}
and 
\begin{equation}
\lbrack i\frac \partial {\partial \tau }-\overline{H}_0(\overrightarrow{X},%
\overrightarrow{x})]\chi _{P\varsigma }(X,x)=\int d^4Yd^4y\overline{K}%
(X-Y,x,y)\chi _{P\varsigma }(Y,y)  \eqnum{2.39}
\end{equation}
where the subscript $\varsigma $ in the B-S amplitude designates the other
quantum numbers of a bound state. In the above derivation, the fact that the
functions $S(X-Y,x,y)$ and $\overline{S}(X-Y,x,y)$ have no the bound state
poles has been considered. Eqs. (2.38) and (2.39) are just the wanted D-S
equations satisfied by the B-S amplitudes. Eq. (2.38) describes the
evolution of the $q\overline{q}$ bound state with the center of mass time $t$%
, while, Eq. (2.39) describes the evolution of the $q\overline{q}$ bound
state with the relative time $\tau .$ Clearly, both of the equations are all
the first-order differential equations whose solutions are only determined
by the initial amplitudes at the origin of times.

Considering the translation-invariance of the B-S amplitude and the kernels
as shown in Eq. (2.37) and in the following 
\begin{equation}
\begin{array}{c}
K(X-Y,x,y)=\int \frac{d^4Q}{(2\pi )^4}e^{-iQ(X-Y)}K(Q,x,y) \\ 
\overline{K}(X-Y,x,y)=\int \frac{d^4Q}{(2\pi )^4}e^{-iQ(X-Y)}\overline{K}%
(Q,x,y)
\end{array}
\eqnum{2.40}
\end{equation}
one can obtain from Eqs. (2.38) and (2.39) the equations satisfied by the
amplitude which describes the internal motion of the $q\overline{q}$ bound
system 
\begin{equation}
\lbrack E-H_0(\overrightarrow{P},\overrightarrow{x})]\chi _{P\varsigma
}(x)=\int d^4yK(P,x,y)\chi _{P\varsigma }(y)  \eqnum{2.41}
\end{equation}
\begin{equation}
\lbrack i\frac \partial {\partial \tau }-\overline{H}_0(\overrightarrow{P},%
\overrightarrow{x})]\chi _{P\varsigma }(x)=\int d^4y\overline{K}(P,x,y)\chi
_{P\varsigma }(y)  \eqnum{2.42}
\end{equation}
Furthermore, in view of the Fourier transformation 
\begin{equation}
\begin{array}{c}
\chi _{P\varsigma }(x)=\int \frac{d^4q}{(2\pi )^4}e^{-iqx}\chi _{P\varsigma
}(q) \\ 
K(P,x,y)=\int \frac{d^4q}{(2\pi )^4}\frac{d^4k}{(2\pi )^4}%
e^{-iqx+iky}K(P,q,k),
\end{array}
\eqnum{2.43}
\end{equation}
Eqs. (2.41) and (2.42) will immediately be transformed into the momentum
space 
\begin{equation}
\lbrack E-H_0(\overrightarrow{P},\overrightarrow{q})]\chi _{P\varsigma
}(q)=\int \frac{d^4k}{(2\pi )^4}K(P,q,k)\chi _{P\varsigma }(k)  \eqnum{2.44}
\end{equation}
\begin{equation}
\lbrack q_0-\overline{H}_0(\overrightarrow{P},\overrightarrow{q})]\chi
_{P\varsigma }(q)=\int \frac{d^4k}{(2\pi )^4}\overline{K}(P,q,k)\chi
_{P\varsigma }(k)  \eqnum{2.45}
\end{equation}
where 
\begin{equation}
\begin{array}{c}
H_0(\overrightarrow{P},\overrightarrow{q})=h^{(1)}(\overrightarrow{p_1}%
)+h^{(2)}(\overrightarrow{p_2}) \\ 
\overline{H}_0(\overrightarrow{P},\overrightarrow{q})=\eta _2h^{(1)}(%
\overrightarrow{p}_1)-\eta _1h^{(2)}(\overrightarrow{p}_2)
\end{array}
\eqnum{2.46}
\end{equation}
in which 
\begin{equation}
\begin{array}{c}
h^{(i)}(\overrightarrow{p_i})=\overrightarrow{p_i}\cdot \overrightarrow{%
\alpha _i}+m_i\gamma _i^0,\text{ }i=1,2, \\ 
P=p_1+p_2,\text{ }q=\eta _2p_1-\eta _1P_{2,}\text{ }k=\eta _2q_1-\eta _1q_2
\end{array}
\eqnum{2.47}
\end{equation}
here $q_i$ and $p_{i\text{ }}$ are the initial state and the final state
momenta for i-th single particle. Both of the above equations are identified
themselves with the eigenvalue equations for the $q\overline{q}$ bound
system.

The D-S equations given in Eqs. (2.41) and (2.42) or (2.44) and (2.45) are
Lorentz-covariant. They allow us to investigate the $q\overline{q}$ bound
states in any coordinate frame. Based on the space-like property of a bound
state, it is admissible to discuss the bound state in a special equal-time
Lorentz frame. In this frame, the $q\overline{q}$ four-point Green's
function becomes 
\begin{equation}
{\cal G}(\overrightarrow{x_1}_{,}\overrightarrow{x_2};\overrightarrow{y_1},%
\overrightarrow{y_2};t_1-t_2)=\langle 0^{+}\left| T\{N[{\bf \psi }(%
\overrightarrow{x_1},t_1){\bf \psi }^c(\overrightarrow{x_2},t_1)]N[\overline{%
{\bf \psi }}(\overrightarrow{y_1},t_2)\overline{{\bf \psi }}^c(%
\overrightarrow{y_2},t_2)]\}\right| 0^{-}\rangle  \eqnum{2.48}
\end{equation}
The equation in Eq. (2.28) is now reduced to 
\begin{equation}
\begin{tabular}{l}
$\lbrack i\frac \partial {\partial t_1}-H_0(\overrightarrow{x_1}_{,}%
\overrightarrow{x_2})]{\cal G}(\overrightarrow{x_1}_{,}\overrightarrow{x_2};%
\overrightarrow{y_1},\overrightarrow{y_2};t_1-t_2)$ \\ 
$=\delta (t_1-t_2)S(\overrightarrow{x_1}_{,}\overrightarrow{x_2};%
\overrightarrow{y_1},\overrightarrow{y_2})+\int d^3z_1d^3z_2dt_zK(%
\overrightarrow{x_1}_{,}\overrightarrow{x_2};\overrightarrow{z_1},%
\overrightarrow{z_2};t_1-t_z){\cal G}(\overrightarrow{z_1}_{,}%
\overrightarrow{z_2};\overrightarrow{y_1},\overrightarrow{y_2};t_z-t_2)$%
\end{tabular}
\eqnum{2.49}
\end{equation}
where 
\begin{equation}
S(\overrightarrow{x_1}_{,}\overrightarrow{x_2};\overrightarrow{y_1},%
\overrightarrow{y_2})=\delta ^3(\overrightarrow{x_1}-\overrightarrow{y_1}%
)\gamma _1^0S_F^c(\overrightarrow{x_2}-\overrightarrow{y_2})+\delta ^3(%
\overrightarrow{x_2}-\overrightarrow{y_2})\gamma _2^0S_F(\overrightarrow{x_1}%
-\overrightarrow{y_1})  \eqnum{2.50}
\end{equation}
in which 
\begin{equation}
\begin{array}{c}
S_F(\overrightarrow{x_1}-\overrightarrow{y_1})=\frac 1i\langle 0^{+}\left| T[%
{\bf \psi }(\overrightarrow{x_1},t_1)[\overline{{\bf \psi }}(\overrightarrow{%
y_1},t_1)]\right| 0^{-}\rangle \\ 
S_F^c(\overrightarrow{x_2}-\overrightarrow{y_2})=\langle 0^{+}\left| T[{\bf %
\psi }^c(\overrightarrow{x_2},t_2)\overline{{\bf \psi }}^c(\overrightarrow{%
y_2},t_2)]\right| 0^{-}\rangle
\end{array}
\eqnum{2.51}
\end{equation}
are the equal-time quark and antiquark propagators respectively which are
actually independent of time due to the translation-invariance property. By
the Fourier transformations 
\begin{equation}
\begin{array}{c}
{\cal G}(\overrightarrow{x_1}_{,}\overrightarrow{x_2};\overrightarrow{y_1},%
\overrightarrow{y_2};t_1-t_2)=\int_{-\infty }^{+\infty }\frac{dE}{2\pi }%
e^{iE(t_1-t_2)}{\cal G}(\overrightarrow{x_1}_{,}\overrightarrow{x_2};%
\overrightarrow{y_1},\overrightarrow{y_2};E) \\ 
K(\overrightarrow{x_1}_{,}\overrightarrow{x_2};\overrightarrow{z_1},%
\overrightarrow{z_2};t_1-t_2)=\int_{-\infty }^{+\infty }\frac{dE}{2\pi }%
e^{iE(t_1-t_2)}K(\overrightarrow{x_1}_{,}\overrightarrow{x_2};%
\overrightarrow{z_1},\overrightarrow{z_2};E)
\end{array}
\eqnum{2.52}
\end{equation}
Eq. (2.49) will be represented as [15] 
\begin{equation}
\begin{tabular}{l}
$\lbrack E-H_0(\overrightarrow{x_1}_{,}\overrightarrow{x_2})]{\cal G}(%
\overrightarrow{x_1}_{,}\overrightarrow{x_2};\overrightarrow{y_1},%
\overrightarrow{y_2};E)$ \\ 
$=S(\overrightarrow{x_1}_{,}\overrightarrow{x_2};\overrightarrow{y_1},%
\overrightarrow{y_2})+\int d^3z_1d^3z_2K(\overrightarrow{x_1}_{,}%
\overrightarrow{x_2};\overrightarrow{z_1},\overrightarrow{z_2};E){\cal G}(%
\overrightarrow{z_1}_{,}\overrightarrow{z_2};\overrightarrow{y_1},%
\overrightarrow{y_2};E)$%
\end{tabular}
\eqnum{2.53}
\end{equation}
This just is the three-dimensional equation satisfied by the $q\overline{q}$
four-point Green's function defined in the equal-time Lorentz frame.

In the equal-time frame, the relative time of the $q\overline{q}$ system is
zero. Therefore, the equation in Eq. (2.32) becomes meaningless. We are left
with only the equation given in Eq. (2.53). The Lehmann representation of
the Green's function ${\cal G}(\overrightarrow{x_1}_{,}\overrightarrow{x_2};%
\overrightarrow{y_1},\overrightarrow{y_2};t)$ is still represented in Eq.
(2.36) except that the four-dimensional relative coordinates $x$ and $y$ in
the B-S amplitudes are now replaced by the three-dimensional ones $%
\overrightarrow{x}$ and $\overrightarrow{y}$. Substituting such a Lehmann
representation into Eq. (2.53), by the same procedure as stated in Eqs.
(2.36)-(2.38), one may obtain a D-S equation represented in the
three-dimensional coordinate space such that 
\begin{equation}
\lbrack E-H_0(\overrightarrow{X},\overrightarrow{x})]\chi _{P\varsigma }(%
\overrightarrow{X},\overrightarrow{x})=\int d^3Yd^3yK(\overrightarrow{X}-%
\overrightarrow{Y},\overrightarrow{x},\overrightarrow{y})\chi _{P\varsigma }(%
\overrightarrow{Y},\overrightarrow{y})  \eqnum{2.54}
\end{equation}
where 
\begin{equation}
\chi _{P\varsigma }(\overrightarrow{X},\overrightarrow{x})=e^{i%
\overrightarrow{P}\cdot \overrightarrow{X}}\chi _{P\varsigma }(%
\overrightarrow{x})  \eqnum{2.55}
\end{equation}
Apparently, in the momentum space, Eq. (2.54) becomes [15] 
\begin{equation}
\lbrack E-H_0(\overrightarrow{P},\overrightarrow{q})]\chi _{P\varsigma }(%
\overrightarrow{q})=\int \frac{d^3k}{(2\pi )^3}K(\overrightarrow{P},%
\overrightarrow{q},\overrightarrow{k};E)\chi _{P\varsigma }(\overrightarrow{k%
})  \eqnum{2.56}
\end{equation}
This is precisely the three-dimensional D-S equation satisfied by the
amplitude $\chi _{P\varsigma }(\overrightarrow{k})$ which describes the
internal motion of the $q\overline{q}$ bound system and may be used to solve
the eigenvalue problem for the system. This equation is rigorous because the
retardation effect is completely included in the kernel of the equation.

\section{Derivation of the interaction kernel}

In this section, the interaction kernel in the D-S equation will be derived
by making use of equations of motion which describe the variation of the
Green's functions with coordinates $y_1$ and $y_2$. The equations satisfied
by the Green's function $G(x_{1,}x_2;y_1,y_2)$ are derived in Appendix A and
can directly be written out from Eqs. (A.22) and (A.31) by setting the
source $J$ to be zero. The result is 
\begin{equation}
\begin{tabular}{l}
$G(x_{1,}x_2;y_1,y_2)_{\alpha \beta \tau \sigma }(i\overleftarrow{{\bf %
\partial }}_{y_1}+m_1)_{\tau \rho }=-\delta _{\alpha \rho }\delta
^4(x_1-y_1)S_F^c(x_2-y_2)_{\beta \sigma }$ \\ 
$-C_{\rho \sigma }\delta ^4(y_1-y_2)S_F^{*}(x_1-x_2)_{\alpha \beta }+G_\mu
^a(y_1\mid x_1,x_2;y_1,y_2)_{\alpha \beta \tau \sigma }(\Gamma ^{a\mu
})_{\tau \rho }$%
\end{tabular}
\eqnum{3.1}
\end{equation}
\begin{equation}
\begin{tabular}{l}
$G(x_{1,}x_2;y_1,y_2)_{\alpha \beta \rho \delta }(i\overleftarrow{{\bf %
\partial }}_{y_2}+m_2)_{\delta \sigma }=-\delta _{\beta \sigma }\delta
^4(x_2-y_2)S_F(x_1-y_1)_{\alpha \rho }$ \\ 
$-C_{\rho \sigma }\delta ^4(y_1-y_2)S_F^{*}(x_1-x_2)_{\alpha \beta }+G_\nu
^b(y_2\mid x_1,x_2;y_1,y_2)_{\alpha \beta \rho \delta }(\overline{\Gamma }%
^{b\nu })_{\delta \sigma }$%
\end{tabular}
\eqnum{3.2}
\end{equation}
where $G_\mu ^a(y_i\mid x_1,x_2;y_1,y_2)$ $($ $i=1,2)$ were defined in Eq.
(2.13) with the replacement of $x_i$ by $y_i$.

Substituting the relation in Eq. (2.4) into the above two equations and
employing the following equations obeyed by the propagator $\overline{S}%
_F^{*}(y_1-y_2)$ whose derivation was mentioned in Appendix A 
\begin{equation}
\begin{array}{c}
\overline{S}_F^{*}(y_1-y_2)_{\tau \sigma }(i\overleftarrow{{\bf \partial }}%
_{y_1}+m_1)_{\tau \rho }=C_{\rho \sigma }\delta ^4(y_1-y_2)+\overline{%
\Lambda }_\nu ^{b*}(y_1\mid y_1,y_2)_{\tau \sigma }(\Gamma ^{b\nu })_{\tau
\rho } \\ 
\overline{S}_F^{*}(y_1-y_2)_{\rho \delta }(i\overleftarrow{{\bf \partial }}%
_{y_2}+m_2)_{\delta \sigma }=C_{\rho \sigma }\delta ^4(y_1-y_2)+\overline{%
\Lambda }_\nu ^{b*}(y_2\mid y_1,y_2)_{\rho \delta }(\overline{\Gamma }^{b\nu
})_{\delta \sigma }
\end{array}
\eqnum{3.3}
\end{equation}
where 
\begin{equation}
\overline{\Lambda }_\nu ^{b*}(x_i\mid y_1,y_2)_{\tau \sigma }=\frac 1i%
\left\langle 0^{+}\left| T\{{\bf A}_\nu ^b(x_i)\overline{{\bf \psi }}_\tau
(y_1)\overline{{\bf \psi }}_\sigma ^c(y_2)\}\right| 0^{-}\right\rangle , 
\eqnum{3.4}
\end{equation}
it is easy to find 
\begin{equation}
\begin{tabular}{l}
${\cal G}(x_{1,}x_2;y_1,y_2)_{\alpha \beta \tau \sigma }(i\overleftarrow{%
{\bf \partial }}_{y_1}+m_1)_{\tau \rho }=-\delta _{\alpha \rho }\delta
^4(x_1-y_1)S_F^c(x_2-y_2)_{\beta \sigma }$ \\ 
$+{\cal G}_\mu ^a(y_1\mid x_1,x_2;y_1,y_2)_{\alpha \beta \tau \sigma
}(\Gamma ^{a\mu })_{\tau \rho }$%
\end{tabular}
\eqnum{3.5}
\end{equation}
and 
\begin{equation}
\begin{tabular}{l}
${\cal G}(x_{1,}x_2;y_1,y_2)_{\alpha \beta \rho \delta }(i\overleftarrow{%
{\bf \partial }}_{y_2}+m_2)_{\delta \sigma }=-\delta _{\beta \sigma }\delta
^4(x_2-y_2)S_F(x_1-y_1)_{\alpha \rho }$ \\ 
$+{\cal G}_\nu ^b(y_2\mid x_1,x_2;y_1,y_2)_{\alpha \beta \rho \delta }(%
\overline{\Gamma }^{b\nu })_{\delta \sigma }$%
\end{tabular}
\eqnum{3.6}
\end{equation}
where 
\begin{equation}
\begin{tabular}{l}
${\cal G}_\mu ^a(y_i\mid x_1,x_2;y_1,y_2)_{\alpha \beta \rho \sigma
}=\left\langle 0^{+}\left| T\{N[{\bf \psi }_\alpha (x_1){\bf \psi }_\beta
^c(x_2)]N[{\bf A}_\mu ^a(y_i)\overline{{\bf \psi }}_\tau (y_1)\overline{{\bf %
\psi }}_\sigma ^c(y_2)]\}\right| 0^{-}\right\rangle $ \\ 
$=G_\mu ^a(y_i\mid x_1,x_2;y_1,y_2)_{\alpha \beta \rho \sigma
}+S_F^{*}(x_1-x_2)_{\alpha \beta }\overline{\Lambda }_\mu ^{a*}(y_i\mid
y_1,y_2)_{\rho \sigma }$%
\end{tabular}
\eqnum{3.7}
\end{equation}
here $i=1,2$.

To derive the interaction kernel, we also need equations obeyed by the
Green's functions ${\cal G}_\mu ^a(x_i\mid x_1,x_2;y_1,y_2)$. According to
the description given in Appendix A, the equations for the Green's functions 
$G_\mu ^a(x_i\mid x_1,x_2;y_1,y_2)$ can be derived by differentiating Eqs.
(A.22) and (A.31) with respect to the source $J^{a\mu }(x_i)$ and then
setting $J=0$. The result is 
\begin{equation}
\begin{tabular}{l}
$G_\mu ^a(x_i\mid x_{1,}x_2;y_1,y_2)_{\alpha \beta \tau \sigma }(i%
\overleftarrow{{\bf \partial }}_{y_1}+m_1)_{\tau \rho }=-\delta _{\alpha
\rho }\delta ^4(x_1-y_1)\Lambda _\mu ^{{\bf c}a}(x_i\mid x_2,y_2)_{\beta
\sigma }$ \\ 
$-C_{\rho \sigma }\delta ^4(y_1-y_2)\Lambda _\mu ^{a*}(x_i\mid
x_1,x_2)_{\alpha \beta }+G_{\mu \nu }^{ab}(x_i,y_1\mid
x_1,x_2;y_1,y_2)_{\alpha \beta \tau \sigma }(\Gamma ^{a\mu })_{\tau \rho }$%
\end{tabular}
\eqnum{3.8}
\end{equation}
and 
\begin{equation}
\begin{tabular}{l}
$G_\mu ^a(x_i\mid x_{1,}x_2;y_1,y_2)_{\alpha \beta \rho \delta }(i%
\overleftarrow{{\bf \partial }}_{y_2}+m_2)_{\delta \sigma }=-\delta _{\beta
\sigma }\delta ^4(x_2-y_2)\Lambda _\mu ^a(x_i\mid x_1,y_1)_{\alpha \rho }$
\\ 
$-C_{\rho \sigma }\delta ^4(y_1-y_2)\Lambda _\mu ^{a*}(x_i\mid
x_1,x_2)_{\alpha \beta }+G_{\mu \nu }^{ab}(x_i,y_2\mid
x_1,x_2;y_1,y_2)_{\alpha \beta \rho \delta }(\overline{\Gamma }^{b\nu
})_{\delta \sigma }$%
\end{tabular}
\eqnum{3.9}
\end{equation}
where 
\begin{equation}
\begin{array}{c}
\Lambda _\mu ^a(x_i\mid x_1,y_1)_{\gamma \rho }=\frac 1i\langle 0^{+}\left|
T[{\bf A}_\mu ^a(x_i){\bf \psi }_\gamma (x_1)\overline{{\bf \psi }}_\rho
(y_1)]\right| 0^{-}\rangle , \\ 
\Lambda _\mu ^{{\bf c}a}(x_i\mid x_2,y_2)_{\lambda \sigma }=\frac 1i\langle
0^{+}\left| T[{\bf A}_\mu ^a(x_i){\bf \psi }_\lambda ^c(x_2)\overline{{\bf %
\psi }}_\sigma ^c(y_2)]\right| 0^{-}\rangle , \\ 
\Lambda _\mu ^{a*}(x_i\mid y_1,y_2)_{\tau \sigma }=\frac 1i\left\langle
0^{+}\left| T[{\bf A}_\mu ^a(x_i){\bf \psi }_\tau (y_1){\bf \psi }_\sigma
^c(y_2)]\right| 0^{-}\right\rangle
\end{array}
\eqnum{3.10}
\end{equation}
and 
\begin{equation}
G_{\mu \nu }^{ab}(x_i,y_j\mid x_1,x_2;y_1,y_2)_{\alpha \beta \rho \sigma
}=\langle 0^{+}\left| T\{{\bf A}_\mu ^a(x_i){\bf A}_\nu ^b(y_j){\bf \psi }%
_\alpha (x_1){\bf \psi }_\beta ^c(x_2)\overline{{\bf \psi }}_\rho (y_1)%
\overline{{\bf \psi }}_\sigma ^c(y_2)\}\right| 0^{-}\rangle  \eqnum{3.11}
\end{equation}
here $i,j=1,2.$ On inserting the relation in Eq. (2.18) into Eqs. (3.8) and
(3.9) and utilizing the equations in Eq. (3.3), we are led to 
\begin{equation}
\begin{tabular}{l}
${\cal G}_\mu ^a(x_i\mid x_{1,}x_2;y_1,y_2)_{\alpha \beta \tau \sigma }(i%
\overleftarrow{{\bf \partial }}_{y_1}+m_1)_{\tau \rho }=-\delta _{\alpha
\rho }\delta ^4(x_1-y_1)\Lambda _\mu ^{{\bf c}a}(x_i\mid x_2,y_2)_{\beta
\sigma }$ \\ 
$+{\cal G}_{\mu \nu }^{ab}(x_i,y_1\mid x_1,x_2;y_1,y_2)_{\alpha \beta \tau
\sigma }(\Gamma ^{a\mu })_{\tau \rho }$%
\end{tabular}
\eqnum{3.12}
\end{equation}
and 
\begin{equation}
\begin{tabular}{l}
${\cal G}_\mu ^a(x_i\mid x_{1,}x_2;y_1,y_2)_{\alpha \beta \rho \delta }(i%
\overleftarrow{{\bf \partial }}_{y_2}+m_2)_{\delta \sigma }=-\delta _{\beta
\sigma }\delta ^4(x_2-y_2)\Lambda _\mu ^a(x_i\mid x_1,y_1)_{\alpha \rho }$
\\ 
$+{\cal G}_{\mu \nu }^{ab}(x_i,y_2\mid x_1,x_2;y_1,y_2)_{\alpha \beta \rho
\delta }(\overline{\Gamma }^{b\nu })_{\delta \sigma }$%
\end{tabular}
\eqnum{3.13}
\end{equation}
where 
\begin{equation}
\begin{tabular}{l}
${\cal G}_{\mu \nu }^{ab}(x_i,y_j\mid x_1,x_2;y_1,y_2)_{\alpha \beta \rho
\sigma }$ \\ 
=$\langle 0^{+}\left| T\{N[{\bf A}_\mu ^a(x_i){\bf \psi }_\alpha (x_1){\bf %
\psi }_\beta ^c(x_2)]N[{\bf A}_\nu ^b(y_j)\overline{{\bf \psi }}_\rho (y_1)%
\overline{{\bf \psi }}_\sigma ^c(y_2)]\}\right| 0^{-}\rangle $ \\ 
$=G_{\mu \nu }^{ab}(x_i,y_j\mid x_1,x_2;y_1,y_2)_{\alpha \beta \rho \sigma
}+\Lambda _\mu ^{a*}(x_i\mid x_1,x_2)_{\alpha \beta }\overline{\Lambda }_\nu
^{b*}(y_j\mid y_1,y_2)_{\rho \sigma }$%
\end{tabular}
\eqnum{3.14}
\end{equation}

Now, let us multiply Eqs. (3.5) and (3.6) respectively with $\gamma _1^0$
and $\gamma _2^0$ from the right and sum up the both equations thus
obtained. In this way, writing in the matrix from, we obtain the following
equation 
\begin{equation}
\begin{tabular}{l}
${\cal G}(x_1,x_2;y_1,y_2)[i\overleftarrow{\frac \partial {\partial
t^{\prime }}}+\overleftarrow{H_0}(y_1,y_2)]$ \\ 
$=-S(x_1,x_2;y_1,y_2)-\sum\limits_{i=1}^2{\cal G}^{(i)}(y_i\mid
x_1,x_2;y_1,y_2)$%
\end{tabular}
\eqnum{3.15}
\end{equation}
where 
\begin{equation}
\frac \partial {\partial t^{\prime }}=\frac \partial {\partial y_1^0}+\frac 
\partial {\partial y_2^0}=\frac \partial {\partial Y^0},  \eqnum{3.16}
\end{equation}
\begin{equation}
H_0(y_1,y_2)=h^{(1)}(y_1)+h^{(2)}(y_2)  \eqnum{3.17}
\end{equation}
here $h^{(1)}(y_1)$ and $h^{(2)}(y_2)$ were represented in Eq. (2.21), 
\begin{equation}
\begin{array}{c}
{\cal G}^{(1)}(y_1\mid x_1,x_2;y_1,y_2)_{\alpha \beta \rho \sigma }={\cal G}%
_\mu ^a(y_1\mid x_1,x_2;y_1,y_2)_{\alpha \beta \tau \sigma }(\overline{%
\Omega }_1^{a\mu })_{\tau \rho } \\ 
{\cal G}^{(2)}(y_2\mid x_1,x_2;y_1,y_2)_{\alpha \beta \rho \sigma }={\cal G}%
_\mu ^a(y_2\mid x_1,x_2;y_1,y_2)_{\alpha \beta \rho \delta }(\overline{%
\Omega }_2^{a\mu })_{\delta \sigma }
\end{array}
\eqnum{3.18}
\end{equation}
in which 
\begin{equation}
\overline{\Omega }_1^{a\mu }=-g\gamma _1^\mu \gamma _1^0T^a=-g\gamma
_1^0\gamma _1^{\mu +}T^a,\text{ }\overline{\Omega }_2^{a\mu }=-g\gamma
_2^\mu \gamma _2^0\overline{T}^a=--g\gamma _2^0\gamma _2^{\mu +}\overline{T}%
^a  \eqnum{3.19}
\end{equation}
and $S(x_1,x_2;y_1,y_2)$ was defined in Eq. (2.29).

Similarly, on multiplying Eqs. (3.12) and (3.13) respectively with $\gamma
_1^0$ and $\gamma _2^0$ from the right and summing up the both equations
thus obtained, one can get 
\begin{equation}
\begin{tabular}{l}
${\cal G}_\mu ^a(x_i\mid x_1,x_2;y_1,y_2)_{\alpha \beta \lambda \tau }[i%
\overleftarrow{\frac \partial {\partial t^{\prime }}}+\overleftarrow{H_0}%
(y_1,y_2)]_{\lambda \tau \rho \sigma }$ \\ 
$=-{\cal R}_\mu ^{(i)a}(x_1,x_2;y_1,y_2)_{\alpha \beta \rho \sigma }-{\cal Q}%
_\mu ^{(i)a}(x_1,x_2;y_1,y_2)_{\alpha \beta \rho \sigma }$%
\end{tabular}
\eqnum{3.20}
\end{equation}
where 
\begin{equation}
\begin{tabular}{l}
${\cal R}_\mu ^{(i)a}(x_1,x_2;y_1,y_2)_{\alpha \beta \rho \sigma }=\delta
^4(x_1-y_1)(\gamma _1^0)_{\alpha \rho }\Lambda _\mu ^{{\bf c}a}(x_i\mid
x_2,y_2)_{\beta \sigma }$ \\ 
$+\delta ^4(x_2-y_2)(\gamma _2^0)_{\beta \sigma }\Lambda _\mu ^a(x_i\mid
x_1,y_1)_{\alpha \rho }$%
\end{tabular}
\eqnum{3.21}
\end{equation}
and 
\begin{equation}
\begin{tabular}{l}
${\cal Q}_\mu ^{(i)a}(x_1,x_2;y_1,y_2)_{\alpha \beta \rho \sigma }={\cal G}%
_{\mu \nu }^{ab}(x_i,y_1\mid x_1,x_2;y_1,y_2)_{\alpha \beta \tau \sigma }(%
\overline{\Omega }_1^{b\nu })_{\tau \rho }$ \\ 
$+{\cal G}_{\mu \nu }^{ab}(x_i,y_2\mid x_1,x_2;y_1,y_2)_{\alpha \beta \rho
\delta }(\overline{\Omega }_2^{b\nu })_{\delta \sigma }$%
\end{tabular}
\eqnum{3.22}
\end{equation}
With the definitions given in Eq. (2.22) and in the following 
\begin{equation}
{\cal R}^{(i)}(x_1,x_2;y_1,y_2)=\Omega _i^{a\mu }{\cal R}_\mu
^{(i)a}(x_1,x_2;y_1,y_2)  \eqnum{3.23}
\end{equation}
\begin{equation}
{\cal Q}^{(i)}(x_1,x_2;y_1,y_2)=\Omega _i^{a\mu }{\cal Q}_\mu
^{(i)a}(x_1,x_2;y_1,y_2)  \eqnum{3.24}
\end{equation}
we can write from Eq. (3.20) the equation satisfied by the function ${\cal G}%
_{\alpha \beta \rho \sigma }^{(i)}(x_i\mid x_1,x_2;y_1,y_2)$. In the matrix
form, it reads 
\begin{equation}
\begin{tabular}{l}
${\cal G}^{(i)}(x_i\mid x_1,x_2;y_1,y_2)[i\overleftarrow{\frac \partial {%
\partial t^{\prime }}}+\overleftarrow{H_0}(y_1,y_2)]$ \\ 
$=-{\cal R}^{(i)}(x_1,x_2;y_1,y_2)-{\cal Q}^{(i)}(x_1,x_2;y_1,y_2)$%
\end{tabular}
\eqnum{3.25}
\end{equation}

Up to the present, we are ready to derive the interaction kernel. Acting on
the both sides of Eq. (2.24) with the operator $i\overleftarrow{\frac 
\partial {\partial t^{\prime }}}+\overleftarrow{H_0}(y_1,y_2)$ from the
right and employing Eqs. (3.15) and (3.25), we have 
\begin{equation}
\begin{tabular}{l}
$\int d^4z_1d^4z_2K^{(i)}(x_1,x_2;z_1,z_2)S(z_1,z_2;y_1,y_2)={\cal R}%
^{(i)}(x_1,x_2;y_1,y_2)+{\cal Q}^{(i)}(x_1,x_2;y_1,y_2)$ \\ 
$-\int d^4z_1d^4z_2K^{(i)}(x_1,x_2;z_1,z_2)\sum\limits_{j=1}^2{\cal G}%
^{(j)}(y_j\mid z_1,z_2;y_1,y_2)$%
\end{tabular}
\eqnum{3.26}
\end{equation}
In order to obtain the kernel, we may operate on the above equation with the
inverse of $S(x_1,x_2;y_1,y_2)$ and the kernel on the RHS of Eq. (3.26) may
be eliminated by the following relation given by acting on Eq. (2.24) with
the inverse of the Green's function ${\cal G}(x_1,x_2;y_1,y_2)$%
\begin{equation}
K^{(i)}(x_1,x_2;y_1,y_2)=\int d^4u_1d^4u_2{\cal G}^{(i)}(x_i\mid
x_1,x_2;u_1,u_2){\cal G}^{-1}(u_1,u_2;y_1,y_2)  \eqnum{3.27}
\end{equation}
With these operations, one may derive from Eq.(3.26) a closed expression of
the kernel $K^{(i)}(x_1,x_2;y_1,y_2)$ such that 
\begin{equation}
\begin{tabular}{l}
$K^{(i)}(x_1,x_2;y_1,y_2)=\int d^4z_1d^4z_2\{{\cal R}^{(i)}(x_1,x_2;z_1,z_2)+%
{\cal Q}^{(i)}(x_1,x_2;z_1,z_2)$ \\ 
$-{\cal D}^{(i)}(x_1,x_2;z_1,z_2)\}S^{-1}(z_1,z_2;y_1,y_2)$%
\end{tabular}
\eqnum{3.28}
\end{equation}
where 
\begin{equation}
\begin{tabular}{l}
${\cal D}^{(i)}(x_1,x_2;z_1,z_2)$ \\ 
$=\int \prod\limits_{k=1}^2d^4u_kd^4v_k{\cal G}^{(i)}(x_i\mid
x_1,x_2;u_1,u_2){\cal G}^{-1}(u_1,u_2;v_1,v_2)\sum\limits_{j=1}^2{\cal G}%
^{(j)}(y_j\mid v_1,v_2;z_1,z_2)$%
\end{tabular}
\eqnum{3.29}
\end{equation}
In the above derivation, existence of the inverses ${\cal G}%
^{-1}(u_1,u_2;z_1,z_2)$ and $S^{-1}(z_1,z_2;y_1,y_2)$ has been assumed. The
rationality of the assumption will be illustrated later.

Clearly, the total interaction kernel appearing in Eq. (2.38) is given by
the sum 
\begin{equation}
\begin{tabular}{l}
$K(x_1,x_2;y_1,y_2)=\sum\limits_{j=1}^2K^{(i)}(x_1,x_2;y_1,y_2)=\int
d^4z_1d^4z_2\{{\cal R}(x_1,x_2;z_1,z_2)$ \\ 
$+{\cal Q}(x_1,x_2;z_1,z_2)\ -{\cal D}(x_1,x_2;z_1,z_2)%
\}S^{-1}(z_1,z_2;y_1,y_2)$%
\end{tabular}
\eqnum{3.30}
\end{equation}
where 
\begin{equation}
{\cal R}(x_1,x_2;z_1,z_2)=\sum_{i=1}^2{\cal R}^{(i)}(x_1,x_2;z_1,z_2) 
\eqnum{3.31}
\end{equation}
\begin{equation}
{\cal Q}(x_1,x_2;z_1,z_2)=\sum_{i=1}^2{\cal Q}^{(i)}(x_1,x_2;z_1,z_2)=%
\sum_{i,j=1}^2\Omega _i^{a\mu }{\cal G}_{\mu \nu }^{ab}(x_i,z_j\mid
x_1,x_2;z_1,z_2)\overline{\Omega }_j^{b\nu }  \eqnum{3.32}
\end{equation}
and 
\begin{equation}
\begin{tabular}{l}
${\cal D}(x_1,x_2;z_1,z_2)$ \\ 
$=\int \prod\limits_{k=1}^2d^4u_kd^4v_k\sum\limits_{i,j=1}^2{\cal G}%
^{(i)}(x_i\mid x_1,x_2;u_1,u_2){\cal G}^{-1}(u_1,u_2;v_1,v_2){\cal G}%
^{(j)}(y_j\mid v_1,v_2;z_1,z_2)$%
\end{tabular}
\eqnum{3.33}
\end{equation}

We would like here to discuss the role played by the last term in Eq.
(3.30). In view of the relation in Eq. (2.24) and the following relation 
\begin{equation}
{\cal G}^{(j)}(y_j\mid x_1,x_2;y_1,y_2)=\int d^4z_1d^4z_2{\cal G}%
(x_1,x_2;z_1,z_2)K^{(j)}(z_1,z_2;y_1,y_2)  \eqnum{3.34}
\end{equation}
which also follows from the B-S reducibility of the Green's function ${\cal G%
}^{(j)}(y_j\mid x_1,x_2;y_1,y_2)$ and considering 
\begin{equation}
\int d^4z_1d^4z_2{\cal G}(x_1,x_2;z_1,z_2){\cal G}^{-1}(z_1,z_2;y_1,y_2)=%
\delta ^4(x_1-y_1)\delta ^4(x_2-y_2)  \eqnum{3.35}
\end{equation}
the function ${\cal D}(x_1,x_2;z_1,z_2)$ in Eq. (3.33) may be represented as 
\begin{equation}
\ {\cal D}(x_1,x_2;z_1,z_2)=\int \prod_{k=1}^2d^4u_kd^4v_kK(x_1,x_2;u_1,u_2)%
{\cal G}(u_1,u_2;v_1,v_2)K(v_1,v_2;z_1,z_2)  \eqnum{3.36}
\end{equation}
which manifests itself the typical structure of the B-S reducibility of the
kernel. Therefore, the last term in Eq. (3.30) just plays the role of
cancelling the B-S reducible part of the remaining terms in Eq. (3.30). As a
result of the cancellation, the interaction kernel given in Eq. (3.30) is
really B-S irreducible, consistent with the ordinary concept for the kernel
in a two-body relativistic equation. Inserting Eq. (3.36) into Eq. (3.30),
writing in the operator form, we have 
\begin{equation}
KS={\cal R+Q-}K{\cal G}K  \eqnum{3.37}
\end{equation}
This can be regarded as the integral equation satisfied by the kernel $K.$

Analogously, In accord with the definition in Eq. (2.35) and the expression
in Eq. (3.28), one may write out an explicit expression of the kernel
occurring in Eq. (2.39) 
\begin{equation}
\begin{tabular}{l}
$\overline{K}(x_1,x_2;y_1,y_2)=\eta _2K^{(1)}(x_1,x_2;y_1,y_2)-\eta
_1K^{(2)}(x_1,x_2;y_1,y_2)$ \\ 
$=\int d^4z_1d^4z_2\{\overline{{\cal R}}(x_1,x_2;z_1,z_2)+\overline{{\cal Q}}%
(x_1,x_2;z_1,z_2)\ $ \\ 
$-\overline{{\cal D}}(x_1,x_2;z_1,z_2)\}S^{-1}(z_1,z_2;y_1,y_2)$%
\end{tabular}
\eqnum{3.38}
\end{equation}
in which 
\begin{equation}
\overline{{\cal A}}(x_1,x_2;z_1,z_2)=\eta _2{\cal A}^{(1)}(x_1,x_2;z_1,z_2)-%
\eta _1{\cal A}^{(2)}(x_1,x_2;z_1,z_2)  \eqnum{3.39}
\end{equation}
where ${\cal A}$ stands for ${\cal R},{\cal Q}$ or ${\cal D}$.

Now, we are in a position to write out the interaction kernel appearing in
the three-dimensional D-S equation. As demonstrated in Ref. [15], the kernel
in Eq. (2.54) can be derived by the same procedure as for the
four-dimensional kernel. The expression of the three-dimensional kernel
shown below formally is the same as the four-dimensional one described in
Eq. (3.30) except that it is now represented in the three-dimensional space. 
\begin{equation}
\begin{tabular}{l}
$K(\overrightarrow{x_1}_{,}\overrightarrow{x_2};\overrightarrow{y_1},%
\overrightarrow{y_2};E)=\int d^3z_1d^3z_2\{{\cal R}(\overrightarrow{x_1}_{,}%
\overrightarrow{x_2};\overrightarrow{z_1},\overrightarrow{z_2})$ \\ 
$+{\cal Q}(\overrightarrow{x_1}_{,}\overrightarrow{x_2};\overrightarrow{z_1},%
\overrightarrow{z_2};E)-{\cal D}(\overrightarrow{x_1}_{,}\overrightarrow{x_2}%
;\overrightarrow{z_1},\overrightarrow{z_2};E)\}S^{-1}(\overrightarrow{z_1},%
\overrightarrow{z_2};\overrightarrow{y}_1,\overrightarrow{y}_2)$%
\end{tabular}
\eqnum{3.40}
\end{equation}
where ${\cal R}(\overrightarrow{x_1}_{,}\overrightarrow{x_2};\overrightarrow{%
z_1},\overrightarrow{z_2}),{\cal Q}(\overrightarrow{x_1}_{,}\overrightarrow{%
x_2};\overrightarrow{z_1},\overrightarrow{z_2};E)$ and ${\cal D}(%
\overrightarrow{x_1}_{,}\overrightarrow{x_2};\overrightarrow{z_1},%
\overrightarrow{z_2};E)$ can be written out from Eqs. (3.31)-(3.33) as
follows: 
\begin{equation}
{\cal R}(\overrightarrow{x_1}_{,}\overrightarrow{x_2};\overrightarrow{z_1},%
\overrightarrow{z_2})=\sum_{i=1}^2\Omega _i^{a\mu }{\cal R}_\mu ^{(i)a}(%
\overrightarrow{x_1}_{,}\overrightarrow{x_2};\overrightarrow{z_1},%
\overrightarrow{z_2})  \eqnum{3.41}
\end{equation}
in which 
\begin{equation}
{\cal R}_\mu ^{(i)a}(\overrightarrow{x_1}_{,}\overrightarrow{x_2};%
\overrightarrow{z_1},\overrightarrow{z_2})=\delta ^3(\overrightarrow{x_1}-%
\overrightarrow{z_1})\gamma _1^0\Lambda _\mu ^{{\bf c}a}(\overrightarrow{x_i}%
\mid \overrightarrow{x_2},\overrightarrow{z_2})+\delta ^3(\overrightarrow{x_2%
}-\overrightarrow{z_2})\gamma _2^0\Lambda _\mu ^a(\overrightarrow{x_i}\mid 
\overrightarrow{x_1},\overrightarrow{z_1})  \eqnum{3.42}
\end{equation}
here $\Lambda _\mu ^a(\overrightarrow{x_i}\mid \overrightarrow{x_1},%
\overrightarrow{y_1})$ and $\Lambda _\mu ^{{\bf c}a}(\overrightarrow{x_i}%
\mid \overrightarrow{x_2},\overrightarrow{y_2})$ are defined as in Eq.
(3.10) except that the time variables in all the field operators are now
taken to be the same and therefore they are time-independent due to the
translation-invariance property of the Green's functions, 
\begin{equation}
{\cal Q}(\overrightarrow{x_1}_{,}\overrightarrow{x_2};\overrightarrow{z_1},%
\overrightarrow{z_2};E)=\sum_{i,j=1}^2\Omega _i^{a\mu }{\cal G}_{\mu \nu
}^{ab}(\overrightarrow{x_i},\overrightarrow{z_j}\mid \overrightarrow{x_1}_{,}%
\overrightarrow{x_2};\overrightarrow{z_1},\overrightarrow{z_2};E)\overline{%
\Omega }_j^{b\nu }  \eqnum{3.43}
\end{equation}
in which ${\cal G}_{\mu \nu }^{ab}(\overrightarrow{x_i},\overrightarrow{z_j}%
\mid \overrightarrow{x_1}_{,}\overrightarrow{x_2};\overrightarrow{z}_1,%
\overrightarrow{z}_2;E)$ is the Fourier transform of the Green's function
defined by 
\begin{equation}
\begin{tabular}{l}
${\cal G}_{\mu \nu }^{ab}(\overrightarrow{x_i},\overrightarrow{z_j}\mid 
\overrightarrow{x_1}_{,}\overrightarrow{x_2};\overrightarrow{z}_1,%
\overrightarrow{z}_2;t_1-t_2)$ \\ 
$=\langle 0^{+}\left| T\{N[{\bf A}_\mu ^a(\overrightarrow{x_i},t_1){\bf \psi 
}(\overrightarrow{x_1},t_1){\bf \psi }^c(\overrightarrow{x_2},t_1)]N[{\bf A}%
_\nu ^b(\overrightarrow{z_j},t_2)\overline{{\bf \psi }}(\overrightarrow{z_1}%
,t_2)\overline{{\bf \psi }}^c(\overrightarrow{z_2},t_2)]\}\right|
0^{-}\rangle $%
\end{tabular}
\eqnum{3.44}
\end{equation}
and 
\begin{equation}
\begin{tabular}{l}
${\cal D}(\overrightarrow{x_1}_{,}\overrightarrow{x_2};\overrightarrow{z_1},%
\overrightarrow{z_2};E)=\int
\prod\limits_{k=1}^2d^3u_kd^3v_k\sum\limits_{i,j=1}^2\Omega _i^{a\mu }{\cal G%
}_\mu ^{(i)a}(\overrightarrow{x_i}\mid \overrightarrow{x_1}_{,}%
\overrightarrow{x_2};\overrightarrow{u_1},\overrightarrow{u_2};E)$ \\ 
$\times {\cal G}^{-1}(\overrightarrow{u_1}_{,}\overrightarrow{u_2};%
\overrightarrow{v}_1,\overrightarrow{v}_2;E){\cal G}_\nu ^{(j)b}(%
\overrightarrow{z_j}\mid \overrightarrow{v}_1,\overrightarrow{v}_2;%
\overrightarrow{z_1},\overrightarrow{z_2};E)\overline{\Omega }_j^{b\nu }$%
\end{tabular}
\eqnum{3.45}
\end{equation}
in which ${\cal G}_\mu ^{(i)a}(\overrightarrow{x_i}\mid \overrightarrow{x_1}%
_{,}\overrightarrow{x_2};\overrightarrow{u_1},\overrightarrow{u_2};E)$ and $%
{\cal G}_\nu ^{(j)b}(\overrightarrow{z_j}\mid \overrightarrow{v}_1,%
\overrightarrow{v}_2;\overrightarrow{z_1},\overrightarrow{z_2};E)$ are the
Fourier transforms of the following Green's functions 
\begin{equation}
\begin{tabular}{l}
${\cal G}_\mu ^{(i)a}(\overrightarrow{x_i}\mid \overrightarrow{x_1}_{,}%
\overrightarrow{x_2};\overrightarrow{u_1},\overrightarrow{u_2};t_1-t_2)$ \\ 
$=\langle 0^{+}\left| T\{N[{\bf A}_\mu ^a(\overrightarrow{x_i},t_1){\bf \psi 
}(\overrightarrow{x_1},t_1){\bf \psi }^c(\overrightarrow{x_2},t_1)]N[%
\overline{{\bf \psi }}(\overrightarrow{u_1},t_2)\overline{{\bf \psi }}^c(%
\overrightarrow{u_2},t_2)]\}\right| 0^{-}\rangle $%
\end{tabular}
\eqnum{3.46}
\end{equation}
and 
\begin{equation}
\begin{tabular}{l}
${\cal G}_\nu ^{(j)b}(\overrightarrow{z_j}\mid \overrightarrow{v}_1,%
\overrightarrow{v}_2;\overrightarrow{z_1},\overrightarrow{z_2};t_1-t_2)$ \\ 
$=\langle 0^{+}\left| T\{N[{\bf \psi }(\overrightarrow{v_1},t_1){\bf \psi }%
^c(\overrightarrow{v_2},t_1)]N[{\bf A}_\nu ^b(\overrightarrow{z_j},t_2)%
\overline{{\bf \psi }}(\overrightarrow{z_1},t_2)\overline{{\bf \psi }}^c(%
\overrightarrow{z_2},t_2)]\}\right| 0^{-}\rangle $%
\end{tabular}
\eqnum{3.47}
\end{equation}
and $S^{-1}(\overrightarrow{z_1}_{,}\overrightarrow{z_2};\overrightarrow{y_1}%
,\overrightarrow{y_2})$ is the inverse of the function in Eq. (2.50).

\section{Another derivation of the interaction kernel}

The aim of this section is to give a different expression of the interaction
kernel in Eqs. (2.38) and (2.39) which will be derived by means of the
irreducible decomposition of the Green's functions ${\cal G}_\mu ^a(x_i\mid
x_1,x_2;y_1,y_2)$ denoted in Eq. (2.18). First, we start from the relation
between the full $q\overline{q}$ four-point Green's function $%
G(x_{1,}x_2;y_1,y_2)$ and its connected one $G_c(x_{1,}x_2;y_1,y_2)$ which
is derived in the beginning of Appendix B [18, 33] 
\begin{equation}
\begin{tabular}{l}
$G(x_{1,}x_2;y_1,y_2)=G_c(x_{1,}x_2;y_1,y_2)+S_F(x_1-y_1)S_F^c(x_2-y_2)$ \\ 
$-S_F^{*}(x_1-x_2)\overline{S}_F^{*}(y_1-y_2)$%
\end{tabular}
\eqnum{4.1}
\end{equation}
where all the fermion propagators were defined in Eqs. (2.6)-(2.9).
Substituting Eq. (4.1) into Eq. (2.4), we get 
\begin{equation}
{\cal G}(x_{1,}x_2;y_1,y_2)=G_c(x_{1,}x_2;y_1,y_2)+S_F(x_1-y_1)S_F^c(x_2-y_2)
\eqnum{4.2}
\end{equation}
where the last term is the unconnected part of the function ${\cal G}%
(x_{1,}x_2;y_1,y_2).$

From Eq. (B.5) given in Appendix B, we obtain by setting the source $J=0$
that [18, 33] 
\begin{equation}
\begin{tabular}{l}
$G_\mu ^a(x_i\mid x_{1,}x_2;y_1,y_2)=G_{c\mu }^a(x_i\mid
x_{1,}x_2;y_1,y_2)+\Lambda _\mu ^a(x_i\mid x_1;y_1)S_F^c(x_1-y_1)$ \\ 
$+S_F(x_1-y_1)\Lambda _\mu ^{{\bf c}a}(x_i\mid x_2;y_2)-\Lambda _\mu
^{a*}(x_i\mid x_1,x_2)\overline{S}_F^{*}(y_1-y_2)-S_F^{*}(x_1-x_2)\overline{%
\Lambda }_\mu ^{a*}(x_i\mid y_1,y_2)$%
\end{tabular}
\eqnum{4.3}
\end{equation}
where $i=1,2$, $G_{c\mu }^a(x_i\mid x_{1,}x_2;y_1,y_2)$ is the connected
part of the Green's function $G_\mu ^a(x_i\mid x_{1,}x_2;y_1,y_2)$ and $%
\Lambda _\mu ^{a*}(x_i\mid x_1,x_2)$, $\Lambda _\mu ^a(x_i\mid x_1;y_1)$, $%
\Lambda _\mu ^{{\bf c}a}(x_i\mid x_2;y_2)$ and $\overline{\Lambda }_\mu
^{a*}(x_i\mid y_1,y_2)$ are the three-point Green's functions which are
given in Eqs. (2.15) and (3.10). On inserting the decomposition in Eq. (4.3)
into Eq. (2.18), we see that the last unconnected term in Eq. (2.18) is
cancelled out. Thus, we have 
\begin{equation}
\begin{tabular}{l}
${\cal G}_\mu ^a(x_i\mid x_1,x_2;y_1,y_2)=G_{c\mu }^a(x_i\mid
x_{1,}x_2;y_1,y_2)+\Lambda _\mu ^a(x_i\mid x_1;y_1)S_F^c(x_2-y_2)$ \\ 
$+S_F(x_1-y_1)\Lambda _\mu ^{{\bf c}a}(x_i\mid x_2;y_2)-S_F^{*}(x_1-x_2)%
\overline{\Lambda }_\mu ^{a*}(x_i\mid y_1,y_2)$%
\end{tabular}
\eqnum{4.4}
\end{equation}
where $i=1,2$. Substituting the above expression in Eq. (2.22), we will
obtain the decomposition of the function ${\cal G}^{(i)}(x_i\mid
x_1,x_2;y_1,y_2).$

In the following, we are devoted to analyzing the terms on the RHS of Eq.
(4.4) through the one-particle-irreducible decompositions of the connected
Green's functions included in those terms. The decompositions have been
carried out in Appendix B.

\subsection{The t-channel one-gluon exchange kernel}

First we focus our attention on the second and third terms in Eq. (4.4).
According to the decomposition in Eq. (B.15), the three-point gluon-quark
Green's functions $\Lambda _\mu ^a(x_i\mid x_j;y_k)$ which is fully
connected can be represented in the form 
\begin{equation}
\Lambda _\mu ^a(x_i\mid x_j;y_k)=\int d^4z_1\Sigma _\mu ^a(x_i\mid
x_j;z_1)S_F(z_1-y_k)  \eqnum{4.5}
\end{equation}
where 
\begin{equation}
\Sigma _\mu ^a(x_i\mid x_j;z_1)=\int d^4u_1d^4u_2\Delta _{\mu \nu
}^{ab}(x_i-u_1)S_F(x_j-u_2)\Gamma ^{b\nu }(u_1\mid u_2,z_1)  \eqnum{4.6}
\end{equation}
in which 
\begin{equation}
\Delta _{\mu \nu }^{ab}(x_i-u_j)=\frac 1i\left\langle 0^{+}\left| T[{\bf A}%
_\mu ^a(x_i){\bf A}_\nu ^b(u_j)]\right| 0^{-}\right\rangle  \eqnum{4.7}
\end{equation}
is the exact gluon propagator and $\Gamma ^{b\nu }(u_1\mid u_2,z_1)$ is the
gluon-quark three-line proper vertex as defined in Eq. (B.17). The
one-particle-irreducible decompositions of the three-point gluon-antiquark
Green's functions $\Lambda _\mu ^{{\bf c}a}(x_i\mid x_j;y_k)$ can be
obtained from Eqs. (4.5) and (4.6) by the charge conjugation transformation.
The result is 
\begin{equation}
\Lambda _\mu ^{{\bf c}a}(x_i\mid x_j;y_k)=\int d^4z_2\Sigma _\mu ^{{\bf c}%
a}(x_i\mid x_j;z_2)S_F^c(z_2-y_k)  \eqnum{4.8}
\end{equation}
where 
\begin{equation}
\Sigma _\mu ^{ca}(x_i\mid x_j;z_2)=\int d^4u_1d^4u_2\Delta _{\mu \nu
}^{ab}(x_i-u_1)S_F^c(x_j-u_2)\Gamma _c^{b\nu }(u_1\mid u_2,z_2)  \eqnum{4.9}
\end{equation}
in which $\Gamma _c^{b\nu }(u_1\mid u_2,z_2)$ is the gluon-antiquark
three-line proper vertex as defined in Eq. (B.18).

When we set $i=j$ in Eqs. (4.5) and (4.8), it is found that the $\Sigma _\mu
^a(x_1\mid x_1;z_1)$ in Eq. (4.6) and the $\Sigma _\mu ^{{\bf c}a}(x_2\mid
x_2;z_2)$ in Eq. (4.9) are respectively related to the quark and antiquark
self-energies in such a way 
\begin{equation}
\begin{tabular}{l}
$\Omega _1^{a\mu }\Sigma _\mu ^a(x_1\mid x_1;z_1)=-\gamma _1^0\Sigma
(x_1,z_1)=\widehat{\Sigma }(x_1,z_1),$ \\ 
$\Omega _2^{a\mu }\Sigma _\mu ^{{\bf c}a}(x_2\mid x_2;z_2)=-\gamma
_2^0\Sigma ^c(x_2,z_2)=\widehat{\Sigma }^c(x_2,z_2)$%
\end{tabular}
\eqnum{4.10}
\end{equation}
Thus, from Eqs. (4.5), (4.8) and (4.10), we have 
\begin{equation}
\begin{tabular}{l}
$\Omega _1^{a\mu }\Lambda _\mu ^a(x_1\mid x_1;y_1)S_F^c(x_2-y_2)+\Omega
_2^{a\mu }\Lambda _\mu ^{{\bf c}a}(x_2\mid x_2;z)S_F(x_1-y_1)$ \\ 
$=\int d^4z_1d^4z_2\Sigma (x_1,x_2;z_1,z_2)S_F(z_1-y_1)S_F^c(z_2-y_2)$%
\end{tabular}
\eqnum{4.11}
\end{equation}
where 
\begin{equation}
\Sigma (x_1,x_2;z_1,z_2)=\widehat{\Sigma }(x_1,z_1)\delta ^4(x_2-z_2)+%
\widehat{\Sigma }^c(x_2,z_2)\delta ^4(x_1-z_1)  \eqnum{4.12}
\end{equation}
which is the total self-energy of the $q\overline{q}$ system. According to
the definitions given in Eqs. (4.4), (2.22)-(2.24), (2.31) and (4.2), we
see, $\Sigma (x_1,x_2;z_1,z_2)$ as a self energy term to appear in the
interaction kernel.

In the case of $i\neq j$, from Eqs. (4.5) and (4.8), it can be written 
\begin{equation}
\begin{tabular}{l}
$\Omega _2^{a\mu }\Lambda _\mu ^a(x_2\mid x_1;y_1)S_F^c(x_2-y_2)+\Omega
_1^{a\mu }\Lambda _\mu ^{{\bf c}a}(x_1\mid x_2;y_2)S_F(x_1-y_1)$ \\ 
$=\int d^4z_1d^4z_2K_t(x_1,x_2;z_1,z_2)S_F(z_1-y_1)S_F^c(z_2-y_2)$%
\end{tabular}
\eqnum{4.13}
\end{equation}
where 
\begin{equation}
K_t(x_1,x_2;z_1,z_2)=\Omega _2^{a\mu }\Sigma _\mu ^a(x_2\mid x_1;z_1)\delta
^4(x_2-z_2)+\Omega _1^{a\mu }\Sigma _\mu ^{{\bf c}a}(x_1\mid x_2;z_2)\delta
^4(x_1-z_1)  \eqnum{4.14}
\end{equation}
Based on Eqs. (4.4), (2.22)-(2.24), (2.31) and (4.2), it is clear that the $%
K_t(x_1,x_2;z_1,z_2)$ acts as a part of the interaction kernel to appear in
the D-S equation. As will be seen in section 6, this part is precisely the
kernel of t-channel one-gluon exchange interaction..

\subsection{The s-channel one-gluon exchange kernel}

Next, we turn to the last term in Eq. (4.4). The one-particle irreducible
decomposition of the three-point Green's function in this term may also be
found from Eqs. (4.5) and (4.6) by the charge conjugation transformation.
The result is 
\begin{equation}
\overline{\Lambda }_\mu ^{a*}(x_i\mid y_1,y_2)=-\int d^4z_1d^4z_2d^4z\Delta
_{\mu \nu }^{ab}(x_i-z)\overline{\Gamma }^{b\nu *}(z\mid
z_1,z_2)S_F(z_1-y_1)S_F^c(z_2-y_2)  \eqnum{4.15}
\end{equation}
where $\overline{\Gamma }^{b\nu *}(z\mid z_1,z_2)$ is the
gluon-quark-antiquark proper vertex defined in Eq. (B.19). With the
decomposition given above, according to Eqs. (2.22)-(2.24) and (2.31), the
contribution of the last term in Eq. (4.4) to the kernel $K(x_1,x_2;y_1,y_2)$
can be found from the sum 
\begin{equation}
-\sum_{i=1}^2\Omega _i^{a\mu }S_F^{*}(x_1-x_2)\overline{\Lambda }_\mu
^{a*}(x_i\mid y_1,y_2)=\int
d^4z_1d^4z_2K_s(x_1,x_2;z_1,z_2)S_F(z_1-y_1)S_F^c(z_2-y_2)  \eqnum{4.16}
\end{equation}
where 
\begin{equation}
K_s(x_1,x_2;z_1,z_2)=S_F^{*}(x_1-x_2)\int d^4z\sum_{i=1}^2\Omega _i^{a\mu
}\Delta _{\mu \nu }^{ab}(x_i-z)\overline{\Gamma }^{b\nu *}(z\mid z_1,z_2) 
\eqnum{4.17}
\end{equation}
is just the s-channel one-gluon exchange kernel occurring in the D-S
equation which will be discussed in section 6. It is noted here that the
s-channel one-gluon exchange describes the $q\overline{q}$ annihilation and
creation process which takes place between the quark (antiquark) in the
initial state and the antiquark (quark) in the final state as indicated by
the gluon propagator in Eq. (4.17) (not between the quark and the antiquark
both of which are simultaneously related to the initial state or the final
state B-S amplitude for a bound state).

\subsection{The kernel from the Green's function $G_{c\mu }^a(x_i\mid
x_1,x_2;y_1,y_2)$}

Now let us concentrate our attention on the irreducible decomposition of the
first term in Eq. (4.4). As stated in Appendix B, this decomposition may be
derived from the functional differential of the Green's function $%
G_c(x_{1,}x_2;y_1,y_2)$ with respect to the source $J^{a\mu }(x_i)$ by using
the one-particle irreducible decomposition of the function $%
G_c(x_{1,}x_2;y_1,y_2)$. The latter decomposition whose derivation is
sketched in Appendix B is well-known [18, 33] and can be represented in the
form 
\begin{equation}
\begin{tabular}{l}
$G_c(x_{1,}x_2;y_1,y_2)=\int
\prod\limits_{i=1}^2d^4u_id^4v_iS_F(x_1-u_1)S_F^c(x_2-u_2)$ \\ 
$\times \Gamma (u_1,u_2;v_1,v_2)S_F(v_1-y_1)S_F^c(v_2-y_2)$%
\end{tabular}
\eqnum{4.18}
\end{equation}
where 
\begin{equation}
\Gamma (u_1,u_2;v_1,v_2)=\sum_{i=1}^3\Gamma _i(u_1,u_2;v_1,v_2)  \eqnum{4.19}
\end{equation}
in which 
\begin{equation}
\Gamma _1(u_1,u_2;v_1,v_2)=-\int d^4z_1d^4z_2\Gamma ^{b\nu }(z_1\mid
u_1,v_1)D_{\nu \nu ^{\prime }}^{bb^{\prime }}(z_1-z_2)\Gamma _c^{b^{\prime
}\nu ^{\prime }}(z_2\mid u_2,v_2)  \eqnum{4.20}
\end{equation}
\begin{equation}
\Gamma _2(u_1,u_2;v_1,v_2)=\int d^4z_1d^4z_2\Gamma ^{*b\nu }(z_1\mid
u_1,u_2)D_{\nu \nu ^{\prime }}^{bb^{\prime }}(z_1-z_2)\overline{\Gamma }%
^{*b^{\prime }\nu ^{\prime }}(z_2\mid v_1,v_2)  \eqnum{4.21}
\end{equation}
here the three-line vertices are defined in Eqs. (B.17)-(B.19) and (B.21), $%
D_{\nu \nu ^{\prime }}^{bb^{\prime }}(z_1-z_2)=i\Delta _{\nu \nu ^{\prime
}}^{bb^{\prime }}(z_1-z_2)$ with $\Delta _{\nu \nu ^{\prime }}^{bb^{\prime
}}(z_1-z_2)$ defined in Eq. (4.7) and $\Gamma _3(u_1,u_2;v_1,v_2)$ defined
in Eq. (B.22) is the quark-antiquark four-line proper vertex. After
substituting the expressions in Eqs. (4.18)-(4.21), which are now given in
the presence of source $J$, into Eq. (B.6) and completing the
differentiation, the one-particle irreducible decomposition of the Green's
function $G_{c\mu }^a(x_i\mid x_{1,}x_2;y_1,y_2)$ will be found and,
thereby, we can write 
\begin{equation}
\begin{tabular}{l}
$\sum\limits_{i=1}^2G_c^{(i)}(x_i\mid x_{1,}x_2;y_1,y_2)\equiv \Omega
_1^{a\mu }G_{c\mu }^a(x_1\mid x_{1,}x_2;y_1,y_2)+\Omega _2^{a\mu }G_{c\mu
}^a(x_2\mid x_{1,}x_2;y_1,y_2)$ \\ 
$=\sum\limits_{j=1}^3G_j(x_{1,}x_2;y_1,y_2)$%
\end{tabular}
\eqnum{4.22}
\end{equation}
where 
\begin{equation}
\begin{tabular}{l}
$G_1(x_{1,}x_2;y_1,y_2)$ \\ 
$=\int \prod\limits_{j=1}^2d^4u_jd^4v_j\sum\limits_{i=1}^2\Omega _i^{a\mu
}[\Lambda _\mu ^a(x_i\mid x_1;u_1)S_F^c(x_2-u_2)+S_F(x_1-u_1)$ \\ 
$\times \Lambda _\mu ^{{\bf c}a}(x_i\mid x_2;u_2)]\Gamma
(u_1,u_2;v_1,v_2)S_F(v_1-y_1)S_F^c(v_2-y_2)$%
\end{tabular}
\eqnum{4.23}
\end{equation}
\begin{equation}
\begin{tabular}{l}
$G_2(x_{1,}x_2;y_1,y_2)$ \\ 
$=\int \prod\limits_{j=1}^2d^4u_jd^4v_j\sum\limits_{i=1}^2\Omega _i^{a\mu
}S_F(x_1-u_1)S_F^c(x_2-u_2)\Gamma (u_1,u_2;v_1,v_2)$ \\ 
$\times [\Lambda _\mu ^a(x_i\mid v_1;y_1)S_F^c(v_2-y_2)+\Lambda _\mu ^{{\bf c%
}a}(x_i\mid v_2;y_2)S_F(v_1-y_1)]$%
\end{tabular}
\eqnum{4.24}
\end{equation}
and 
\begin{equation}
\begin{tabular}{l}
$G_3(x_{1,}x_2;y_1,y_2)$ \\ 
$=\int \prod\limits_{j=1}^2d^4u_jd^4v_j\sum\limits_{i=1}^2\Omega _i^{a\mu
}S_F(x_1-u_1)S_F^c(x_2-u_2)$ \\ 
$\times \Gamma ^{a\mu }(x_i\mid u_1,u_2;v_1,v_2)S_F(v_1-y_1)S_F^c(v_2-y_2)$%
\end{tabular}
\eqnum{4.25}
\end{equation}
where $\Gamma ^{a\mu }(x_i\mid u_1,u_2;v_1,v_2)$ is a kind of five-line
vertex which is defined in Eq. (B.27) and will be specified soon later.

Before specifying the function $\Gamma ^{a\mu }(x_i\mid u_1,u_2;v_1,v_2)$,
we first analyze the expressions in Eqs. (4.23) and (4.24). According to the
expressions in Eqs. (4.11), (4.13) and (4.18), Eq. (4.23) may be represented
as 
\begin{equation}
G_1(x_{1,}x_2;y_1,y_2)=\int d^4z_1d^4z_2[\Sigma
(x_1,x_2;z_1,z_2)+K_t(x_1,x_2;z_1,z_2)]G_c(z_{1,}z_2;y_1,y_2)  \eqnum{4.26}
\end{equation}
where $\Sigma (x_1,x_2;z_1,z_2)$ and $K_t(x_1,x_2;z_1,z_2)$ were
respectively described in Eqs. (4.12) and (4.14). In view of the
decompositions in Eqs. (4.5) and (4.8), Eq. (4.24) may be written in the
form 
\begin{equation}
G_2(x_{1,}x_2;y_1,y_2)=\int
d^4z_1d^4z_2K_1(x_1,x_2;z_1,z_2)S_F(z_1-y_1)S_F^c(z_2-y_2)  \eqnum{4.27}
\end{equation}
where 
\begin{equation}
\begin{tabular}{l}
$K_1(x_1,x_2;z_1,z_2)=\int
\prod\limits_{j=1}^2d^4u_jd^4v\sum\limits_{i=1}^2\Omega _i^{a\mu
}S_F(x_1-u_1)S_F^c(x_2-u_2)$ \\ 
$\times [\Gamma (u_1,u_2;v,z_2)\Sigma _\mu ^a(x_i\mid v;z_1)+\Gamma
(u_1,u_2;z_1,v)\Sigma _\mu ^{{\bf c}a}(x_i\mid v;z_2)]$%
\end{tabular}
\eqnum{4.28}
\end{equation}
in which $\Sigma _\mu ^a(x_i\mid v;z_1)$ and $\Sigma _\mu ^{{\bf c}%
a}(x_i\mid v;z_2)$ were represented in Eqs (4.6) and (4.9) respectively.

Let us turn to the five-line vertex function contained in Eq. (4.25). This
vertex is two-particle reducible (or say, B-S reducible) although it is
one-particle-irreducible. Therefore, it can be decomposed into a B-S
irreducible part $\Gamma _{IR}^{a\mu }$ and a B-S reducible part $\Gamma
_{RE}^{a\mu }$ 
\begin{equation}
\Gamma ^{a\mu }(x_i\mid u_1,u_2;v_1,v_2)=\Gamma _{IR}^{a\mu }(x_i\mid
u_1,u_2;v_1,v_2)+\Gamma _{RE}^{a\mu }(x_i\mid u_1,u_2;v_1,v_2)  \eqnum{4.29}
\end{equation}
In order to exhibit the above decomposition specifically, as mentioned in
Appendix B, we may insert Eqs. (4.19)- (4.21), which are now given in the
case of presence of the source $J,$ into Eq. (B.27) and complete the
differentiation with respect to the source $J^{a\mu }(x_i)$. After
completing the differentiations shown in Eqs. (B.23) and (B.25), we can
write 
\begin{equation}
\Gamma ^{a\mu }(x_i\mid u_1,u_2;v_1,v_2)=\sum_{j=1}^3\Gamma _j^{a\mu
}(x_i\mid u_1,u_2;v_1,v_2)  \eqnum{4.30}
\end{equation}
where $\Gamma _j^{a\mu }(x_i\mid u_1,u_2;v_1,v_2)$ are given by the
differential of the functions $\Gamma _j(u_1,u_2;v_1,v_2)$ in Eq. (4.19) and
separately shown below 
\begin{equation}
\begin{tabular}{l}
$\Gamma _1^{a\mu }(x_i\mid u_1,u_2;v_1,v_2)$ \\ 
$=-\int d^4zd^4z_1d^4z_2D_{\mu \nu }^{ab}(x_i-z)\{\Gamma _\nu ^{bc,\lambda
}(z,z_1\mid u_1,v_1)D_{\lambda \lambda ^{\prime }}^{cc^{\prime
}}(z_1-z_2)\Gamma _{{\bf c}}^{c^{\prime }\lambda ^{\prime }}(z_2\mid
u_2,v_2) $ \\ 
$+\Gamma ^{c\lambda }(z_1\mid u_1,v_1)[D_{\lambda \lambda ^{\prime
}}^{cc^{\prime }}(z_1-z_2)\Gamma _{{\bf c}\nu }^{bc^{\prime },\lambda
}(z,z_2\mid u_2,v_2)+\Pi _{\nu \lambda \lambda ^{\prime }}^{bcc^{\prime
}}(z,z_1,z_2)\Gamma ^{c^{\prime }\lambda ^{\prime }}(z_2\mid u_2,v_2)]\}$%
\end{tabular}
\eqnum{4.31}
\end{equation}
in which besides the gluon-quark and gluon-antiquark three-line vertices
mentioned before, there occur the gluon-quark four-line proper vertex $%
\Gamma _{\nu \lambda }^{bc}(z,z_1\mid u_1,v_1)$ and the corresponding
gluon-antiquark one $\Gamma _{{\bf c}\nu \lambda }^{bc}(z,z_1\mid u_1,v_1)$
which are defined respectively in Eqs. (B.28) and (B.29) and 
\begin{equation}
\Pi _{\nu \rho \sigma }^{bcd}(z,z_1,z_2)=\int d^4u_1d^4u_2D_{\rho \rho
^{\prime }}^{cc^{\prime }}(z_1-u_1)\Gamma _{bc^{\prime }d^{\prime }}^{\nu
\rho ^{\prime }\sigma ^{\prime }}(z,u_1,u_2)D_{\sigma ^{\prime }\sigma
}^{d^{\prime }d}(u_2-z_2)  \eqnum{4.32}
\end{equation}
in which $\Gamma _{bc^{\prime }d^{\prime }}^{\nu \rho ^{\prime }\sigma
^{\prime }}(z,u_1,u_2)$ is the gluon three-line proper vertex defined in Eq.
(B.24), 
\begin{equation}
\begin{tabular}{l}
$\Gamma _2^{a\mu }(x_i\mid u_1,u_2;v_1,v_2)$ \\ 
$=\int d^4zd^4z_1d^4z_2D_{\mu \nu }^{ab}(x_i-z)\{\Gamma _\nu ^{bc,\lambda
*}(z,z_1\mid u_1,u_2)D_{\lambda \lambda ^{\prime }}^{cc^{\prime }}(z_1-z_2)%
\overline{\Gamma }_{{\bf c}}^{c^{\prime }\lambda ^{\prime }*}(z_2\mid
v_1,v_2)$ \\ 
$+\Gamma ^{c\lambda *}(z_1\mid u_1,u_2)[D_{\lambda \lambda ^{\prime
}}^{cc^{\prime }}(z_1-z_2)\overline{\Gamma }_{{\bf c}\nu }^{bc^{\prime
},\lambda *}(z,z_2\mid v_1,v_2)+\Pi _{\nu \lambda \lambda ^{\prime
}}^{bcc^{\prime }}(z,z_1,z_2)\ \overline{\Gamma }^{c^{\prime }\lambda
^{\prime }*}(z_2\mid v_1,v_2)]\}$%
\end{tabular}
\eqnum{4.33}
\end{equation}
in which $\Gamma _{\nu \lambda }^{bc*}(z,z_1\mid u_1,u_2)$ and $\overline{%
\Gamma }_{{\bf c}\nu \lambda }^{bc^{*}}(z,z_1\mid v_1,v_2)$ are the
gluon-quark-antiquark four-line proper vertices defined in Eqs. (B.30) and
(B.31) and especially 
\begin{equation}
\Gamma _3^{a\mu }(x_i\mid u_1,u_2;v_1,v_2)=\int d^4zD_{\mu \nu }^{ab}(x_i-z)%
\widehat{\Gamma }^{b\nu }(z\mid u_1,u_2;v_1,v_2)  \eqnum{4.34}
\end{equation}
in which $\widehat{\Gamma }^{b\nu }(z\mid u_1,u_2;v_1,v_2)$ stands for the
gluon-quark-antiquark five-line proper vertex defined in Eq. (B.33).

It is pointed out that the vertices formulated in Eqs. (4.31) and (4.33) are
not only one-particle-irreducible, but also B-S irreducible. This point can
be seen more clearly when the vertices are represented by their Feynman
diagrams. From the diagrams, one can find that it is impossible to divide
each of the diagrams into two unconnected parts by cutting two fermion
lines. However, the five-line proper vertex $\widehat{\Gamma }^{b\nu }(z\mid
u_1,u_2;v_1,v_2)$ in Eq. (4.34) is B-S reducible. It can be decomposed into
a B-S reducible part $\widehat{\Gamma }_{RE}^{b\nu }$ and a B-S irreducible
part $\widehat{\Gamma }_{IR}^{b\nu },$%
\begin{equation}
\widehat{\Gamma }^{b\nu }(z\mid u_1,u_2;v_1,v_2)=\widehat{\Gamma }%
_{IR}^{b\nu }(z\mid u_1,u_2;v_1,v_2)+\widehat{\Gamma }_{RE}^{b\nu }(z\mid
u_1,u_2;v_1,v_2)  \eqnum{4.35}
\end{equation}
This enables us to write Eq. (4.34) as follows: 
\begin{equation}
\Gamma _3^{a\mu }(x_i\mid u_1,u_2;v_1,v_2)=\Gamma _{31}^{a\mu }(x_i\mid
u_1,u_2;v_1,v_2)+\Gamma _{32}^{a\mu }(x_i\mid u_1,u_2;v_1,v_2)  \eqnum{4.36}
\end{equation}
where 
\begin{equation}
\Gamma _{31}^{a\mu }(x_i\mid u_1,u_2;v_1,v_2)=\int d^4zD_{\mu \nu
}^{ab}(x_i-z)\widehat{\Gamma }_{IR}^{b\nu }(z\mid u_1,u_2;v_1,v_2) 
\eqnum{4.37}
\end{equation}
and 
\begin{equation}
\Gamma _{32}^{a\mu }(x_i\mid u_1,u_2;v_1,v_2)=\int d^4zD_{\mu \nu
}^{ab}(x_i-z)\widehat{\Gamma }_{RE}^{b\nu }(z\mid u_1,u_2;v_1,v_2) 
\eqnum{4.38}
\end{equation}
From the above statement, it is clearly seen that the B-S irreducible part
of the vertex in Eq. (4.29) is given by the sum 
\begin{equation}
\begin{tabular}{l}
$\Gamma _{IR}^{a\mu }(x_i\mid u_1,u_2;v_1,v_2)=\Gamma _1^{a\mu }(x_i\mid
u_1,u_2;v_1,v_2)$ \\ 
$+\Gamma _2^{a\mu }(x_i\mid u_1,u_2;v_1,v_2)+\Gamma _{31}^{a\mu }(x_i\mid
u_1,u_2;v_1,v_2)$%
\end{tabular}
\eqnum{4.39}
\end{equation}
where the three terms on the RHS of Eq. (4.39) were given in Eqs. (4.31),
(4.33) and (4.37) respectively. While, the B-S reducible part in Eq. (4.29)
is given by Eq. (4.38) 
\begin{equation}
\Gamma _{RE}^{a\mu }(x_i\mid u_1,u_2;v_1,v_2)=\Gamma _{32}^{a\mu }(x_i\mid
u_1,u_2;v_1,v_2)  \eqnum{4.40}
\end{equation}

Based on the decomposition in Eq. (4.29), the function in Eq. (4.25) will be
decomposed into 
\begin{equation}
G_3(x_{1,}x_2;y_1,y_2)=G_3^{(1)}(x_{1,}x_2;y_1,y_2)+G_3^{(2)}(x_{1,}x_2;y_1,y_2)
\eqnum{4.41}
\end{equation}
where 
\begin{equation}
G_3^{(1)}(x_{1,}x_2;y_1,y_2)=\int
d^4z_1d^4z_2K_2(x_1,x_2;z_1,z_2)S_F(z_1-y_1)S_F^c(z_2-y_2)  \eqnum{4.42}
\end{equation}
in which 
\begin{equation}
K_2(x_1,x_2;z_1,z_2)=\int \prod_{j=1}^2d^4u_j\sum_{i=1}^2\Omega _i^{a\mu
}S_F(x_1-u_1)S_F^c(x_2-u_2)\Gamma _{IR}^{a\mu }(x_i\mid u_1,u_2;z_1,z_2) 
\eqnum{4.43}
\end{equation}
and 
\begin{equation}
\begin{tabular}{l}
$G_3^{(2)}(x_{1,}x_2;y_1,y_2)=\int
\prod\limits_{j=1}^2d^4u_jd^4v_j\sum\limits_{i=1}^2\Omega _i^{a\mu
}S_F(x_1-u_1)S_F^c(x_2-u_2)$ \\ 
$\times \Gamma _{RE}^{a\mu }(x_i\mid
u_1,u_2;v_1,v_2)S_F(v_1-y_1)S_F^c(v_2-y_2)$%
\end{tabular}
\eqnum{4.44}
\end{equation}
It is emphasized that due to the B-S irreducibility of the vertex $\Gamma
_{IR}^{a\mu }(x_i\mid u_1,u_2;v_1,v_2)$, the function $%
G_3^{(1)}(x_{1,}x_2;y_1,y_2)$ can only be written in the form as shown in
Eq. (4.42). While, since the vertex $\Gamma _{RE}^{a\mu }(x_i\mid
u_1,u_2;v_1,v_2)$ is B-S reducible, as was similarly done for the Green's
function ${\cal G}^{(i)}(x_1,x_2;y_1,y_2)$, the function $%
G_3^{(2)}(x_{1,}x_2;y_1,y_2),$ as a part of the connected Green's function,
must be represented in the B-S reducible form 
\begin{equation}
G_3^{(2)}(x_{1,}x_2;y_1,y_2)=\int d^4z_1d^4z_2\widetilde{K}%
(x_1,x_2;z_1,z_2)G_c(z_{1,}z_2;y_1,y_2)  \eqnum{4.45}
\end{equation}
where $\widetilde{K}(x_1,x_2;z_1,z_2)$ is a kind of kernel needed to be
determined later.

\subsection{Complete expression of the interaction kernel}

Up to the present, the irreducible decompositions of the functions ${\cal G}%
^{(i)\text{ }}(x_1,x_2;y_1,y_2)$ $(i=1,2)$ appearing in Eq. (2.24) have been
completed. Now, let us first sum up the expressions given in Eqs. (4.11) and
(4.13) which correspond to the second and third terms in Eq. (4.4) and the
expression in Eq. (4.26) for the function $G_1(x_{1,}x_2;y_1,y_2)$ which is
contained in the connected Green's functions $G_{c\mu }^a(x_i\mid
x_{1,}x_2;y_1,y_2)$. The summation yields 
\begin{equation}
\int d^4z_1d^4z_2[\Sigma (x_1,x_2;z_1,z_2)+K_t(x_1,x_2;z_1,z_2)]{\cal G}%
(z_{1,}z_2;y_1,y_2)  \eqnum{4.46}
\end{equation}
where the relation in Eq. (4.2) has been considered. Then, we combine the
expression in Eq. (4.16) which corresponds to the last term in Eq. (4.4) and
the expressions in Eqs. (4.27) and (4.42) which come from the B-S
irreducible part of connected Green's functions $G_{c\mu }^a(x_i\mid
x_{1,}x_2;y_1,y_2)$ and obtain 
\begin{equation}
\int
d^4z_1d^4z_2[K_s(x_1,x_2;z_1,z_2)+K_1(x_1,x_2;z_1,z_2)+K_2(x_1,x_2;z_1,z_2)]S_F(z_1-y_1)S_F^c(z_2-y_2)
\eqnum{4.47}
\end{equation}
The final decomposition of the sum of the functions ${\cal G}^{(1)}(x_1\mid
x_1,x_2;y_1,y_2)$ and ${\cal G}^{(2)}(x_2\mid x_1,x_2;y_1,y_2)$ will be
given by the sum of Eqs. (4.45)-(4.47). Obviously, in order to make the D-S
equation to be closed, the kernel In Eq. (4.45) must be 
\begin{equation}
\widetilde{K}%
(x_1,x_2;z_1,z_2)=K_s(x_1,x_2;z_1,z_2)+K_1(x_1,x_2;z_1,z_2)+K_2(x_1,x_2;z_1,z_2)
\eqnum{4.48}
\end{equation}
Thus, the summation of Eqs. (4.45) and (4.47) gives 
\begin{equation}
\int
d^4z_1d^4z_2[K_s(x_1,x_2;z_1,z_2)+K_1(x_1,x_2;z_1,z_2)+K_2(x_1,x_2;z_1,z_2)]%
{\cal G}(z_{1,}z_2;y_1,y_2)  \eqnum{4.49}
\end{equation}
Combining Eqs. (4.46) and (4.49), we eventually arrive at 
\begin{equation}
\sum_{i=1}^2{\cal G}^{(i)}(x_i\mid x_1,x_2;y_1,y_2)=\int
d^4z_1d^4z_2K(x_1,x_2;z_1,z_2){\cal G}(z_{1,}z_2;y_1,y_2)  \eqnum{4.50}
\end{equation}
where 
\begin{equation}
\begin{tabular}{l}
$K(x_1,x_2;z_1,z_2)=\Sigma (x_1,x_2;z_1,z_2)+K_t(x_1,x_2;z_1,z_2)$ \\ 
$+K_s(x_1,x_2;z_1,z_2)+K_1(x_1,x_2;z_1,z_2)+K_2(x_1,x_2;z_1,z_2)$%
\end{tabular}
\eqnum{4.51}
\end{equation}
This just is the kernel appearing in Eq. (2.38). The five terms on the RHS
of Eq. (4.51), as respectively shown in Eqs. (4.12), (4.14), (4.17), (4.28)
and (4.43), are only represented in terms of the quark, antiquark and gluon
propagators and some kinds of three, four and five-line proper vertices and
therefore exhibits a more specific structure of the kernel. The equivalence
between the both expressions given in the preceding section and this section
for the kernel will be illustrated in section 6 for the one-gluon exchange
kernels. The exact proof of the equivalence has been done by the technique
of irreducible decomposition of the Green's functions. From the proof, we
find that the expression described in this section can surely be obtained
from the expression given in the preceding section.

\section{Pauli-Schr\"odinger equation}

As mentioned in Introduction, the D-S equations formulated in the Dirac
spinor space may be reduced to an equivalent equations represented in the
Pauli spinor space with the help of Dirac spinors. Let us start from the
equation given in Eq. (2.44). The Dirac spinors are defined as [30] 
\begin{equation}
U(\overrightarrow{p})=\sqrt{\frac{\omega +m}{2\omega }}\left( 
\begin{array}{c}
1 \\ 
\frac{\overrightarrow{\sigma }\cdot \overrightarrow{p}}{\omega +m}
\end{array}
\right)  \eqnum{5.1}
\end{equation}
\begin{equation}
V(\overrightarrow{p})=\sqrt{\frac{\omega +m}{2\omega }}\left( 
\begin{array}{c}
-\frac{\overrightarrow{\sigma }\cdot \overrightarrow{p}}{\omega +m} \\ 
1
\end{array}
\right)  \eqnum{5.2}
\end{equation}
where $\overrightarrow{\sigma \text{ }}$ are the Pauli matrices, $\omega =(%
\overrightarrow{p}^2+m^2)^{1/2}$, $U(\overrightarrow{p})$ and $V(%
\overrightarrow{p})$ are the positive energy and negative energy spinors
respectively. They satisfy the orthonornality relations 
\begin{equation}
\begin{array}{c}
U^{+}(\overrightarrow{p})U(\overrightarrow{p})=V^{+}(\overrightarrow{p})V(%
\overrightarrow{p})=1 \\ 
U^{+}(\overrightarrow{p})V(\overrightarrow{p})=V^{+}(\overrightarrow{p})U(%
\overrightarrow{p})=0
\end{array}
\eqnum{5.3}
\end{equation}
and the completeness relation 
\begin{equation}
\Lambda ^{+}(\overrightarrow{p})+\Lambda ^{-}(\overrightarrow{p})=1 
\eqnum{5.4}
\end{equation}
where $\Lambda ^{+}(\overrightarrow{p})$ and $\Lambda ^{-}(\overrightarrow{p}%
)$ are respectively the positive and negative energy state projection
operators defined by 
\begin{equation}
\Lambda ^{+}(\overrightarrow{p})=U(\overrightarrow{p})U^{+}(\overrightarrow{p%
}),\text{ }\Lambda ^{-}(\overrightarrow{p})=V(\overrightarrow{p})V^{+}(%
\overrightarrow{p})  \eqnum{5.5}
\end{equation}
Define 
\begin{equation}
W_a(\overrightarrow{p})=\left\{ 
\begin{array}{c}
U(\overrightarrow{p}),\text{ }if\text{ }a=+, \\ 
V(\overrightarrow{p}),\text{ }if\text{ }a=-,
\end{array}
\right\}  \eqnum{5.6}
\end{equation}
then, the two fermion spinors can be written as 
\begin{equation}
W_{ab}(\overrightarrow{P},\overrightarrow{q})=W_a(\overrightarrow{p_1})W_b(%
\overrightarrow{p_2})  \eqnum{5.7}
\end{equation}
With this definition, the completeness relation for two fermion spinors can
be represented as 
\begin{equation}
\sum_{ab}W_{ab}(\overrightarrow{P},\overrightarrow{q})W_{ab}^{+}(%
\overrightarrow{P},\overrightarrow{q})=1  \eqnum{5.8}
\end{equation}

Premiltiplying Eq. (2.44) with $W_{ab}^{+}(\overrightarrow{P},%
\overrightarrow{q})$, sandwiching Eq. (5.8) between the kernel $K(P,q,k)$
and the amplitude $\chi _{P\varsigma }(k)$ on the RHS of Eq. (2.44) and
applying the Dirac equation 
\begin{equation}
h(\overrightarrow{p})W_a(\overrightarrow{p})=a\omega (\overrightarrow{p})W_a(%
\overrightarrow{p})  \eqnum{5.9}
\end{equation}
we obtain

\begin{equation}
\Delta _{ab}(P,\overrightarrow{q})\phi _{ab}(P,q)=\sum_{cd}\int \frac{d^4k}{%
(2\pi )^4}K_{abcd}(P,q,k)\phi _{cd}(P,k)  \eqnum{5.10}
\end{equation}
where $a,b,c,d=\pm 1$, $P=(E,\overrightarrow{P})$, 
\begin{equation}
\phi _{ab}(P,q)=W_{ab}^{+}(\overrightarrow{P},\overrightarrow{q})\chi
_{P\varsigma }(q)  \eqnum{5.11}
\end{equation}
\begin{equation}
\Delta _{ab}(P,\overrightarrow{q})=E-a\omega _1(\overrightarrow{p_1}%
)-b\omega _2(\overrightarrow{p_2})  \eqnum{5.12}
\end{equation}
\begin{equation}
K_{abcd}(P,q,k)=W_{ab}^{+}(\overrightarrow{P},\overrightarrow{q}%
)K(P,q,k)W_{cd}(\overrightarrow{P},\overrightarrow{k})  \eqnum{5.13}
\end{equation}
Eq. (5.10) is a set of coupled equations satisfied by the amplitudes $\phi
_{ab}(P,q)$ each of which is represented in the Pauli spinor space and of
dimension four. In the infinite-dimensional space of the momentum $q$ or $k$%
, according to $ab=++$ and $ab\neq ++$, Eq. (5.10) may be, in the matrix
form, separately written as 
\begin{equation}
\Delta _{++}(p)\phi _{++}(P)=K_{++++}(P)\phi _{++}(P)+\sum_{cd\neq
++}K_{++cd}(P)\phi _{cd}(P)  \eqnum{5.14}
\end{equation}
and

\begin{equation}
\Delta _{ab}(P)\phi _{ab}(P)=K_{ab++}(P)\phi _{++}(P)+\sum_{cd\neq
++}K_{abcd}(P)\phi _{cd}(P)  \eqnum{5.15}
\end{equation}
where $ab\neq ++$ and the terms related to $\phi _{++}(P)$ have been
separated out from the others. Furthermore, In the three-dimensional spinor
space spanned by $\phi _{ab}(P)$ with $ab\neq ++$, Eqs. (5.14) and (5.15)
may be written in the full matrix form 
\begin{equation}
\Delta _{+}(P)\psi (P)=K_{+}(P)\psi (P)+\overline{K}^t(P)\phi (P) 
\eqnum{5.16}
\end{equation}
and 
\begin{equation}
\phi (P)=\overline{G}(P)\psi (P)+G(P)\phi (P)  \eqnum{5.17}
\end{equation}
where $\psi (P)=\phi _{++}(P)$, $\Delta _{+}(P)=\Delta _{++}(P)$, $%
K_{+}(P)=K_{++++}(P)$, while, $\phi (P)=\{\phi _{ab}(P)\}$, $\overline{K}%
^t(P)=\{K_{++cd}(P)\}$, $\overline{G}(P)=\{K_{ab++}(P)/\Delta _{ab}(P)\}$
and $G(P)=\{K_{abcd}(P)/\Delta _{ab}(P)\}$ represent the matrices in the
three-dimensional spinor space. Solving the equation (5.17), we obtain 
\begin{equation}
\phi (P)=\frac 1{1-G(P)}\overline{G}(P)\psi (P)  \eqnum{5.18}
\end{equation}
Substituting the above expression into Eq. (5.16), we finally arrive at 
\begin{equation}
\Delta _{+}(P)\psi (P)=V(P)\psi (P)  \eqnum{5.19}
\end{equation}
where 
\begin{equation}
V(P)=K_{+}(P)+\overline{K}^t(P)\frac 1{1-G(P)}\overline{G}(P)  \eqnum{5.20}
\end{equation}
which identifies itself with the interaction Hamiltonian. With the
definition 
\begin{equation}
\frac 1{1-G(P)}=\sum_{n=0}^\infty G^{(n)}(P),  \eqnum{5.21}
\end{equation}
Eq. (5.20) can be written as 
\begin{equation}
V(P)=\sum_{n=0}^\infty V^{(n)}(P)  \eqnum{5.22}
\end{equation}
where 
\begin{equation}
\begin{tabular}{l}
$V^{(0)}(P)=K_{+}(P),$ \\ 
$V^{(1)}(P)=\overline{K}^t(P)\overline{G}(P),$ \\ 
$V^{(2)}(P)=\overline{K}^t(P)G(P)\overline{G}(P),$ \\ 
$\cdot \cdot \cdot \cdot \cdot \cdot \cdot $%
\end{tabular}
\eqnum{5.23}
\end{equation}
Written out explicitly, Eq. (5.19) reads 
\begin{equation}
\lbrack E-\omega _1(\overrightarrow{p_1})-\omega _2(\overrightarrow{p_2}%
)]\psi (P,q)=\int \frac{d^4k}{(2\pi )^4}V(P,q,k)\psi (P,k)  \eqnum{5.24}
\end{equation}
The terms in the interaction Hamiltonian in Eq. (5.23) are specified as 
\begin{equation}
V^{(0)}(P,q,k)=K_{++++}(P,q,k)  \eqnum{5.25}
\end{equation}
\begin{equation}
V^{(1)}(P,q,k)=\sum_{ab\neq ++}\int \frac{d^4l}{(2\pi )^4}\frac{%
K_{++ab}(P,q,l)K_{ab++}(P,l,k)}{E-a\overline{\omega }_1(\overrightarrow{l})-b%
\overline{\omega }_2(\overrightarrow{l})}  \eqnum{5.26}
\end{equation}
\begin{equation}
V^{(2)}(P,q,k)=\sum_{ab\neq ++}\sum_{cd\neq ++}\int \frac{d^4l_1}{(2\pi )^4}%
\frac{d^4l_2}{(2\pi )^4}\frac{%
K_{++ab}(P,q,l_1)K_{abcd}(P,l_1,l_2)K_{cd++}(P,l_2,k)}{[E-a\overline{\omega }%
_1(\overrightarrow{l_1})-b\overline{\omega }_2(\overrightarrow{l_1})][E-a%
\overline{\omega }_1(\overrightarrow{l_2})-b\overline{\omega }_2(%
\overrightarrow{l_2})]}  \eqnum{5.27}
\end{equation}
and so on, where for simplicity of representation, we have defined $%
\overline{\omega }_1(\overrightarrow{l})=\omega _1(\eta _1\overrightarrow{P}+%
\overrightarrow{l})$ and $\overline{\omega }_2(\overrightarrow{l})=\omega
_2(\eta _2\overrightarrow{P}-\overrightarrow{l})$. In the center of mass
frame, $\overline{\omega }_i(\overrightarrow{l})=\omega _i(\overrightarrow{l}%
)$ ($i=1,2$). Eq. (5.24) is the equation satisfied by the positive energy
state amplitude $\psi (P,q)$ which is of dimension four in the two-fermion
Pauli spinor space. This is the reason why the above equation is called
Pauli-Schr\"odinger (P-S) equation.

By the same procedure, the D-S equation in Eq. (2.45) may also be reduced to
a corresponding P-S equation as represented in the following 
\begin{equation}
\lbrack q_0-\eta _2\omega _1(\overrightarrow{p_1})+\eta _1\omega _2(%
\overrightarrow{p_2})]\psi (P,q)=\int \frac{d^4k}{(2\pi )^4}\overline{V}%
(P,q,k)\psi (P,k)  \eqnum{5.28}
\end{equation}
where $q_0$ is the relative energy and $\overline{V}(P,q,k)$ is a kind of
interaction Hamiltonian which can be written out from the expression of $%
V(P,q,k)$ by the replacement: $K(P,q,k)\rightarrow \overline{K}(P,q,k)$ and $%
\Delta _{ab}(P,\overrightarrow{q})$ $\rightarrow \overline{\Delta }_{ab}(%
\overrightarrow{P},q)=q_0-a\eta _2\omega _1(\overrightarrow{p_1})+b\eta
_1\omega _2(\overrightarrow{p_2})$. For the three-dimensional D-S equation,
the P-S equation in Eq. (5.28) disappears. We are left only with a
three-dimensional P-S equation derived from Eq. (2.56) such that 
\begin{equation}
\lbrack E-\omega _1(\overrightarrow{p_1})-\omega _2(\overrightarrow{p_2}%
)]\psi (P,\overrightarrow{q})=\int \frac{d^3k}{(2\pi )^3}V(P,\overrightarrow{%
q},\overrightarrow{k})\psi (P,\overrightarrow{k})  \eqnum{5.29}
\end{equation}
where the Hamiltonian $V(P,\overrightarrow{q},\overrightarrow{k})$ formally
has the same expressions as written in Eqs. (5.22) and (5.25)-(5.27) except
that the four-dimensional kernel $K(P,q,k)$ in those expressions is now
replaced by the three-dimensional one $K(P,\overrightarrow{q},%
\overrightarrow{k})$ which is the Fourier transform of the kernel in Eq.
(3.40)

It is worthy to point out that for a given kernel in the D-S equation, there
are a series of terms (the ladder diagrams) to appear in the interaction
Hamiltonian in the P-S equation. If the D-S equation with a given kernel
could be solved, the contribution arising from a series of ladder diagrams
characterized by the series of terms in the Hamiltonian are precisely taken
into account. Another point we would like to stress is that as seen from
Eqs. (5.26) and (5.27), the negative energy state only acts as intermediate
states to appear in the interaction Hamiltonian. Particularly, for the bound
state, the positive energy state does not appear in the intermediate states.
While, for the scattering state P-S equation as discussed in Ref. [30], the
intermediate states in the interaction Hamiltonian must include the positive
energy state. In this case, the series expansion of the interaction
Hamiltonian in Eq. (5.22) has an one-to-one correspondence with the
perturbative expansion of the S-matrix. The above statement reveals an
essential difference between the interactions taking place in the bound
state and the scattering state.

\section{One-gluon exchange kernels}

In this section, we limit ourself to give a brief derivation and description
of the one gluon exchange kernel (OGEK). First we discuss the t-channel OGEK
and the s-channel OGEK appearing in the four-dimensional D-S equation to
illustrate the equivalence between the expressions of the interaction kernel
derived in the sections 3 and 4. Then, we show the OGEK in the
three-dimensional D-S equation and the corresponding Hamiltonian in the P-S
equation.

\subsection{The t-channel one-gluon exchange kernel}

The exact form of the t-channel OGEK was represented in Eq. (4.14) with $%
\Sigma _\mu ^a(x_2\mid x_1;z_1)$ and $\Sigma _\mu ^{{\bf c}a}(x_1\mid
x_2;z_2)$ given in Eqs. (4.6) and (4.9). In the lowest-order approximation,
the propagators and the vertices in Eqs. (4.6) and (4.9) are taken
respectively to be the free ones and the bare ones. The bare vertices are of
the form 
\begin{equation}
\begin{array}{c}
\Gamma ^{b\nu }(u_1\mid u_2,z_1)=-ig\gamma ^\nu T^b\delta ^4(u_1-u_2)\delta
^4(u_2-z_1) \\ 
\Gamma _c^{b\nu }(u_1\mid u_2,z_2)=-ig\gamma ^\nu \overline{T}^b\delta
^4(u_1-u_2)\delta ^4(u_2-z_2)
\end{array}
\eqnum{6.1}
\end{equation}
With the vertices given above, the kernel in Eq. (4.14) becomes 
\begin{equation}
\begin{tabular}{l}
$K_t^0(X-Y,x,y)=ig^2T^a\overline{T}^b\{\Delta _{\mu \nu
}^{ab}(x_2-y_1)S_F(x_1-y_1)\gamma _1^\mu \gamma _2^0\gamma _2^\nu \delta
^4(x_2-y_2)$ \\ 
$+\Delta _{\mu \nu }^{ab}(x_1-y_2)S_F^c(x_2-y_2)\gamma _2^\nu \gamma
_1^0\gamma _1^\mu \delta ^4(x_1-y_1)\}$%
\end{tabular}
\eqnum{6.2}
\end{equation}
From now on, the $S_F(x-y)$ and $\Delta _{\mu \nu }^{ab}(x-y)$ in the above
are understood to be free propagators. By the Fourier transformation, we get
in the momentum space that 
\begin{equation}
K_t^0(P,q,k)=S(P,q)ig^2T^a\overline{T}^b\Delta _{\mu \nu }^{ab}(q-k)\gamma
_1^\mu \gamma _2^\nu  \eqnum{6.3}
\end{equation}
where 
\begin{equation}
S(P,q)=S_F(p_1)\gamma _2^0+S_F^c(p_2)\gamma _1^0=[\widehat{S}_F(p_1)+%
\widehat{S}_F^c(p_2)]\gamma _1^0\gamma _2^0  \eqnum{6.4}
\end{equation}
in which 
\begin{equation}
S_F(p)=\widehat{S}_F(p)\gamma ^0  \eqnum{6.5}
\end{equation}
\begin{equation}
\widehat{S}_F(p)=\frac 1{p_0-h(\overrightarrow{p})+i\varepsilon }=\frac{%
\Lambda ^{+}(\overrightarrow{p})}{p_0-\omega (\overrightarrow{p}%
)+i\varepsilon }+\frac{\Lambda ^{-}(\overrightarrow{p})}{p_0+\omega (%
\overrightarrow{p})-i\varepsilon }  \eqnum{6.6}
\end{equation}
here $h(\overrightarrow{p})$ is the free fermion Hamiltonian, $\Lambda ^{+}(%
\overrightarrow{p})$ and $\Lambda ^{-}(\overrightarrow{p})$ were defined in
Eq. (5.5).

Let us turn to derive the above kernel from the closed expression presented
in Eq. (3.30). In the perturbative approximation of order $g^2$, only the
first and second terms in Eq. (3.30) can contribute to the OGEK. In the
first term which was defined in Eqs. (3.31), (3.23) and (3.21), there are
four terms: two represent the self-energies of quark and antiquark and the
other two are related to the t-channel one-gluon exchange interaction which
is concerned here only. The $\Lambda _\mu ^a(x_2\mid x_1,z_1)$ and $\Lambda
_\mu ^{{\bf c}a}(x_1\mid x_2,z_2)$ in Eq. (3.21) was respectively
represented in Eqs. (4.5), (4.6), (4.8) and (4.9). When the vertices are
taken to be the bare ones shown in Eq. (6.1), the terms contained in the $%
{\cal R}(x_1,x_2;z_1,z_2)$ which contributes to the OGEK can be written as 
\begin{equation}
\begin{tabular}{l}
$H_1^t(x_1,x_2;z_1,z_2)=ig^2T^a\overline{T}^b\int d^4u\{\Delta _{\mu \nu
}^{ab}(x_2-u)\gamma _2^0\gamma _2^\mu \gamma _2^0S_F(x_1-u)\gamma _1^\nu
S_F(u-z_1)\delta ^4(x_2-z_2)$ \\ 
$+\Delta _{\mu \nu }^{ab}(x_1-u)\gamma _1^0\gamma _1^\mu \gamma
_1^0S_F^c(x_2-u)\gamma _2^\nu S_F^c(u-z_2)\delta ^4(x_1-z_1)\}$%
\end{tabular}
\eqnum{6.7}
\end{equation}

Next, we turn to the second term in Eq. ( 3.30) which was defined in Eqs.
(3.32) and (3.14). Through a perturbative calculation of the Green's
function ${\cal G}_{\mu \nu }^{ab}(x_i,z_j\mid x_1,x_2;z_1,z_2)$ defined in
Eq. (3.14) or performing a decomposition of the Green's function into the
connected ones, one may find that there are a function $i\Delta _{\mu \nu
}^{ab}(x_i-z_j)S_F(x_1-z_1)S_F^c(x_2-z_2)$ included in the function ${\cal G}%
_{\mu \nu }^{ab}(x_i,z_j\mid x_1,x_2;z_1,z_2)$ which just is related to the
t-channel OGEK when $i\neq j$. Thus, according to Eq. (3.32), the term
included in the ${\cal Q}(x_1,x_2;z_1,z_2)$ which contributes to the OGEK
may be written as: 
\begin{equation}
\begin{tabular}{l}
$H_2^t(x_1,x_2;z_1,z_2)=ig^2T^a\overline{T}^b\{\Delta _{\mu \nu
}^{ab}(x_1-z_2)\gamma _1^0\gamma _1^\mu S_F(x_1-z_1)S_F^c(x_2-z_2)\gamma
_2^\nu \gamma _2^0$ \\ 
$+\Delta _{\mu \nu }^{ab}(x_2-z_1)\gamma _2^0\gamma _2^\mu
S_F^c(x_2-z_2)S_F(x_1-z_1)\gamma _1^\nu \gamma _1^0\}$%
\end{tabular}
\eqnum{6.8}
\end{equation}
Substituting Eqs. (6.7) and (6.8) into Eq. (3.30), we will obtain the
expression of the t-channel OGEK $K_t^0(X-Y,x,y)$. By Fourier
transformation, its expression given in the momentum space may be found to
be 
\begin{equation}
K_t^0(P,q,k)=\sum\limits_{i=1}^2H_i^t(P,q,k)S^{-1}(P,k)  \eqnum{6.9}
\end{equation}
where 
\begin{equation}
\begin{tabular}{l}
$H_1^t(P,q,k)=ig^2T^a\overline{T}^b\Delta _{\mu \nu }^{ab}(q-k)\{\gamma
_2^0\gamma _2^\mu \gamma _2^0S_F(p_1)\gamma _1^\nu S_F(q_1)$ \\ 
$+\gamma _1^0\gamma _1^\mu \gamma _1^0S_F^c(p_2)\gamma _2^\nu S_F^c(q_2)\}$%
\end{tabular}
\eqnum{6.10}
\end{equation}
and 
\begin{equation}
\begin{tabular}{l}
$H_2^t(P,q,k)=ig^2T^a\overline{T}^b\Delta _{\mu \nu }^{ab}(q-k)\{\gamma
_1^0\gamma _1^\mu S_F(q_1)S_F^c(p_2)\gamma _2^\nu \gamma _2^0$ \\ 
$+\gamma _2^0\gamma _2^\mu S_F^c(q_2)S_F(p_1)\gamma _1^\nu \gamma _1^0\}$%
\end{tabular}
\eqnum{6.11}
\end{equation}
Employing the representation of fermion propagator denoted in Eqs. (6.5) and
(6.6) and noticing 
\begin{equation}
S^{-1}(P,k)=\gamma _1^0\gamma _2^0[\widehat{S}_F(q_1)+\widehat{S}%
_F^c(q_2)]^{-1}  \eqnum{6.12}
\end{equation}
one may exactly obtain from Eqs. (6.9)-(6.11) the expression denoted in Eq.
(6.3). Thus, the equivalence between the both expressions of the D-S kernel
derived in sections 3 and 4 is proved in the lowest order approximation.

Now let us focus on the three-dimensional t-channel OGEK which was derived
for the first time in Ref. [15, 16]. For comparison with the
four-dimensional kernel, it is necessary to give this kernel a further
description based on the closed expression formulated in Eqs. (3.40)-(3.47).
Analogous to the four-dimensional case, in the lowest order approximation,
only the first two terms in Eq. (3.40) can contribute to the
three-dimensional t-channel OGEK. Therefore, we can write 
\begin{equation}
K_t^0(\overrightarrow{x_1},\overrightarrow{x_2};\overrightarrow{y_1},%
\overrightarrow{y_2};t_1-t_2)=\int d^3z_1d^3z_2\sum_{i=1}^2H_i^t(%
\overrightarrow{x_1},\overrightarrow{x_2};\overrightarrow{z_1},%
\overrightarrow{z_2};t_1-t_2)S^{-1}(\overrightarrow{z_1},\overrightarrow{z_2}%
;\overrightarrow{y_1},\overrightarrow{y_2})  \eqnum{6.13}
\end{equation}
where $H_1^t(\overrightarrow{x_1},\overrightarrow{x_2};\overrightarrow{z_1},%
\overrightarrow{z_2};t_1-t_2)$ arises from Eqs. (3.41) and (3.42) with the
three-point Green's functions in Eq. (3.42 ) being given by 
\begin{equation}
\begin{tabular}{l}
$\Lambda _\mu ^a(\overrightarrow{x_2}\mid \overrightarrow{x_1},%
\overrightarrow{z_1};t_1-t_2)=-ig\int d^3udu_0\Delta _{\mu \nu }^{ab}(%
\overrightarrow{x_2}-\overrightarrow{u};t_1-u_0)$ \\ 
$\times S_F(\overrightarrow{x_1}-\overrightarrow{u};t_1-u_0)\gamma ^\nu
T^bS_F(\overrightarrow{u}-\overrightarrow{z_1};u_0-t_2)\}$%
\end{tabular}
\eqnum{6.14}
\end{equation}
\begin{equation}
\begin{tabular}{l}
$\Lambda _\mu ^{{\bf c}a}(\overrightarrow{x_1}\mid \overrightarrow{x_2},%
\overrightarrow{z_2};t_1-t_2)=-ig\int d^3udu_0\Delta _{\mu \nu }^{ab}(%
\overrightarrow{x_1}-\overrightarrow{u};t_1-u_0)$ \\ 
$\times S_F^c(\overrightarrow{x_2}-\overrightarrow{u};t_1-u_0)\gamma ^\nu 
\overline{T}^bS_F^c(\overrightarrow{u}-\overrightarrow{z_2};u_0-t_2)\}$%
\end{tabular}
\eqnum{6.15}
\end{equation}
and $H_2^t(\overrightarrow{x_1},\overrightarrow{x_2};\overrightarrow{z_1},%
\overrightarrow{z_2};t_1-t_2)$ is derived from Eqs. (3.43) and (3.44) when
the terms $i\Delta _{\mu \nu }^{ab}(\overrightarrow{x_i}-\overrightarrow{z_j}%
;t_1-t_2)S_F(\overrightarrow{x_1}-\overrightarrow{z_1};t_1-t_2)S_F^c(%
\overrightarrow{x_2}-\overrightarrow{z_2};t_1-t_2)$ included in ${\cal G}%
_{\mu \nu }^{ab}(\overrightarrow{x_i},\overrightarrow{z_j}\mid 
\overrightarrow{x_1}_{,}\overrightarrow{x_2};\overrightarrow{z_1},%
\overrightarrow{z_2};E)$ with $i\neq j$ are taken into account only. By the
Fourier transformation, it is not difficult to derive the following
expression 
\begin{equation}
K_t^0(\overrightarrow{P},\overrightarrow{q},\overrightarrow{k}%
,E)=\sum_{i=1}^2H_i^t(\overrightarrow{P},\overrightarrow{q},\overrightarrow{k%
},E)S^{-1}(\overrightarrow{P},\overrightarrow{k})  \eqnum{6.16}
\end{equation}
where 
\begin{equation}
\begin{tabular}{l}
$H_1^t(\overrightarrow{P},\overrightarrow{q},\overrightarrow{k},E)$ \\ 
$=-ig\int \frac{dq_0}{2\pi }\frac{dk_0}{2\pi }\{\Omega _1^{a\mu }\gamma
_1^0\Delta _{\mu \nu }^{ab}(\overrightarrow{q}-\overrightarrow{k}%
;q_0-k_0)S_F^c(\overrightarrow{p_2},q_0)\gamma ^\nu \overline{T}^bS_F^c(%
\overrightarrow{q_2},k_0)$ \\ 
$+\Omega _2^{a\mu }\gamma _2^0\Delta _{\mu \nu }^{ab}(\overrightarrow{k}-%
\overrightarrow{q};k_0-q_0)S_F(\overrightarrow{p_1},q_0)\gamma ^\nu T^bS_F(%
\overrightarrow{q_1},k_0)\}$%
\end{tabular}
\eqnum{6.17}
\end{equation}
and 
\begin{equation}
\begin{tabular}{l}
$H_2^t(\overrightarrow{P},\overrightarrow{q},\overrightarrow{k},E)$ \\ 
$=i\int \frac{dq_0}{2\pi }\frac{dk_0}{2\pi }\{\Omega _1^{a\mu }\Delta _{\mu
\nu }^{ab}(\overrightarrow{q}-\overrightarrow{k};E-q_0-k_0)S_F(%
\overrightarrow{q_1},k_0)S_F^c(\overrightarrow{p_2},q_0)\overline{\Omega }%
_2^{b\nu }$ \\ 
$+\Omega _2^{a\mu }\Delta _{\mu \nu }^{ab}(\overrightarrow{k}-%
\overrightarrow{q};E-k_0-q_0)S_F(\overrightarrow{p_1},q_0)S_F^c(%
\overrightarrow{q_2},k_0)\overline{\Omega }_1^{b\nu }\}$%
\end{tabular}
\eqnum{6.18}
\end{equation}
The integrals over $q_0$ and $k_0$ can easily be calculated by applying the
Cauchy theorem in the complex planes of $q_0$ and $k_0$. Since QCD is an
unitary theory, the matrix element of the kernel between the spinor wave
functions is independent of the gauge parameter. Therefore, we only need to
show the result given in the Feynman gauge. In this gauge, noticing the
representation of the gluon propagator 
\begin{equation}
\begin{tabular}{l}
$\Delta _{\mu \nu }^{ab}(Q)=-\frac{\delta _{ab}g_{\mu \nu }}{Q_0^2-%
\overrightarrow{Q}^2+i\varepsilon }$ \\ 
$=-\frac{\delta _{ab}g_{\mu \nu }}{2\left| \overrightarrow{Q}\right| }[\frac 
1{Q_0-\left| \overrightarrow{Q}\right| +i\varepsilon }-\frac 1{Q_0+\left| 
\overrightarrow{Q}\right| -i\varepsilon }]$%
\end{tabular}
\eqnum{6.19}
\end{equation}
and the expression shown in Eqs. (6.5) and (6.6), it can be found that 
\begin{equation}
\begin{tabular}{l}
$H_1^t(\overrightarrow{P},\overrightarrow{q},\overrightarrow{k},E)=\frac{%
ig^2T_1^a\overline{T}_2^a}{2\left| \overrightarrow{q}-\overrightarrow{k}%
\right| }\{\frac 1{\omega (\overrightarrow{p_2})+\omega (\overrightarrow{q_2}%
)+\left| \overrightarrow{q}-\overrightarrow{k}\right| }\gamma _1^0\gamma
_1^\mu [\Lambda ^{+}(\overrightarrow{p_2})\gamma _2^0\gamma _{2\mu }\Lambda
^{-}(\overrightarrow{q_2})$ \\ 
$+\Lambda ^{-}(\overrightarrow{p_2})\gamma _2^0\gamma _{2\mu }\Lambda ^{+}(%
\overrightarrow{q_2})]+\frac 1{\omega (\overrightarrow{p_1})+\omega (%
\overrightarrow{q_1})+\left| \overrightarrow{q}-\overrightarrow{k}\right| }$
\\ 
$\times \gamma _2^0\gamma _2^\mu [\Lambda ^{+}(\overrightarrow{p_1})\gamma
_1^0\gamma _{1\mu }\Lambda ^{-}(\overrightarrow{q_1})+\Lambda ^{-}(%
\overrightarrow{p_1})\gamma _1^0\gamma _{1\mu }\Lambda ^{+}(\overrightarrow{%
q_1})]\}\gamma _1^0\gamma _2^0$%
\end{tabular}
\eqnum{6.20}
\end{equation}
and 
\begin{equation}
\begin{tabular}{l}
$H_2^t(\overrightarrow{P},\overrightarrow{q},\overrightarrow{k},E)=\frac{%
ig^2T_1^a\overline{T}_2^a}{2\left| \overrightarrow{q}-\overrightarrow{k}%
\right| }\{\frac{\Lambda ^{+}(\overrightarrow{p_2})\gamma _2^0\gamma _{2\mu
}\gamma _1^0\gamma _1^\mu \Lambda ^{+}(\overrightarrow{q_1})}{E-\omega (%
\overrightarrow{p_2})-\omega (\overrightarrow{q_1})-\left| \overrightarrow{q}%
-\overrightarrow{k}\right| }$ \\ 
$+\frac{\Lambda ^{+}(\overrightarrow{p_1})\gamma _1^0\gamma _1^\mu \gamma
_2^0\gamma _{2\mu }\Lambda ^{+}(\overrightarrow{q_2})}{E-\omega (%
\overrightarrow{p_1})-\omega (\overrightarrow{q_2})-\left| \overrightarrow{q}%
-\overrightarrow{k}\right| }-\frac{\Lambda ^{-}(\overrightarrow{p_2})\gamma
_2^0\gamma _{2\mu }\gamma _1^0\gamma _1^\mu \Lambda ^{-}(\overrightarrow{q_1}%
)}{E+\omega (\overrightarrow{p_2})+\omega (\overrightarrow{q_1})+\left| 
\overrightarrow{q}-\overrightarrow{k}\right| }$ \\ 
$-\frac{\Lambda ^{-}(\overrightarrow{p_1})\gamma _1^0\gamma _1^\mu \gamma
_2^0\gamma _{2\mu }\Lambda ^{-}(\overrightarrow{q_2})}{E+\omega (%
\overrightarrow{p_1})+\omega (\overrightarrow{q_2})+\left| \overrightarrow{q}%
-\overrightarrow{k}\right| }\}\gamma _1^0\gamma _2^0$%
\end{tabular}
\eqnum{6.21}
\end{equation}
It is seen that $H_1^t(\overrightarrow{P},\overrightarrow{q},\overrightarrow{%
k},E)$ is actually independent of the energy E. We would like to emphasize
that the expressions in Eqs. (6.20) and (6.21) can more directly be obtained
from Eqs. (6.10) and (6.11) by the following integration 
\begin{equation}
H_i^t(\overrightarrow{P},\overrightarrow{q},\overrightarrow{k},E)=\int \frac{%
dq_0}{2\pi }\frac{dk_0}{2\pi }H_i^t(P,q,k),i=1,2.  \eqnum{6.22}
\end{equation}

Now, we discuss the inverse of the function $S(\overrightarrow{P},%
\overrightarrow{k})$ which is the Fourier transform of the function in Eq.
(2.50). The equal-time propagators can be defined in such a manner [34] 
\begin{equation}
\begin{tabular}{l}
$S_F(\overrightarrow{x}-\overrightarrow{y})=\frac 1{2i}\left\langle
0^{+}\left| \psi (\overrightarrow{x},t)\overline{\psi }(\overrightarrow{y}%
,t)-\overline{\psi }^T(\overrightarrow{y},t)\psi ^T(\overrightarrow{x}%
,t)\right| 0^{-}\right\rangle $ \\ 
$=\int \frac{d^3p}{(2\pi )^3}S_F(\overrightarrow{p})e^{i\overrightarrow{p}%
\cdot (\overrightarrow{x}-\overrightarrow{y})}$%
\end{tabular}
\eqnum{6.23}
\end{equation}
where 
\begin{equation}
S_F(\overrightarrow{p})=\frac 1{2i}\frac{h(\overrightarrow{p})}{\omega (%
\overrightarrow{p})}\gamma ^0  \eqnum{6.24}
\end{equation}
With this representation, the function $S(\overrightarrow{P},\overrightarrow{%
k})$ and its inverse, i.e. the three-dimensional counterparts of those in
Eqs. (6.4) and (6.12) will be written as 
\begin{equation}
S(\overrightarrow{P},\overrightarrow{k})=\frac 1{2i}[\frac{h(\overrightarrow{%
q_1})}{\omega (\overrightarrow{q_1})}+\frac{h(\overrightarrow{q_2})}{\omega (%
\overrightarrow{q_2})}]\gamma _1^0\gamma _2^0  \eqnum{6.25}
\end{equation}
and 
\begin{equation}
S^{-1}(\overrightarrow{P},\overrightarrow{k})=2i\gamma _1^0\gamma _2^0[\frac{%
h(\overrightarrow{q_1})}{\omega (\overrightarrow{q_1})}+\frac{h(%
\overrightarrow{q_2})}{\omega (\overrightarrow{q_2})}]^{-1}  \eqnum{6.26}
\end{equation}

When Eqs. (6.20), (6.21) and (6.26) are substituted into Eq. (6.16), one may
write out explicitly the expression of the three-dimensional t-channel OGEK.
On inserting this kernel into the first term of the effective interaction
Hamiltonian denoted in Eq. (5.25) and employing the orthogonality relations
of Dirac spinors and the Dirac equation, we are led to [15, 16] 
\begin{equation}
V_t^{(0)}(\overrightarrow{P},\overrightarrow{q},\overrightarrow{k;E}%
)=g^2T_1^a\overline{T}_2^a\Delta (\overrightarrow{q}-\overrightarrow{k};E)%
\overline{U}(\overrightarrow{p_1})\gamma _1^\mu U(\overrightarrow{q_1})%
\overline{U}(\overrightarrow{p_2})\gamma _{2\mu }U(\overrightarrow{q_2}) 
\eqnum{6.27}
\end{equation}
where $U(\overrightarrow{q})$ was represented in Eq. (5.1) and 
\begin{equation}
\begin{tabular}{l}
$\Delta (\overrightarrow{q}-\overrightarrow{k};E)=-\frac 1{2\left| 
\overrightarrow{q}-\overrightarrow{k}\right| }[\frac 1{E-\omega (%
\overrightarrow{p_2})-\omega (\overrightarrow{q_1})-\left| \overrightarrow{q}%
-\overrightarrow{k}\right| }$ \\ 
$+\frac 1{E-\omega (\overrightarrow{p_1})-\omega (\overrightarrow{q_2}%
)-\left| \overrightarrow{q}-\overrightarrow{k}\right| }]$%
\end{tabular}
\eqnum{6.28}
\end{equation}
is just the exact three-dimensional gluon propagator given in the Feynman
gauge which is off-shell because the energy E is off-shell. It is noted here
that the lowest order interaction Hamiltonian $V^{(0)}(\overrightarrow{P},%
\overrightarrow{q},\overrightarrow{k;E})$ is only given by the function $%
H_2^t(\overrightarrow{P},\overrightarrow{q},\overrightarrow{k},E)$ in Eq.
(6.21) because the function $H_1^t(\overrightarrow{P},\overrightarrow{q},%
\overrightarrow{k},E)$ in Eq. (6.20) gives a vanishing contribution to the
lowest order Hamiltonian.

\subsection{The s-channel one-gluon exchange kernel}

The four-dimensional s-channel OGEK was represented in Eq. (4.17). By means
of the charge conjugation of the quark field, the vertex in Eq. (4.17) can
be expressed as 
\begin{equation}
\overline{\Gamma }^{b\nu *}(z\mid z_1,z_2)_{\rho \sigma }=-C_{\sigma \lambda
}\Gamma ^{b\nu }(z\mid z_2,z_1)_{\lambda \rho }  \eqnum{6.29}
\end{equation}
With this relation and the expression shown in Eq. (6.1), the kernel in the
lowest-order approximation can be written as 
\begin{equation}
\begin{tabular}{l}
$K_s^0(x_1,x_2;z_1,z_2)_{\alpha \beta \rho \sigma }=ig\{\Delta _{\mu \nu
}^{ab}(x_1-z_1)(\Omega _1^{a\mu })_{\alpha \gamma }S_F^{*}(x_1-x_2)_{\gamma
\beta }$ \\ 
$+\Delta _{\mu \nu }^{ab}(x_2-z_2)(\Omega _2^{a\mu })_{\beta \lambda
}S_F^{*}(x_1-x_2)_{\alpha \lambda }\}(C\gamma ^\nu T^b)_{\rho \sigma }\delta
^4(z_1-z_2)$%
\end{tabular}
\eqnum{6.30}
\end{equation}
In the momentum space, it reads 
\begin{equation}
K_s^0(P,q,k)_{\alpha \beta \rho \sigma }=ig\Delta _{\mu \nu }^{ab}(P)L^{a\mu
}(P,q)_{\alpha \beta }(C\gamma ^\nu T^b)_{\rho \sigma }  \eqnum{6.31}
\end{equation}
where 
\begin{equation}
L^{a\mu }(P,q)_{\alpha \beta }=(\Omega _1^{a\mu })_{\alpha \gamma
}S_F^{*}(-p_2)_{\gamma \beta }+(\Omega _2^{a\mu })_{\beta \lambda
}S_F^{*}(p_1)_{\alpha \lambda }  \eqnum{6.32}
\end{equation}
in which 
\begin{equation}
S_F^{*}(p)=\int d^4xS_F^{*}(x)e^{-iqx}=S_F(p)C^{-1}=C^{-1}S_F^c(-p)^T 
\eqnum{6.33}
\end{equation}

Now, let us derive the above kernel from the closed expression in Eq.
(3.30). From the perturbative calculation, It can be found that in the
lowest order approximation, only the second term in Eq. (3.30) can
contribute to the s-channel OGEK because in the perturbative expansion of
the Green's function ${\cal G}_{\mu \nu }^{ab}(x_i,z_j\mid x_1,x_2;z_1,z_2)$%
, there is a term $-i\Delta _{\mu \nu }^{ab}(x_i-z_j)S_F^{*}(x_1-x_2)%
\overline{S}_F^{*}(z_1-z_2)$ which is merely related to the s-channel OGEK.
Thus, the terms in the ${\cal Q}(x_1,x_2;z_1,z_2)$ which contribute to the
s-channel OGEK, according to Eq. (3.32) can be written as 
\begin{equation}
H^s(x_1,x_2;z_1,z_2)=-\sum_{i,j=1}^2i\Delta _{\mu \nu }^{ab}(x_i-z_j)\Omega
_i^{a\mu }S_F^{*}(x_1-x_2)\overline{S}_F^{*}(z_1-z_2)\overline{\Omega }%
_j^{b\nu }  \eqnum{6.34}
\end{equation}
Substituting the above expression into Eq. (3.30), in the momentum space, we
have 
\begin{equation}
K_s^0(P,q,k)=H^s(P,q,k)S^{-1}(P,k)  \eqnum{6.35}
\end{equation}
where 
\begin{equation}
H^s(P,q,k)_{\alpha \beta \lambda \delta }=i\Delta _{\mu \nu }^{ab}(P)L^{a\mu
}(P,q)_{\alpha \beta }\overline{L}^{b\nu }(P,k)_{\lambda \delta } 
\eqnum{6.36}
\end{equation}
in which $L^{a\mu }(P,q)_{\alpha \beta }$ was given in (6.32) and 
\begin{equation}
\overline{L}^{b\nu }(P,k)_{\lambda \delta }=-[\overline{S}%
_F^{*}(-q_1)_{\lambda \tau }(\overline{\Omega }_2^{a\mu })_{\tau \delta }+%
\overline{S}_F^{*}(q_2)_{\tau \delta }(\overline{\Omega }_1^{a\mu })_{\tau
\lambda }]  \eqnum{6.37}
\end{equation}
here 
\begin{equation}
\overline{S}_F^{*}(q)=\int d^4x\overline{S}%
_F^{*}(x)e^{iqx}=C^{-1}S_F^c(q)=S_F^T(-q)C^{-1}  \eqnum{6.38}
\end{equation}
In light of the charge conjugation for the $\gamma $-matrix and for the
propagators shown in Eqs. (6.33) and (6.38), it is easy to find 
\begin{equation}
\overline{L}^{b\nu }(P,k)_{\lambda \delta }S^{-1}(P,k)_{\lambda \delta \rho
\sigma }=g(C\gamma ^\nu T^b)_{\rho \sigma }  \eqnum{6.39}
\end{equation}
With this relation, we see, the kernel in Eq. (6.35) is exactly equal to the
one written in Eq. (6.31). This gives a further proof of the equivalence
between the both expressions of the D-S kernel derived in sections 3 and 4.
By the charge conjugation, it is not difficult to find 
\begin{equation}
L^{a\mu }(P,q)_{\alpha \beta }=g\widehat{S}(P,q)_{\alpha \beta \lambda \tau
}(C\gamma ^\mu T^a)_{\lambda \tau }  \eqnum{6.40}
\end{equation}
where 
\begin{equation}
\widehat{S}(P,q)=\widehat{S}_F(p_1)+\widehat{S}_F^c(p_2)  \eqnum{6.41}
\end{equation}
Therefore, the kernel in Eq. (6.31) can be expressed as 
\begin{equation}
K_s^0(P,q,k)=ig^2\Delta _{\mu \nu }^{ab}(P)\widehat{S}(P,q)_{\alpha \beta
\lambda \tau }(C\gamma ^\mu T^a)_{\lambda \tau }(C\gamma ^\nu T^b)_{\rho
\sigma }  \eqnum{6.42}
\end{equation}

At the last of this section, we would like to discuss the three-dimensional
form of the s-channel OGEK. In accordance with Eq. (3.40), this kernel is
represented as 
\begin{equation}
K_s^0(\overrightarrow{x_1},\overrightarrow{x_2};\overrightarrow{y_1},%
\overrightarrow{y_2};t_1-t_2)=\int d^3z_1d^3z_2H^s(\overrightarrow{x_1},%
\overrightarrow{x_2};\overrightarrow{z_1},\overrightarrow{z_2}%
;t_1-t_2)S^{-1}(\overrightarrow{z_1},\overrightarrow{z_2};\overrightarrow{y_1%
},\overrightarrow{y_2})  \eqnum{6.43}
\end{equation}
where $H^s(\overrightarrow{x_1},\overrightarrow{x_2};\overrightarrow{z_1},%
\overrightarrow{z_2};t_1-t_2)$ can be written from Eq. (6.34) by setting $%
x_1^0=x_2^0=t_1$ and $z_1^0=z_2^0=t_2$ in the equal-time frame, that is 
\begin{equation}
H^s(\overrightarrow{x_1},\overrightarrow{x_2};\overrightarrow{z_1},%
\overrightarrow{z_2};t_1-t_2)=-i\sum_{i,j=1}^2\Delta _{\mu \nu }^{ab}(%
\overrightarrow{x}_i-\overrightarrow{z}_j;t_1-t_2)\Omega _i^{a\mu }S_F^{*}(%
\overrightarrow{x}_1-\overrightarrow{x}_2)\overline{S}_F^{*}(\overrightarrow{%
z}_1-\overrightarrow{z}_2)\overline{\Omega }_j^{b\nu }  \eqnum{6.44}
\end{equation}
In the momentum space, it is of the form 
\begin{equation}
H^s(\overrightarrow{P},\overrightarrow{q},\overrightarrow{k};E)_{\alpha
\beta \lambda \delta }=i\Delta _{\mu \nu }^{ab}(\overrightarrow{P},E)L^{a\mu
}(\overrightarrow{P},\overrightarrow{q})_{\alpha \beta }\overline{L}^{b\nu }(%
\overrightarrow{P},\overrightarrow{k})_{\lambda \delta }  \eqnum{6.45}
\end{equation}
where 
\begin{equation}
L^{a\mu }(\overrightarrow{P},\overrightarrow{q})_{\alpha \beta }=(\Omega
_1^{a\mu })_{\alpha \gamma }S_F^{*}(-\overrightarrow{p}_2)_{\gamma \beta
}+(\Omega _2^{a\mu })_{\beta \gamma }S_F^{*}(\overrightarrow{p}_1)_{\alpha
\gamma }  \eqnum{6.46}
\end{equation}
and 
\begin{equation}
\overline{L}^{b\nu }(\overrightarrow{P},\overrightarrow{k})_{\lambda \delta
}=-[\overline{S}_F^{*}(-\overrightarrow{q}_1)_{\lambda \tau }(\overline{%
\Omega }_2^{a\mu })_{\tau \delta }+\overline{S}_F^{*}(\overrightarrow{q}%
_2)_{\tau \delta }(\overline{\Omega }_1^{a\mu })_{\tau \lambda }] 
\eqnum{6.47}
\end{equation}
which are the three-dimensional form of the functions in Eqs. (6.32) and
(6.37). It is emphasized that in Eq. (6.45), only the gluon propagator is
dependent on energy E, while, the fermion propagators are
energy-independent. By the same charge conjugation transformations as shown
in Eqs. (6.33) and (6.38), one may obtain a kernel similar to Eq. (6.42) 
\begin{equation}
K_s^0(\overrightarrow{P},\overrightarrow{q},\overrightarrow{k};E)_{\alpha
\beta \rho \sigma }=ig^2\Delta _{\mu \nu }^{ab}(\overrightarrow{P},E)%
\widehat{S}(\overrightarrow{P},\overrightarrow{q})_{\alpha \beta \lambda
\tau }(C\gamma ^\mu T^a)_{\lambda \tau }(C\gamma ^\nu T^b)_{\rho \sigma } 
\eqnum{6.48}
\end{equation}
which may also be represented as 
\begin{equation}
K_s^0(\overrightarrow{P},\overrightarrow{q},\overrightarrow{k};E)_{\alpha
\beta \rho \sigma }=ig^2\Delta _{\mu \nu }^{ab}(\overrightarrow{P},E)S(%
\overrightarrow{P},\overrightarrow{q})_{\alpha \beta \lambda \tau }(\gamma
^\mu CT^a)_{\lambda \tau }(C\gamma ^\nu T^b)_{\rho \sigma }  \eqnum{6.49}
\end{equation}
where 
\begin{equation}
S(\overrightarrow{P},\overrightarrow{q})=\widehat{S}(\overrightarrow{P},%
\overrightarrow{q})\gamma _1^0\gamma _2^0  \eqnum{6.50}
\end{equation}
which is the three-dimensional form of Eq. (6.4).

In the P-S equation, similar to Eq. (6.27), the lowest order interaction
Hamiltonian given by the kernel in Eq. (6.48), according to Eq. (5.25), will
be written as 
\begin{equation}
V_s^{(0)}(\overrightarrow{P},\overrightarrow{q},\overrightarrow{k}%
;E)=g^2\Delta _{\mu \nu }^{ab}(\overrightarrow{P},E)U_\alpha (%
\overrightarrow{p_1})^{+}U_\beta (\overrightarrow{p_2})^{+}(C\gamma ^\mu
T^a)_{\alpha \beta }(C\gamma ^\nu T^b)_{\rho \sigma }U_\rho (\overrightarrow{%
q_1})U_\sigma (\overrightarrow{q_2})  \eqnum{6.51}
\end{equation}
It should be noted that the positive energy state Dirac spinors used here
were defined in Eq. (5.1). The negative energy state spinor may be given by
the charge conjugation relation: $V(\overrightarrow{p})=CU(\overrightarrow{p}%
)$ here $C=\gamma ^5\gamma ^0$ [30]. However, the matrix $C$ in Eq. (6.51)
is defined by $C=i\gamma ^2\gamma ^0$ [18]. Correspondingly, the charge
conjugation relation between the spinor wave functions is given by $v^s(%
\overrightarrow{p})=C\overline{u}^s(\overrightarrow{p})^T$ where $\overline{u%
}^s(\overrightarrow{p})=u^s(\overrightarrow{p})^{+}\gamma ^0$ with $u^s(%
\overrightarrow{p})$ and $v^s(\overrightarrow{p})$ being the positive and
negative energy spinor wave functions respectively [18] and represented as 
\begin{equation}
u^s(\overrightarrow{p})=\widetilde{U}(\overrightarrow{p})\varphi ^s(%
\overrightarrow{p}),v^s(\overrightarrow{p})=\widetilde{V}(\overrightarrow{p}%
)\chi ^s(\overrightarrow{p})  \eqnum{6.52}
\end{equation}
here $\varphi ^s(\overrightarrow{p})$ and $\chi ^s(\overrightarrow{p})$ are
the spin wave functions and 
\begin{equation}
\widetilde{U}(\overrightarrow{p})=\sqrt{\frac \omega m}U(\overrightarrow{p}),%
\widetilde{V}(\overrightarrow{p})=-\sqrt{\frac \omega m}\gamma ^0V(%
\overrightarrow{p})  \eqnum{6.53}
\end{equation}
Usually, the S-matrix element given by the kernel in Eq. (6.48) is
represented by 
\begin{equation}
T_s(\overrightarrow{P},\overrightarrow{q},\overrightarrow{k;E})=u_\alpha
^{s_1}(\overrightarrow{p_1})^{+}u_\beta ^{s_2}(\overrightarrow{p_2}%
)^{+}K_s^0(\overrightarrow{P},\overrightarrow{q},\overrightarrow{k}%
;E)_{\alpha \beta \rho \sigma }u_\rho ^{r_1}(\overrightarrow{q_1})u_\sigma
^{r_2}(\overrightarrow{q_2})  \eqnum{6.54}
\end{equation}
For later derivation, it is more convenient to use the expression of the
kernel written in Eq. (6.49). On inserting Eq. (6.49) into Eq. (6.54) and
noticing 
\begin{equation}
u_\alpha ^{s_1}(\overrightarrow{p_1})^{+}u_\beta ^{s_2}(\overrightarrow{p_2}%
)^{+}S(\overrightarrow{P},\overrightarrow{q})_{\alpha \beta \lambda \tau }=-i%
\overline{u}_\lambda ^{s_1}(\overrightarrow{p_1})\overline{u}_\tau ^{s_2}(%
\overrightarrow{p_2})  \eqnum{6.55}
\end{equation}
which is obtained by applying the Dirac equation and 
\begin{equation}
\begin{array}{c}
\overline{u}_\beta ^{s_2}(\overrightarrow{p_2})(C\gamma ^\mu T^a)_{\alpha
\beta }=(\gamma ^\mu T^a)_{\alpha \beta }v_\beta ^{s_2}(\overrightarrow{p_2}%
), \\ 
(C\gamma ^\nu T^b)_{\rho \sigma }u_\rho ^{r_1}(\overrightarrow{q_1})=%
\overline{v}_\rho ^{r_1}(\overrightarrow{q_1})(\gamma ^\nu T^b)_{\rho \sigma
},
\end{array}
\eqnum{6.56}
\end{equation}
one can get 
\begin{equation}
T_s(\overrightarrow{P},\overrightarrow{q},\overrightarrow{k;E})=g^2\Delta
_{\mu \nu }^{ab}(\overrightarrow{P},E)\overline{u}^{s_1}(\overrightarrow{p_1}%
)\gamma ^\mu T^av^{s_2}(\overrightarrow{p_2})\overline{v}^{r_1}(%
\overrightarrow{q_1})\gamma ^\nu T^bu^{r_2}(\overrightarrow{q_2}) 
\eqnum{6.57}
\end{equation}
This just is the S-matrix element for the one-gluon exchange interaction
taking place in the s-channel. It is easy to verify that the above matrix
element is independent of the gauge parameter. Therefore, we only need to
work in the Feynman gauge. In this gauge, 
\begin{equation}
\Delta _{\mu \nu }^{ab}(\overrightarrow{P},E)=\frac{\delta _{ab}g_{\mu \nu }%
}{E^2-\overrightarrow{P}^2+i\varepsilon }  \eqnum{6.58}
\end{equation}
With this propagator, as shown in Ref. [35], by the charge conjugation and
Fierz transformation, Eq. (6.57) can be represented in the form 
\begin{equation}
T_s(\overrightarrow{P},\overrightarrow{q},\overrightarrow{k;E})=\left( \frac{%
\omega _1\omega _2}{m_1m_2}\right) ^{^{1/2}}\varphi _{s_1}^{+}\varphi
_{s_2}^{+}V_s^{(0)}(\overrightarrow{P},\overrightarrow{q},\overrightarrow{k;E%
})\varphi _{r_1}\varphi _{r_2}  \eqnum{6.59}
\end{equation}
where 
\begin{equation}
V_s^{(0)}(\overrightarrow{P},\overrightarrow{q},\overrightarrow{k;E})=-\frac{%
g^2\widehat{C}_s\widehat{F}_s}{E^2-\overrightarrow{P}^2+i\varepsilon }%
\overline{U}(\overrightarrow{p_1})\overline{U}(\overrightarrow{p_2})\Gamma
_{12}U(\overrightarrow{q_1})U(\overrightarrow{q_2})  \eqnum{6.60}
\end{equation}
is the interaction Hamiltonian occurring in the P-S equation in which the
spinor is still defined in Eq. (5.1), $\widehat{C}_s$ is the color matrix 
\begin{equation}
\widehat{C}_s=\frac 1{24}(\lambda _1^a-\lambda _2^{a*})^2  \eqnum{6.61}
\end{equation}
with $\lambda _i^a$ being the Gell-Mann matrices, $\widehat{F}_s$ is the
flavor matrix which has an expression for flavor SU(2) such that

\begin{equation}
\widehat{F}_s=\frac 12(1-\overrightarrow{\tau _1}\cdot \overrightarrow{\tau
_2})  \eqnum{6.62}
\end{equation}
here $\overrightarrow{\tau _i}$ are isospin Pauli matrices and 
\begin{equation}
\Gamma _{12}=-I_1I_2+\gamma _1^5\gamma _2^5-\frac 12\gamma _1^\mu \gamma
_{2\mu }+\frac 12(\gamma _1^5\gamma _1^\mu )(\gamma _2^5\gamma _{2\mu }) 
\eqnum{6.63}
\end{equation}

In the end, it is pointed out that since the matrix element of the color
operator $\widehat{C}_s$ in the $q\overline{q}$ color singlet vanishes, the
s-channel OGEK contributes nothing to the $q\overline{q}$ bound states.
However, for many-quark-antiquark systems such as $\pi \pi $, $K\overline{K}$%
, $\pi N$, $KN$ systems and etc., the contribution of the s-channel OGEK is
not negligible and plays an important role to the interations taking place
in those systems.[36- 38].

\section{Discussions and remarks}

In this paper, the D-S equation satisfied by the $q\overline{q}$ bound
states has been derived from QCD and, especially, the interaction kernel in
the equation has been given two equivalent closed expressions which were
respectively derived by making use of the equations of motion obeyed by the
Green's functions and the irreducible decomposition of the Green's
functions. Since the B-S equation is commonly viewed as the correct equation
for the bound state problem, it is natural to ask what is the relation
between the D-S equation and the B-S equation? As shown in Ref. [5], the B-S
equation may be derived from the D-S equation. In fact, when applying the
Lehmann representation in Eq. (2.36) to Eqs. (2.16) and (2.17), by the same
procedure stated in section.2, one may obtain two D-S equations as follows 
\begin{equation}
(i{\bf \partial }_{x_1}{\bf -}m_1)\chi _{P\varsigma }(x_1,x_2)=\int
d^4y_1d^4y_2\widehat{K}_1(x_1,x_2;y_1,y_2)\chi _{P\varsigma }(y_1,y_2) 
\eqnum{7.1}
\end{equation}
\begin{equation}
(i{\bf \partial }_{x_2}{\bf -}m_2)\chi _{P\varsigma }(x_1,x_2)=\int
d^4y_1d^4y_2\widehat{K}_2(x_1,x_2;y_1,y_2)\chi _{P\varsigma }(y_1,y_2) 
\eqnum{7.2}
\end{equation}
where 
\begin{equation}
\widehat{K}_i(x_1,x_2;y_1,y_2)=\gamma _i^0K_i(x_1,x_2;y_1,y_2),\text{ }i=1,2
\eqnum{7.3}
\end{equation}
in which $K_i(x_1,x_2;y_1,y_2)$ was given in Eq. (3.28). The D-S equations
shown in Eqs. (2.38) and (2.39) may directly be written out from Eqs. (7.1)
and (7.2).

Operating on Eq. (7.1) with $(i{\bf \partial }_{x_2}{\bf -}m_2)$ or on Eq.
(7.2) with $(i{\bf \partial }_{x_1}{\bf -}m_1)$, we have 
\begin{equation}
(i{\bf \partial }_{x_1}{\bf -}m_1)(i{\bf \partial }_{x_2}{\bf -}m_2)\chi
_{P\varsigma }(x_1,x_2)=\int d^4y_1d^4y_2K_B(x_1,x_2;y_1,y_2)\chi
_{P\varsigma }(y_1,y_2)  \eqnum{7.4}
\end{equation}
where 
\begin{equation}
K_B(x_1,x_2;y_1,y_2)=(i{\bf \partial }_{x_1}{\bf -}m_1)\widehat{K}%
_2(x_1,x_2;y_1,y_2)=(i{\bf \partial }_{x_2}{\bf -}m_2)\widehat{K}%
_1(x_1,x_2;y_1,y_2)  \eqnum{7.5}
\end{equation}
is the B-S interaction kernel whose explicit expression was derived in Ref.
[5]. Acting on Eq. (7.4) with the inverse of the operator $(i{\bf \partial }%
_{x_1}{\bf -}m_1)(i{\bf \partial }_{x_2}{\bf -}m_2)$, the B-S equation will
be recast in an integral equation 
\begin{equation}
\chi _{P\varsigma }(x_1,x_2)=\int
d^4y_1d^4y_2d^4z_1d^4z_2S_F^{(0)}(x_1-z_1)S_F^{c(0)}(x_2-z_2)K_B(z_1,z_2;y_1,y_2)\chi _{P\varsigma }(y_1,y_2)
\eqnum{7.6}
\end{equation}
where $S_F^{(0)}(x_1-z_1)$ and $S_F^{c(0)}(x_2-z_2)$ are the free
propagators of quark and antiquark respectively. In the momentum space, it
becomes 
\begin{equation}
\chi _{P\varsigma }(q)=S_F^{(0)}(p_1)S_F^{c(0)}(p_2)\int \frac{d^4k}{(2\pi
)^4}K_B(P,q,k)\chi _{P\varsigma }(k)  \eqnum{7.7}
\end{equation}
Conversely, if we act on Eq. (7.6) with the operators $(i{\bf \partial }%
_{x_2}{\bf -}m_2)$ and $(i{\bf \partial }_{x_1}{\bf -}m_1)$ respectively,
the D-S equations in Eqs. (7.1) and (7.2) will be recovered. This shows that
to get the B-S equation, we need merely to consider the D-S equation. It
should be noted that the four-dimensional B-S equation can not directly be
transformed into the three-dimensional D-S equation. In order to obtain a
three-dimensional equation from the four-dimensional B-S equation, it is
necessary to introduce a certain constraint condition on the relative time
(or relative energy) as was done in an approximate manner such as the
instantaneous approximation or the quasipotential approaches [7-12].

As stated above, the four-dimensional D-S equation and the corresponding B-S
equation can be derived from one another. But, this does not mean that the
D-S equation and the B-S equation are fully equivalent to each other,
similar to the Dirac equation and the K-G equation which can also be derived
from each other. As seen from Eq. (7.4), the B-S equation is a kind of
second-order differential equation in the position space. Therefore, a
solution to the equation depends on not only the amplitude at time origin,
but also the time-differential of the amplitude at the time origin as in the
case for K-G equation. This probably is the origin that causes the B-S
equation to have the unphysical solutions of negative norm. In order to
exclude the unphysical solutions, as mentioned before, the common procedure
is to recast the four-dimensional B-S equation in a three-dimensional form
by eliminating the relative time (or the relative energy) from the equation.
Certainly, the three-dimensional equation, particularly, the exact version
of the equation presented in the sections 2 and 3 is much convenient to use
in solving the eigenvalue problem. However, since the three-dimensional
equation loses the Lorentz-covariance of a relativistic dynamics, it is
sometimes not suitable for carrying out extensive theoretical analyses, for
instance, to perform the irreducible decomposition of the Green's functions
contained in the kernel given in Eq. (3.30). The decomposition can readily
be done in the four-dimensional form as shown in section 4. At this point,
we may ask whether the relativistic bound state problem can be solved
Lorentz-covariantly in the Minkowski space without occurrence of the
unphysical solutions? The answer should be positive because the
Lorentz-covariance of the equation implies that one may work in any Lorentz
frame and gets the same result. Let us turn to the D-S equations shown in
Eqs. (2.41) and (2.42) which are represented in the position space. Clearly,
the equation in Eq.(2.42) is a first-order differential equation of
Schr\"odinger-type. One may first solve this equation to get an amplitude
which describes the evolution of the amplitude with the relative time $\tau $%
. and then substitute this amplitude into Eq.(2.41) to solve the eigenvalue $%
E$ and the amplitude $\chi _{P\varsigma }(x)$. In solving these equations,
we only need the initial conditions of the amplitude at the time origin
without concerning the initial conditions of the time-differentials of the
amplitude. Therefore, the unphysical solutions would not appear in this
case. In this sense, we can say, the D-S equation derived in this paper, as
it provides a new formulation of the relativistic equation for the two
fermion bound system, gives a suitable prescription to solve the
four-dimensional equation. Moreover, based on the relation denoted in Eq.
(7.5), the B-S kernel may conveniently be evaluated from the D-S kernel. In
comparison with the closed expression of the B-S kernel derived in Ref. [5],
the D-S kernel shown in Eq. (3.30) is rather simpler. The main contribution
to the D-S kernel is given by the Green's function ${\cal G}_{\mu \nu
}^{ab}(x_i,y_j\mid x_1,x_2;y_1,y_2)$ written in Eq. (3.14). While, the B-S
kernel concerns more complicated Green's functions such as [5] 
\begin{equation}
\begin{tabular}{l}
${\cal G}_{\mu \nu \lambda \tau }^{abcd}(x_1,x_2,y_1,y_2\mid
x_1,x_2;y_1,y_2) $ \\ 
=$\left\langle 0^{+}\left| T\{N[{\bf A}_\mu ^a(x_1){\bf A}_\nu ^b(x_2){\bf %
\psi }(x_1){\bf \psi }^c(x_2)]N[{\bf A}_\lambda ^c(y_1){\bf A}_\tau ^d(y_2)%
\overline{{\bf \psi }}(y_1)\overline{{\bf \psi }}^c(y_2)]\right|
0^{-}\right\rangle $%
\end{tabular}
\eqnum{7.8}
\end{equation}
which gives the major contribution to the B-S kernel. In particular, in
comparison of the four-dimensional kernel represented in Eqs. (3.30)-(3.33)
with the three-dimensional counterpart written in Eqs. (3.40-)-(3.43), we
see, there is an one-to-one correspondence between the both kernels.
Therefore, to calculate the three-dimensional kernel, one may first
calculate the four-dimensional one and then convert it to the
three-dimensional form according the correspondence relation between the
both of them. Since the four-dimensional D-S equation is Lorentz-covariant,
its kernel can conveniently be analyzed and calculated by means of the
familiar technique developed in the covariant quantum field theory.

In the end, we would like to address that unlike the Dyson-Schwinger
equation [39, 40] which contains an infinite set of equations, the D-S
equation\ derived in this paper is of a closed form with a closed expression
of the kernel as given in section 3 or section 4. The kernel can easily be
calculated by the perturbation method. For example, in the perturbative
calculation of the kernel given in section 3 which contains only a few types
of Green's functions, we only need the familiar perturbative expansions of
the Green's functions without concerning the calculation of other more-point
Green's functions as it is necessary to be done for the Dyson-Schwinger
equation. Especially, each of the Green's functions can be represented in
the form of functional integral and is possible to be evaluated by a
nonperturbative method as suggested, for example, by the lattice gauge
theory. Therefore, the expression of the kernel given in this paper provides
a new formalism for exploring the QCD nonperturbative effect and the quark
confinement which are important for the formation of a $q\overline{q}$ bound
state. In the ordinary quark potential model [3, 41], the quark confinement
is usually simulated by a linear potential which was suggested by the
lattice computation of a Wilson loop and by the area law [42-44]. Obviously,
this simulation is oversimplified. For the purpose of investigating the
quark confinement, it is appropriate to start from the kernel given in this
paper for the case that the quark and the antiquark have different flavors.
In this case, all the Green's functions become the ordinary ones as shown in
Eqs. (2.5), (2.13) and (3.11). Since the kernel derived in this paper
contains all the interactions taking place in the bound state and includes
the color-spin matrices $\Omega _i^{a\mu }$ defined in Eq. (2.23) in it, it
is anticipated that a nonperturbative calculation of this kernel would give
a sophisticated confining potential which includes not only its spatial
form, but also its spin and color structures. This just is the advantage of
the formalism of D-S equation presented in this paper.

\section{\bf Acknowledgment}

This project was supposed by National Natural Science Foundation of China.

\section{Appendix A: Derivation of equations of motion satisfied by the
Green's functions}

This appendix is used to derive the equations of motion satisfied by the
quark-antiquark two and four-point Green's functions. These equations may be
derived from the following QCD generating functional [18]. 
\begin{equation}
Z[J,\overline{\eta },\eta ,\overline{\xi },\xi ]=\frac 1N\int {\cal D(}A,%
\overline{\psi },\psi ,\overline{C},C)e^{iI}  \eqnum{A.1}
\end{equation}
where 
\begin{equation}
I=\int d^4x[{\cal L}+J^{a\mu }A_\mu ^a+\overline{\eta }\psi +\overline{\psi }%
\eta +\overline{\xi }C+\overline{C}\xi ]  \eqnum{A.2}
\end{equation}
in which ${\cal L}$ is the effective Lagrangian of QCD 
\begin{equation}
{\cal L}=\overline{\psi }(i{\bf \partial -}m+g{\bf A)}\psi -\frac 14F^{a\mu
\nu }F_{\mu \nu }^a-\frac 1{2\alpha }(\partial ^\mu A_\mu ^a)^2+\overline{C}%
^a\partial ^\mu (D_\mu ^{ab}C^b)  \eqnum{A.3}
\end{equation}
here ${\bf A=}\gamma ^\mu T^aA_\mu ^a$ with $A_\mu ^a$ being the vector
potentials of gluon fields, 
\begin{equation}
F_{\mu \nu }^a=\partial _\mu A_\nu ^a-\partial _\nu A_\mu ^a+gf^{abc}A_\mu
^bA_\nu ^c  \eqnum{A.4}
\end{equation}
are the strength tensors of the gluon field, 
\begin{equation}
D_\mu ^{ab}=\delta ^{ab}\partial _\mu -gf^{abc}A_\mu ^c  \eqnum{A.5}
\end{equation}
are the covariant derivatives, $\overline{C}^a,C^b$ represent the ghost
fields, and $J^{a\mu },\overline{\eta },\eta ,\overline{\xi }$ and $\xi $
denote the external sources coupled to the gluon, quark and ghost fields
respectively. By the charge conjugation transformations shown in Eq. (2.2)
for the quark fields and in the following for the external sources 
\begin{equation}
\eta ^c=C\overline{\eta }^T,\overline{\eta }^c=-\eta ^TC^{-1}  \eqnum{A.6}
\end{equation}
it is easy to prove the relation 
\begin{equation}
\overline{\psi }(i{\bf \partial -}m+g{\bf A)}\psi +\overline{\eta }\psi +%
\overline{\psi }\eta =\overline{\psi }^c(i{\bf \partial -}m+g\overline{{\bf A%
}}{\bf )}\psi ^c+\overline{\eta }^c\psi ^c+\overline{\psi }^c\eta ^c 
\eqnum{A.7}
\end{equation}
where $\overline{{\bf A}}=\gamma ^\mu \overline{T}^aA_\mu ^a$ .

\subsection{Equations of motion with respect to the coordinate $x_1$}

Upon taking the functional derivative of the generating functional in Eq.
(A.1) with respect to the field function $\overline{\psi }_\alpha (x_1)$ and
considering 
\begin{equation}
\frac{\delta Z}{\delta \overline{\psi }_\alpha (x_1)}=0  \eqnum{A.8}
\end{equation}
and 
\begin{equation}
\frac{\delta I}{\delta \overline{\psi }_\alpha (x_1)}=\eta _\alpha (x_1)+[(i%
{\bf \partial }_{x_1}-m_1)_{\alpha \gamma }+g{\bf A(}x_1)_{\alpha \gamma
}]\psi _\gamma (x_1)  \eqnum{A.9}
\end{equation}
it can be found that 
\begin{equation}
\{\eta _\alpha (x_1)+[(i{\bf \partial }_{x_1}-m_1)_{\alpha \gamma }+(\Gamma
^{a\mu })_{\alpha \gamma }\frac \delta {i\delta J^{a\mu }(x_1)}]\frac \delta
{i\delta \overline{\eta }_\gamma (x_1)}\}Z=0  \eqnum{A.10}
\end{equation}
where the fields $A_\mu ^a(x_1)$ and $\psi _\gamma (x_1)$ have been replaced
by the derivatives of the generating functional with respect to the sources $%
J^{a\mu }(x_1)$ and $\overline{\eta }_\gamma (x_1)$ and each of the
subscripts $\alpha ,\beta $ and $\gamma $ marks the components of color,
flavor and spinor. Differentiating Eq. (A.10) with respect to the source $%
\eta _\rho (y_1)$ and then setting all the sources to vanish, we obtain the
equation satisfied by the quark propagator [18] 
\begin{equation}
\lbrack (i{\bf \partial }_{x_1}-m_1+\Sigma )S_F]_{\alpha \rho
}(x_1,y_1)=\delta _{\alpha \rho }\delta ^4(x_1-y_1)  \eqnum{A.11}
\end{equation}
where 
\begin{equation}
\begin{tabular}{l}
$(\Sigma S_F)_{\alpha \rho }(x_1,y_1)\equiv \int d^4z_1\Sigma
(x_1,z_1)_{\alpha \gamma }S_F(z_1-y_1)_{\gamma \rho }$ \\ 
$=(\Gamma ^{a\mu })_{\alpha \gamma }\Lambda _\mu ^a(x_1\mid x_1,y_1)_{\gamma
\beta }$%
\end{tabular}
\eqnum{A.12}
\end{equation}
here $\Sigma (x_1,z_1)$ stands for the quark proper self-energy and $\Lambda
_\mu ^a(x_1\mid x_1,y_1)_{\gamma \rho }$ was defined in the first equality
of Eq. (3.11).

Let us turn to derive the equation of motion satisfied by the Green's
function defined in Eq. (2.1). In doing this, we need first to derive the
equations of motion obeyed by the Green's function defined in Eq. (2.5). By
successively differentiating Eq. (A.10) with respect to the sources $%
\overline{\eta }_\beta ^c(x_2),\eta _\rho (y_1)$ and $\eta _\sigma ^c(y_2)$,
noticing the equality in Eq. (A.7) and the following nonvanishing derivatives

\begin{equation}
\begin{tabular}{l}
$\frac{\delta \eta _\alpha ^c(x)}{\delta \overline{\eta }_\beta (y)}%
=C_{\alpha \beta }\delta ^4(x-y),\frac{\delta \overline{\eta }_\alpha ^c(x)}{%
\delta \eta _\beta (y)}=(C^{-1})_{\alpha \beta }\delta ^4(x-y),$ \\ 
$\frac{\delta \overline{\eta }_\alpha (x)}{\delta \eta _\beta ^c(y)}%
=(C^{-1})_{\alpha \beta }\delta ^4(x-y),\frac{\delta \eta _\alpha (x)}{%
\delta \overline{\eta }_\beta ^c(y)}=C_{\alpha \beta }\delta ^4(x-y)$%
\end{tabular}
\eqnum{A.13}
\end{equation}
we have 
\begin{equation}
\begin{tabular}{l}
$\{C_{\alpha \beta }\delta ^4(x_1-x_2)\frac \delta {i^3\delta \eta _\rho
(y_1)\delta \eta _\sigma ^c(y_2)}+\delta _{\alpha \rho }\delta ^4(x_1-y_1)%
\frac{\delta ^2}{i\delta \overline{\eta }_\beta ^c(x_2)\delta \eta _\sigma
^c(y_2)}$ \\ 
$+\eta _\alpha (x_1)\frac{\delta ^3}{i\delta \overline{\eta }_\beta
^c(x_2)\delta \eta _\rho (y_1)\delta \eta _\sigma ^c(y_2)}-[(i{\bf \partial }%
_{x_1}-m_1)_{\alpha \gamma }+(\Gamma ^{a\mu })_{\alpha \gamma }\frac \delta {%
i\delta J^{a\mu }(x_1)}]$ \\ 
$\times \frac{\delta ^4}{\delta \overline{\eta }_\gamma (x_1)\delta 
\overline{\eta }_\beta ^c(x_2)\delta \eta _\rho (y_1)\delta \eta _\sigma
^c(y_2)}\}Z=0$%
\end{tabular}
\eqnum{A.14}
\end{equation}
When all the sources are set to be zero, one immediately derives Eq. (2.10)
from Eq. (A.14).

It is noted that the first equation in Eq. (2.14) may directly derived from
Eq. (A.11) by the charge conjugation transformation represented in Eq. (2.2)
or by differentiating Eq. (A.10) with respect to the source $\overline{\eta }%
_\beta ^c(x_2)$.

\subsection{ Equations of motion with respect to the coordinate $x_{2}$}

When taking the derivative of the generating functional in Eq. (A.1) with
respect to the field variable $\overline{\psi }_\beta ^c(x_2)$ and noticing
the relation in Eq. (A.7), by the same procedure as described in Eqs.
(A.8)-(A.10), one may obtain 
\begin{equation}
\{\eta _\beta ^c(x_2)+[(i{\bf \partial }_{x_2}-m_2)_{\beta \lambda }+(%
\overline{\Gamma }^{b\nu })_{\beta \lambda }\frac \delta {i\delta J^{b\nu
}(x_2)}]\frac \delta {i\delta \overline{\eta }_\lambda ^c(x_2)}\}Z=0 
\eqnum{A.15}
\end{equation}
Differentiating Eq. (A.15) with respect to $\eta _\sigma ^c(y_2)$ and then
setting all the sources to be zero, one can get the equation for the
antiquark propagator 
\begin{equation}
\lbrack (i{\bf \partial }_{x_2}-m_2+\Sigma ^c)S_F^c]_{\beta \sigma
}(x_2,y_2)=\delta _{\beta \sigma }\delta ^4(x_2-y_2)  \eqnum{A.16}
\end{equation}
where 
\begin{equation}
\begin{tabular}{l}
$(\Sigma ^cS_F^c)_{\beta \sigma }(x_2,y_2)\equiv \int d^4z_2\Sigma
^c(x_2,z_2)_{\beta \lambda }S_F^c(z_2-y_2)_{\lambda \sigma }$ \\ 
$=(\overline{\Gamma }^{b\nu })_{\beta \lambda }\Lambda _\nu ^{cb}(x_2\mid
x_2,y_2)_{\lambda \sigma }$%
\end{tabular}
\eqnum{A.17}
\end{equation}
here $\Sigma ^c(x_2,z_2)$ is the antiquark proper self-energy and $\Lambda
_\nu ^{{\bf c}b}(x_2\mid x_2,y_2)_{\lambda \sigma }$ was represented in the
second equality of Eq. (3.12).

Similarly, Upon differentiating Eq. (A.15) with respect to the sources $%
\overline{\eta }_\alpha (x_1),\eta _\rho (y_1)$ and $\eta _\sigma ^c(y_2)$,
one gets 
\begin{equation}
\begin{tabular}{l}
$\{C_{\alpha \beta }\delta ^4(x_1-x_2)\frac \delta {i^3\delta \eta _\rho
(y_1)\delta \eta _\sigma ^c(y_2)}+\delta _{\beta \sigma }\delta ^4(x_2-y_2)%
\frac{\delta ^2}{i\delta \overline{\eta }_\alpha (x_1)\delta \eta _\rho (y_2)%
}$ \\ 
$+\eta _\beta ^c(x_2)\frac{\delta ^3}{i\delta \overline{\eta }_\alpha
(x_1)\delta \eta _\rho (y_1)\delta \eta _\sigma ^c(y_2)}-[(i{\bf \partial }%
_{x_2}-m_2)_{\beta \lambda }+(\overline{\Gamma }^{b\nu })_{\beta \lambda }%
\frac \delta {i\delta J^{b\nu }(x_2)}]$ \\ 
$\times \frac{\delta ^4}{\delta \overline{\eta }_\alpha (x_1)\delta 
\overline{\eta }_\lambda ^c(x_2)\delta \eta _\rho (y_1)\delta \eta _\sigma
^c(y_2)}\}Z=0$%
\end{tabular}
\eqnum{A.18}
\end{equation}
Setting all the sources to vanish, we directly obtain Eq. (2.11) from the
above equation.

It is mentioned that the second equation in Eq. (2.14) may directly be
derived from Eq. (A.16) by the charge conjugation transformation or by
differentiating Eq. (A.15) with respect to the source $\overline{\eta }%
_\alpha (x_1)$.

\subsection{Equations of motion with respect to the coordinate $y_{1}$}

By taking the derivative of the generating functional in Eq. (A.1) with
respect to the field variable $\psi _\rho (y_1),$ following the same
procedure as deriving Eq. (A.10), one may get 
\begin{equation}
\{\overline{\eta }_\rho (y_1)+\frac \delta {i\delta \eta _\tau (y_1)}[(i%
\overleftarrow{{\bf \partial }}_{y_1}+m_1)_{\tau \rho }-(\Gamma ^{a\mu
})_{\tau \rho }\frac \delta {i\delta J^{a\mu }(y_1)}]\}Z=0  \eqnum{A.19}
\end{equation}
On differentiating the above equation with respect to $\overline{\eta }%
_\alpha (x_1)$ and then turning off all the sources, we arrive at 
\begin{equation}
\lbrack S_F(i\overleftarrow{{\bf \partial }}_{y_1}+m_1-\Sigma )]_{\alpha
\rho }(x_1,y_1)=-\delta _{\alpha \rho }\delta ^4(x_1-y_1)  \eqnum{A.20}
\end{equation}
where 
\begin{equation}
\begin{tabular}{l}
$(S_F\Sigma )_{\alpha \rho }(x_1,y_1)\equiv \int d^4z_1S_F(x_1-z_1)_{\alpha
\tau }\Sigma (z_1,y_1)_{\tau \rho }$ \\ 
$=\Lambda _\mu ^a(y_1\mid x_1,y_1)_{\alpha \tau }(\Gamma ^{a\mu })_{\tau
\rho }$%
\end{tabular}
\eqnum{A.21}
\end{equation}

If we differentiate Eq. (A.19) with respect to $\eta _\sigma ^c(y_2)$, after
letting the sources to be vanishing, we get an equation satisfied by the
propagator $\overline{S}_F^{*}(y_1-y_2)$ as written in the first equation in
Eq. (3.3).

Now let us differentiate Eq. (A.19) with respect to $\overline{\eta }_\alpha
(x_1),\overline{\eta }_\beta ^c(x_2)$ and $\eta _\sigma ^c(y_2)$ and then
set all the sources but the source $J$ to be zero. By these operations, we
get 
\begin{equation}
\begin{tabular}{l}
$G(x_{1,}x_2;y_1,y_2)_{\alpha \beta \tau \sigma }^J(i\overleftarrow{{\bf %
\partial }}_{y_1}+m_1)_{\tau \rho }=-\delta _{\alpha \rho }\delta
^4(x_1-y_1)S_F^c(x_2-y_2)_{\beta \sigma }^J$ \\ 
$-C_{\rho \sigma }\delta ^4(y_1-y_2)S_F^{*}(x_1-x_2)_{\alpha \beta }^J+G_\mu
^a(y_1\mid x_1,x_2;y_1,y_2)_{\alpha \beta \tau \sigma }^J(\Gamma ^{a\mu
})_{\tau \rho }$%
\end{tabular}
\eqnum{A.22}
\end{equation}
where 
\begin{equation}
G(x_{1,}x_2;y_1,y_2)^J=\frac{\delta ^4Z[J,\overline{\eta },\eta ,\overline{%
\xi },\xi ]}{\delta \overline{\eta }(x_1)\delta \overline{\eta }%
^c(x_2)\delta \eta (y_1)\delta \eta ^c(y_2)}\mid _{\overline{\eta }=%
\overline{\eta }^c=\eta =\eta ^c=0}  \eqnum{A.23}
\end{equation}
\begin{equation}
S_F^c(x_2-y_2)_{\beta \sigma }^J=\frac{\delta ^2Z[J,\overline{\eta },\eta ,%
\overline{\xi },\xi ]}{i\delta \overline{\eta }_\beta ^c(x_2)\delta \eta
_\sigma ^c(y_2)}\mid _{\overline{\eta }=\eta =\overline{\xi }=\xi =0} 
\eqnum{A.24}
\end{equation}
\begin{equation}
S_F^{*}(x_1-x_2)_{\alpha \beta }^J==\frac{\delta ^2Z[J,\overline{\eta },\eta
,\overline{\xi },\xi ]}{i^3\delta \overline{\eta }_\alpha (x_1)\delta 
\overline{\eta }_\beta ^c(x_2)}\mid _{\overline{\eta }=\eta =\overline{\xi }%
=\xi =0}  \eqnum{A.25}
\end{equation}
and 
\begin{equation}
G_\mu ^a(y_i\mid x_{1,}x_2;y_1,y_2)^J=\frac \delta {i\delta J^{a\mu }(y_i)}%
G(x_{1,}x_2;y_1,y_2)^J,i=1,2  \eqnum{A.26}
\end{equation}
Once we set $J=0$, Eq. (A.22) will give rise to Eq. (3.1). Furthermore, if
we differentiate Eq. (A.22) with respect to $J^{a\mu }(x_i)$ and
subsequently set $J=0$, noticing 
\begin{equation}
G_{\mu \nu }^{ab}(x_i,y_j\mid x_{1,}x_2;y_1,y_2)=\frac \delta {i\delta
J^{a\mu }(x_i)i\delta J^{b\nu }(y_j)}G(x_{1,}x_2;y_1,y_2)^J\mid _{J=0} 
\eqnum{A.27}
\end{equation}
the equations in Eq. (3.8) will immediately be derived.

\subsection{Equations of motion with respect to the coordinate $y_{2}$}

To derive the equations of motion with respect to $y_2,$ we need to
differentiate the generating functional with respect to the field $\psi
_\sigma ^c(y_2).$ By the same procedure as formulated in Eqs. (A.8)-(A.10),
we get 
\begin{equation}
\{\overline{\eta }_\sigma ^c(y_2)+\frac \delta {i\delta \eta _\delta ^c(y_2)}%
[(i\overleftarrow{{\bf \partial }}_{y_2}+m_2)_{\delta \sigma }-(\overline{%
\Gamma }^{b\nu })_{\delta \sigma }\frac \delta {i\delta J^{b\nu }(y_2)}]\}Z=0
\eqnum{A.28}
\end{equation}
Then, the differentiation of the above equation with respect to $\overline{%
\eta }_\beta ^c(x_2)$ with setting all the sources to vanish subsequently
will lead us to 
\begin{equation}
\lbrack S_F^c(i\overleftarrow{{\bf \partial }}_{y_2}+m_2-\Sigma ^c)]_{\beta
\sigma }(x_2,y_2)=-\delta _{\beta \sigma }\delta ^4(x_2-y_2)  \eqnum{A.29}
\end{equation}
where 
\begin{equation}
\begin{tabular}{l}
$(S_F^c\Sigma ^c)_{\beta \sigma }(x_2,y_2)\equiv \int
d^4z_2S_F^c(x_2-z_2)_{\beta \delta }\Sigma ^c(z_2,y_2)_{\delta \sigma }$ \\ 
$=\Lambda _\nu ^{{\bf c}b}(y_2\mid x_2,y_2)_{\beta \delta }(\overline{\Gamma 
}^{b\nu })_{\delta \sigma }$%
\end{tabular}
\eqnum{A.30}
\end{equation}
Upon differentiating Eq. (A.28) with respect to $\eta _\rho (y_1)$ and then
turning off all the sources, one can obtain the second equation in Eq. (3.3)
for the propagator $\overline{S}_F^{*}(y_1-y_2)$

Let us differentiate Eq. (A.28) with respect to $\overline{\eta }_\alpha
(x_1),\overline{\eta }_\beta ^c(x_2)$ and $\eta _\rho (y_1)$ and set all the
sources except for the $J$ to vanish. As a result, we get 
\begin{equation}
\begin{tabular}{l}
$G(x_{1,}x_2;y_1,y_2)_{\alpha \beta \rho \delta }^J(i\overleftarrow{{\bf %
\partial }}_{y_2}+m_2)_{\delta \sigma }=-\delta _{\beta \sigma }\delta
^4(x_2-y_2)S_F(x_1-y_1)_{\alpha \rho }^J$ \\ 
$-C_{\rho \sigma }\delta ^4(y_1-y_2)S_F^{*}(x_1-x_2)_{\alpha \beta }^J+G_\nu
^b(y_2\mid x_1,x_2;y_1,y_2)_{\alpha \beta \rho \delta }^J(\overline{\Gamma }%
^{b\nu })_{\delta \sigma }$%
\end{tabular}
\eqnum{A.31}
\end{equation}
where 
\begin{equation}
S_F(x_1-y_1)_{\gamma \rho }^J=\frac{\delta ^2Z[J,\overline{\eta },\eta ,%
\overline{\xi },\xi ]}{i\delta \overline{\eta }_\gamma (x_1)\delta \eta
_\rho (y_1)}\mid _{\overline{\eta }=\eta =\overline{\xi }=\xi =0} 
\eqnum{A.32}
\end{equation}
and the other Green's functions in the presence of source $J$ were defined
before. When we set $J=0$, Eq. (A.31) straightforwardly yields Eq. (3.2).
Finally, on differentiating Eq. (A.31) with respect to $J^{a\mu }(x_i)$ and
subsequently setting $J=0$, Eq. (3.9) will be derived.

\section{Appendix B: One-particle irreducible decompositions of the
connected Green's functions}

Let us begin with the relation between the generating functional for full
Green's functions $Z[J,\overline{\eta },\eta ,\overline{\xi },\xi ]$ and the
one for connected Green's functions $W[J,\overline{\eta },\eta ,\overline{%
\xi },\xi ]$ [18, 33] 
\begin{equation}
Z[J,\overline{\eta },\eta ,\overline{\xi },\xi ]=\exp \{iW[J,\overline{\eta }%
,\eta ,\overline{\xi },\xi ]\}  \eqnum{B.1}
\end{equation}
Taking the derivatives of Eq. (B.1) with respect to the sources $\overline{%
\eta }(x_1),\overline{\eta }^c(x_2),\eta (y_1)$ and $\eta ^c(y_2)$ and then
setting all the sources except for the source $J$ to be zero, one may obtain
the following decomposition 
\begin{equation}
\begin{tabular}{l}
$%
G(x_{1,}x_2;y_1,y_2)^J=G_c(x_{1,}x_2;y_1,y_2)^J+S_F(x_1-y_1)^JS_F^c(x_2-y_2)^J 
$ \\ 
$-S_F^{*}(x_1-x_2)^J\overline{S}_F^{*}(y_1-y_2)^J$%
\end{tabular}
\eqnum{B.2}
\end{equation}
where $G(x_{1,}x_2;y_1,y_2)^J,$ $S_F^c(x_2-y_2)^J$, $S_F^{*}(x_1-x_2)^J$ and 
$S_F(x_1-y_1)^J$ were defined in Eqs. (A.23)-(A.25) and (A.32) respectively,
while, $G_c(x_{1,}x_2;y_1,y_2)^J$ and $\overline{S}_F^{*}(y_1-y_2)^J$ are
defined by 
\begin{equation}
G_c(x_{1,}x_2;y_1,y_2)^J=i\frac{\delta ^4W[J,\overline{\eta },\eta ,%
\overline{\xi },\xi ]}{\delta \overline{\eta }(x_1)\delta \overline{\eta }%
^c(x_2)\delta \eta (y_1)\delta \eta ^c(y_2)}\mid _{\overline{\eta }=%
\overline{\eta }^c=\eta =\eta ^c=0},  \eqnum{B.3}
\end{equation}
and 
\begin{equation}
\overline{S}_F^{*}(y_1-y_2)^J=\frac{\delta ^2W[J,\overline{\eta },\eta ,%
\overline{\xi },\xi ]}{i^2\delta \eta (y_1)\delta \eta ^c(y_2)}\mid _{%
\overline{\eta }=\eta =\overline{\xi }=\xi =0}  \eqnum{B.4}
\end{equation}
When we set $J=0$, Eq. (B.2) will go over to the decomposition shown in Eq.
(4.1). Differentiating Eq. (B.2) with respect to the source $J^{a\mu }(x_i)$%
, we have 
\begin{equation}
\begin{tabular}{l}
$G_\mu ^a(x_i\mid x_{1,}x_2;y_1,y_2)^J=G_{c\mu }^a(x_i\mid
x_{1,}x_2;y_1,y_2)^J$ \\ 
$+\Lambda _\mu ^a(x_i\mid x_1;y_1)^JS_F^c(x_1-y_1)^J+S_F(x_1-y_1)^J\Lambda
_\mu ^{{\bf c}a}(x_i\mid x_2;y_2)^J$ \\ 
$-\Lambda _\mu ^{a*}(x_i\mid x_1,x_2)^J\overline{S}%
_F^{*}(y_1-y_2)^J-S_F^{*}(x_1-x_2)^J\overline{\Lambda }_\mu ^{a*}(x_i\mid
y_1,y_2)^J$%
\end{tabular}
\eqnum{B.5}
\end{equation}
where $G_\mu ^a(x_i\mid x_{1,}x_2;y_1,y_2)^J$ was defined in Eq. (A.26) with 
$y_i$ being replaced by $x_i$ and the other functions are defined by 
\begin{equation}
G_{c\mu }^a(x_i\mid x_{1,}x_2;y_1,y_2)^J=\frac \delta {i\delta J^{a\mu }(x_i)%
}G_c(x_{1,}x_2;y_1,y_2)^J  \eqnum{B.6}
\end{equation}
\begin{equation}
\Lambda _\mu ^a(x_i\mid x_1;y_1)^J=\frac \delta {i\delta J^{a\mu }(x_i)}%
S_F(x_1-y_1)^J  \eqnum{B.7}
\end{equation}
\begin{equation}
\Lambda _\mu ^{{\bf c}a}(x_i\mid x_1;y_1)^J=\frac \delta {i\delta J^{a\mu
}(x_i)}S_F^c(x_1-y_1)^J  \eqnum{B.8}
\end{equation}
\begin{equation}
\Lambda _\mu ^{a*}(x_i\mid x_1,x_2)^J=\frac \delta {i\delta J^{a\mu }(x_i)}%
S_F^{*}(x_1-x_2)^J  \eqnum{B.9}
\end{equation}
and 
\begin{equation}
\overline{\Lambda }_\mu ^{a*}(x_i\mid y_1,y_2)^J=\frac \delta {i\delta
J^{a\mu }(x_i)}\overline{S}_F^{*}(y_1-y_2)^J  \eqnum{B.10}
\end{equation}
Upon setting $J=0,$ Eq. (B.5) immediately gives rise to the decomposition in
Eq. (4.3).

Now, let us proceed to carry out one-particle-irreducible decompositions of
the connected Green's functions on the RHS of Eq. (4.4). The decompositions
are easily performed with the help of the Legendre transformation which is
described by the relation between the generating functional of proper
vertices $\Gamma $ and the one for connected Green's functions $W$ [18, 33] 
\begin{equation}
\Gamma [A_\mu ^a,\overline{\psi },\psi ,\overline{C}^a,C^a]=W[J,\overline{%
\eta },\eta ,\overline{\xi },\xi ]-\int d^4x[J^{a\mu }A_\mu ^a+\overline{%
\eta }\psi +\overline{\psi }\eta +\overline{\xi }C+\overline{C}\xi ] 
\eqnum{B.11}
\end{equation}
and the relations between the field functions and the external sources 
\begin{equation}
\psi (x)=\frac{\delta W}{\delta \overline{\eta }(x)},\overline{\psi }(x)=-%
\frac{\delta W}{\delta \eta (x)},A_\mu ^a(x)=\frac{\delta W}{\delta J^{a\mu
}(x)},C^a(x)=\frac{\delta W}{\delta \overline{\xi }^a(x)},\overline{C}^a(x)=-%
\frac{\delta W}{\delta \xi ^a(x)}  \eqnum{B.12}
\end{equation}
\begin{equation}
\eta (x)=-\frac{\delta \Gamma }{\delta \overline{\psi }(x)},\overline{\eta }%
(x)=\frac{\delta \Gamma }{\delta \psi (x)},J_\mu ^a(x)=-\frac{\delta \Gamma 
}{\delta A^{a\mu }(x)},\xi ^a(x)=-\frac{\delta \Gamma }{\delta \overline{C}%
^a(x)},\overline{\xi }^a(x)=\frac{\delta \Gamma }{\delta C^a(x)} 
\eqnum{B.13}
\end{equation}
where the field functions in Eq. (B.12) are all functionals of the external
sources in Eq. (B.13) and, simultaneously, the sources in Eq. (B.13) are all
functionals of the field functions in Eq. (B.12).

Taking the derivative of the both sides of the first equality in Eq. (B.12)
with respect to $\psi (y)$ and employing the first relation in Eq. (B.13),
one may get 
\begin{equation}
\int d^4z\frac{\delta ^2\Gamma }{\delta \psi (y)\delta \overline{\psi }(z)}%
\frac{\delta ^2W}{\delta \eta (z)\delta \overline{\eta }(x)}=\int d^4z\frac{%
\delta ^2W}{\delta \overline{\eta }(x)\delta \eta (z)}\frac{\delta ^2\Gamma 
}{\delta \overline{\psi }(z)\delta \psi (y)}=-\delta ^4(x-y)  \eqnum{B.14}
\end{equation}
where we only keep the term on the RHS of Eq. (B.14) which is nonvanishing
when the sources are set to vanish. In order to find the
one-particle-irreducible decomposition for the quark-gluon three-point
Green's functions, one may differentiate Eq. (B.14) with respect to the
source $J^{a\mu }(x_i)$ and then using Eq. (B.14) once again. By this
procedure, it can be derived that 
\begin{equation}
\begin{tabular}{l}
$\frac{\delta ^3W}{\delta J^{a\mu }(x_i)\delta \overline{\eta }(x_j)\delta
\eta (y_k)}=\int d^4zd^4u_1d^4u_2\frac{\delta ^2W}{\delta J^{a\mu
}(x_i)\delta J^{b\nu }(u_1)}\frac{\delta ^2W}{\delta \overline{\eta }%
(x_j)\delta \eta (u_2)}$ \\ 
$\frac{\delta ^3\Gamma }{\delta A_\nu ^b(u_1)\delta \overline{\psi }%
(u_2)\psi (z)}\frac{\delta ^2W}{\delta \overline{\eta }(z)\delta \eta (y_k)}$%
\end{tabular}
\eqnum{B.15}
\end{equation}
where the coordinates in Eq. (B.14) have been appropriately changed. When
all the sources are set to be zero, noticing the definitions given in Eq.
(A.32) where the $Z$ is replaced by $iW$ and in Eq. (B.7) as well as 
\begin{equation}
\Delta _{\mu \nu }^{ab}(x_i-y_j)=\frac{\delta ^2W}{i^2\delta J^{a\mu
}(x_i)\delta J^{b\nu }(y_j)}\mid _{J=0}  \eqnum{B.16}
\end{equation}
\begin{equation}
\Gamma ^{b\nu }(u_1\mid u_2,z)=i\frac{\delta ^3\Gamma }{\delta A_\nu
^b(u_1)\delta \overline{\psi }(u_2)\delta \psi (z)}\mid _{A=\overline{\psi }%
=\psi =0}  \eqnum{B.17}
\end{equation}
the decomposition shown in Eqs. (4.5) and (4.6) straightforwardly follows
from Eq. (B.15). Analogously, if we replace $\overline{\eta }(x_j)$ and $%
\eta (y_k)$ by $\overline{\eta }^c(x_j)$ and $\eta ^c(y_k)$ in Eq. (B.15)
and noticing 
\begin{equation}
\Gamma _c^{b\nu }(u_1\mid u_2,z)=i\frac{\delta ^3\Gamma }{\delta A_\nu
^b(u_1)\delta \overline{\psi }^c(u_2)\delta \psi ^c(z)}\mid _{A=\overline{%
\psi }=\psi =0}  \eqnum{B.18}
\end{equation}
the decomposition shown in Eq. (4.8) and (4.9) will be derived. This
decomposition may also be derived from Eq. (B.15) by the charge conjugation
transformation for the quark fields. By this transformation, one may readily
derive from Eq. (B.15) the decomposition denoted in Eq. (4.15) in which the
gluon-quark-antiquark vertex is defined by 
\begin{equation}
\overline{\Gamma }^{b\nu *}(z\mid z_1,z_2)=i\frac{\delta ^3\Gamma }{\delta
A_\nu ^b(z)\delta \psi (z_1)\delta \psi ^c(z_2)}\mid _{A=\overline{\psi }%
=\psi =0}  \eqnum{B.19}
\end{equation}

The one-particle-irreducible decomposition of the connected Green's function 
$G_c(x_{1,}x_2;y_1,y_2)$ can be derived by the same procedure as obtaining
Eq. (B.15). On differentiating Eq. (B.14) with respect to $\overline{\eta }%
^c(x_2)$ and $\eta ^c(y_2)$ and setting all the sources but the source $J$
to vanish, one may obtain 
\begin{equation}
\begin{tabular}{l}
$G_c(x_{1,}x_2;y_1,y_2)^J=\int
\prod\limits_{i=1}^2d^4u_id^4v_iS_F(x_1-u_1)^JS_F^c(x_2-u_2)^J$ \\ 
$\times \Gamma (u_1,u_2;v_1,v_2)^JS_F(v_1-y_1)^JS_F^c(v_2-y_2)^J$%
\end{tabular}
\eqnum{B.20}
\end{equation}
where the four-point connected Green's function and the propagators given in
the presence of the sources were defined before and the function $\Gamma
(u_1,u_2;v_1,v_2)^J$ is formally the same as that defined in Eqs.
(4.19)-(4.21). When the source $J$ is turned off, Eq. (B.20) directly goes
over to the decomposition in Eq. (4.18) with the vertices in Eqs.
(4.19)-(4.21) being defined in Eqs. (B.17)-(B.19) and in the following 
\begin{equation}
\Gamma ^{b\nu *}(x_1\mid u_1,v_1)=i\frac{\delta ^3\Gamma }{\delta A_\nu
^b(x_1)\delta \overline{\psi }(u_1)\delta \overline{\psi }^c(v_2)}\mid _{A=%
\overline{\psi }=\psi =0}  \eqnum{B.21}
\end{equation}
which is the charge conjugate to the vertex $\overline{\Gamma }^{b\nu
*}(z\mid z_1,z_2)$ as well as 
\begin{equation}
\Gamma _3(u_1,u_2;v_1,v_2)=i\frac{\delta ^4\Gamma }{\delta \overline{\psi }%
(u_1)\delta \overline{\psi }^c(u_2)\delta \psi (v_1)\delta \psi ^c(v_2)}\mid
_{\overline{\psi }=\psi =\overline{\psi }^c=\psi ^c=0}  \eqnum{B.22}
\end{equation}
which is the quark-antiquark four-line proper vertex. It is emphasized here
that the decomposition of the function $G_c(x_{1,}x_2;y_1,y_2)$ in the
absence of the source $J$ has the same form as that given in the presence of 
$J$. This is because the Green's function is defined only by the
differentials with respect to the fermion fields as indicated in Eq. (B.3).

The one-particle irreducible decomposition of the Green's function $G_{c\mu
}^a(x_i\mid x_{1,}x_2;y_1,y_2)$ may be derived by starting from the
expression given in Eq. (B.15) with $j,$ $k=1$. By differentiating the both
sides of Eq. (B.15) with respect to the sources $\overline{\eta }^c(x_2)$
and $\eta ^c(y_2)$ and then turning off all the external sources, one may
obtain the decomposition of the function $G_{c\mu }^a(x_i\mid
x_{1,}x_2;y_1,y_2)$ as shown in Eqs. (4.22)-(4.25). Alternatively, the
decomposition may also be obtained by starting with the expression written
in Eq. (B.20). Substituting Eq. (B.20) into Eq. (B.6), then completing the
differentiation with respect to the source $J^{a\mu }(x_i)$ and finally
setting the source to vanish, one may also derive the irreducible
decomposition of the function $G_{c\mu }^a(x_i\mid x_{1,}x_2;y_1,y_2)$. In
doing this, it is necessary to perform the differentiations of the fermion
propagators with respect to the source $J^{a\mu }(x_i)$ as shown in Eqs. (
B.7)-(B.10) and use their decompositions presented in Eqs. (4.5)-(4.9). In
addition, we need to carry out the differentiations of the gluon propagator
and some vertices with respect to the source $J^{a\mu }(x_i)$ as shown
below. For the gluon propagator defined in Eq. (B.16), from its
representation in presence of the external source $J$ ,in the same way as
deriving the decomposition represented in Eqs. (B.15), ( 4.5) and (4.6), one
may obtain the one-particle irreducible decomposition of the gluon
three-point Green's function as follows: 
\begin{equation}
\Lambda _{\mu \rho \sigma }^{acd}(x_i,z_1,z_2)=\frac \delta {\delta J^{a\mu
}(x_i))}\Delta _{\rho \sigma }^{cd}(z_1-z_2)^J\mid _{J=0}=\int d^4zD_{\mu
\nu }^{ab}(x_i-z)\Pi _{\rho \sigma }^{bcd,\nu }(z,z_1,z_2)  \eqnum{B.23}
\end{equation}
where $D_{\mu \nu }^{ab}(x_i-z)=i\Delta _{\mu \nu }^{ab}(x_i-z)$ and $\Pi
_{\rho \sigma }^{bcd,\nu }(z,z_1,z_2)$ was represented in Eq. (4.32) with
the gluon three-line proper vertex defined by 
\begin{equation}
\Gamma _{bcd}^{\nu \rho \sigma }(z,u_1,u_2)=i\frac{\delta ^3\Gamma }{\delta
A_\nu ^b(z)\delta A_\rho ^c(u_1)\delta A_\sigma ^d(u_2)}\mid _{A=0} 
\eqnum{B.24}
\end{equation}
For a proper vertex $\Gamma _\alpha (z_1,z_2,\cdot \cdot \cdot )$ with $%
\alpha $ marking the other indices, its derivative with respect to the
source $J^{a\mu }(x_i)$ can be represented as 
\begin{equation}
\frac \delta {i\delta J^{a\mu }(x_i))}\Gamma _\alpha (z_1,z_2,\cdot \cdot
\cdot )^J\mid _{J=0}=\int d^4zD_{\mu \nu }^{ab}(x_i-z)\Gamma _\alpha ^{b\nu
}(z,z_1,z_2,\cdot \cdot \cdot )  \eqnum{B.25}
\end{equation}
where 
\begin{equation}
\Gamma _\alpha ^{b\nu }(z,z_1,z_2,\cdot \cdot \cdot )=\frac \delta {\delta
A_\nu ^b(z)}\Gamma _\alpha (z_1,z_2,\cdot \cdot \cdot )^J\mid _{J=0} 
\eqnum{B.26}
\end{equation}
According to the procedure stated above, it is not difficult to derive the
expressions described in Eqs. (4.30)-(4.33). In the expressions, the
vertices are defined as follows:

\begin{equation}
\Gamma ^{a\mu }(x_i\mid u_1,u_2;v_1,v_2)=\frac{\delta \Gamma
(u_1,u_2;v_1,v_2)^J}{i\delta J^{a\mu }(x_i)}\mid _{J=0}  \eqnum{B.27}
\end{equation}
\begin{equation}
\Gamma _{\nu \lambda }^{bc}(z,z_1\mid u_1,v_1)=i\frac{\delta ^4\Gamma }{%
\delta A^{b\nu }(z)\delta A^{c\lambda }(z_1)\delta \overline{\psi }%
(u_1)\delta \psi (v_1)}\mid _{A=\overline{\psi }=\psi =0}  \eqnum{B.28}
\end{equation}
\begin{equation}
\Gamma _{{\bf c}\nu \lambda }^{bc}(z,z_1\mid u_1,v_1)=i\frac{\delta ^4\Gamma 
}{\delta A^{b\nu }(z)\delta A^{c\lambda }(z_1)\delta \overline{\psi }%
^c(u_1)\delta \psi ^c(v_1)}\mid _{A=\overline{\psi }^c=\psi ^c=0} 
\eqnum{B.29}
\end{equation}
\begin{equation}
\Gamma _{\nu \lambda }^{bc*}(z,z_1\mid u_1,v_1)=i\frac{\delta ^4\Gamma }{%
\delta A^{b\nu }(z)\delta A^{c\lambda }(z_1)\delta \overline{\psi }%
(u_1)\delta \overline{\psi }^c(v_1)}\mid _{A=\overline{\psi }=\overline{\psi 
}^c=0}  \eqnum{B.30}
\end{equation}
\begin{equation}
\overline{\Gamma }_{{\bf c}\nu \lambda }^{bc^{*}}(z,z_1\mid u_1,v_1)=i\frac{%
\delta ^4\Gamma }{\delta A^{b\nu }(z)\delta A^{c\lambda }(z_1)\delta \psi
(u_1)\delta \psi ^c(v_1)}\mid _{A=\psi =\psi ^c=0}  \eqnum{B.31}
\end{equation}
Particularly, by the following differentiation 
\begin{equation}
\Gamma _3^{a\mu }(x_i\mid u_1,u_2;v_1,v_2)=\frac{\delta \Gamma
_3(u_1,u_2;v_1,v_2)^J}{i\delta J^{a\mu }(x_i)}\mid _{J=0}  \eqnum{B.32}
\end{equation}
it is easy to give the expression in Eq. (4.34) in which

\begin{equation}
\widehat{\Gamma }^{b\nu }(z\mid u_1,u_2;v_1,v_2)=i\frac{\delta ^4\Gamma }{%
\delta A^{b\nu }(z)\delta \overline{\psi }(u_1)\delta \overline{\psi }%
^c(u_2)\delta \psi (v_1)\delta \psi ^c(v_2)}\mid _{A=\overline{\psi }=\psi =%
\overline{\psi }^c=\psi ^c=0}  \eqnum{B.33}
\end{equation}
is the gluon-quark-antiquark five-line proper vertex.


\section{References}

\begin{references}
\bibitem{}  E. E. Salpeter and H. A. Bethe, Phys. Rev. 84 (1951) 1232.

\bibitem{}  M. Gell-Mann and F. E. Low, Phys. Rev. 84 (1951) 350.

\bibitem{}  W. Lucha, F. F. Sch\"oberl and D. Gromes, Phys. Rep. 200 (1991)
127, many references concerning the $q\overline{q}$ bound state B-S equation
can be found therein.

\bibitem{}  K. Erkelenz, Phys. Rep. 13 (1974) 191.

\bibitem{}  J. C. Su, Commun. Theor. Phys. 38 (2002) 433.

\bibitem{}  E. E. Salpeter, Phys. Rev. 87 (1952) 328.

\bibitem{}  A. A. Logunov and A. N. Tavkhelidze, Nuovo Cimento 29 (1963) 380.

\bibitem{}  Y. A. Alessandrini and R. L. Omnes, Phys. Rev. 139 (1965) 167.

\bibitem{}  R. Blankenbecler and R. Sugar, Phys. Rev. 142 (1966) 1051.

\bibitem{}  F. Gross, Phys. Rev. 186 (1969) 1448.

\bibitem{}  I. T. Todorov, Phys. Rev. D3 (1971) 2351.

\bibitem{}  J. Bijtebier and J. Broekaert, J. Phys. G: Nucl. Part. Phys. 22
(1996) 559.

\bibitem{}  T. N. Ruan, H. Q. Zhu, T. X. Ho, C. R. Qing and W. Q. Chao,
Phys. Energ. Fortis.\ Phys. Nucl. 5 (1981) 393, 537.

\bibitem{}  S. S. Wu, J. Phys. G: Nucl. Part. Phys. 16 (1990) 1447.

\bibitem{}  J. C. Su and D. Z. Mu, Commun. Theor. Phys. 15 (1991) 437.

\bibitem{}  Y. B. Dong, J. C. Su and S. S. Wu, J. Phys. G: Nucl. Part. Phys.
18 (1992) 75.

\bibitem{}  N. Nakanishi, Prog. Theor. Phys. Suppl. 42 (1959) 1 ; Prog.
Theor. Phys. Suppl. 95 (1988) 1,\ A great deal of references are cited
therein.

\bibitem{}  C. Itzykson and J. B. Zuber, Quantum Field Theory, McGraw-Hill,
New York, 1980.

\bibitem{}  A. Messiah, Quantum Mechanics, vol. II, North-Holland,
Amsterdam, 1962.

\bibitem{}  A. Komar, Phys. Rev. D18 (1978) 1881, 1887.

\bibitem{}  M. King and F. Rohrlich, Phys. Rev. Lett. 44 (1980) 62.

\bibitem{}  L. P. Horowitz and F. Rohrlich, Phys. Rev. D24 (1981) 1528.

\bibitem{}  H. Sazdjian, Phys. Lett. 156B (1985) 381.

\bibitem{}  H. Sadzjian, Phys. Rev. D33 (1986) 3401; 3425; 3435.

\bibitem{}  H. W. Crater and P. Van Alstine, Phys. Rev. Lett. 53 (1984) 1527.

\bibitem{}  H. W. Crater and P. Van Alstine, Phys. Rev. D36 (1987) 3007; D37
(1988) 1982.

\bibitem{}  R. L. Becker, C. Y. Wong and P. Van Alstine, Phys. Rev. D46
(1992) 5117.

\bibitem{}  J. Bijtebier, Nuovo Cimento A 100 (1988) 91; A102 (1989) 1235;
A103 (1990) 317; 639; 669.

\bibitem{}  A. M. Dirac, Lectures on Quantum Mechanics, Yeshiva University,
Belfer Graduate School of Science, New York, 1964.

\bibitem{}  J. C. Su, Commun. Theor. Phys. 15, 229 (1991); 18, 327 (1992).

\bibitem{}  H. Lehmann, Nuovo Cimento 11 (1954) 342.

\bibitem{}  K. Nishijima, Phys. Rev. 111 (1958) 995; R. Haag, Phys. Rev. 112
(1958) 669.

\bibitem{}  E. S. Abers and B. W. Lee, Phys. Rep. C9 (1973) 1.

\bibitem{}  D. Lurie, Particles and Fields, Interscience Publishers, a
divison of John Viley \& Sons, New York, 1968.

\bibitem{}  J. C. Su, Z. Q. Chen and S. S. Wu, Nucl. Phys. A524 (1991) 615.

\bibitem{}  G. Q. Zhao, X. G. Jing and J. C. Su, Phys. Rev.{\it \ }D58{\bf \ 
}(1998) 117503.

\bibitem{}  J. X. Chen, Y. H. Cao and J. C. Su, Phys. Rev. C64{\bf \ }(2001)
065201.

\bibitem{}  H. J. Wang, H. Yang and J. C. Su, Phys. Rev. C68 (2003) 055204.

\bibitem{}  F. J. Dyson, Phys. Rev. 75 (1949) 1736.

\bibitem{}  J. S. Schwinger, Proc. Nat. Acad. Sc. 37 (1951) 452, 455; Phys.
Rev. 125 (1962) 397; 128 (1962) 2425.

\bibitem{}  E. Eichten, K. Gottfried, T. Konishita, K. D. Lane and T. M Yan,
Phys. Rev. Lett. 34 (1975) 369 ; Phys. Rev. D21 (1980) 511 .

\bibitem{}  K. Wilson, Phys. Rev. D10 (1974) 2445.

\bibitem{}  L. S. Brown and W. I. Weisberger, Phys. Rev. D20 (1979) 3239.

\bibitem{}  E. Eichten and F. Feinberg, Phys. Rev. D23 (1981) 2724.
\end{references}
\end{document}